\documentclass[epj,nopacs]{svjour}
\usepackage{graphicx}
\usepackage{amssymb}
\usepackage{amsmath,mathrsfs,threeparttable,cancel,url,subfig}

\usepackage{cite}

\makeatletter
\g@addto@macro\bfseries{\boldmath}
\makeatother

\newlength{\wth}
\newlength{\wthb}
\newcommand{\sixgraphs}[6]{%
 \centering
 \subfloat{
   \includegraphics[width=\wthb,viewport=0 0 550 530,clip]{#1}
 }
 \subfloat{
   \includegraphics[width=\wthb,viewport=0 0 550 530,clip]{#2}
 } \\
 \subfloat{
   \includegraphics[width=\wthb,viewport=0 0 550 530,clip]{#3}
 }
 \subfloat{
   \includegraphics[width=\wthb,viewport=0 0 550 530,clip]{#4}
 } \\
 \subfloat{
   \includegraphics[width=\wthb,viewport=0 0 550 530,clip]{#5}
 }
 \subfloat{
   \includegraphics[width=\wthb,viewport=0 0 550 530,clip]{#6}
 } \\
}

\begin{document}
\title{Should we still believe in constrained supersymmetry?}
\author{Csaba Bal\'azs\inst{1,2,3} 
   \and Andy Buckley\inst{4}
   \and Daniel Carter\inst{1,2}
   \and Benjamin Farmer\inst{1,2}
   \and Martin White\inst{5,6}
}
\institute{School of Physics, Monash University, Melbourne, Victoria 3800 Australia
      \and ARC Centre of Excellence for Particle Physics at the Tera-scale, Monash University, Melbourne, Victoria 3800 Australia
      \and Monash Centre for Astrophysics, Monash University, Melbourne, Victoria 3800 Australia
      \and School of Physics and Astronomy, University of Edinburgh, Edinburgh, EH9 3JZ United Kingdom
      \and School of Physics, The University of Melbourne, Melbourne, Victoria 3010 Australia
      \and ARC Centre of Excellence for Particle Physics at the Tera-scale, The University of Melbourne, Melbourne, Victoria 3010 Australia
}
\date{}
\abstract{%
We calculate partial Bayes factors to quantify how the feasibility of the constrained minimal supersymmetric standard model (CMSSM) has changed in the light of a series of observations.  This is done in the Bayesian spirit where probability reflects a degree of belief in a proposition and Bayes' theorem tells us how to update it after acquiring new information.  Our experimental baseline is the approximate knowledge that was available before LEP, and our comparison model is the Standard Model with a simple dark matter candidate.  To quantify the amount by which experiments have altered our relative belief in the CMSSM since the baseline data we compute the partial Bayes factors that arise from learning in sequence the LEP Higgs constraints, the XENON100 dark matter constraints, the 2011 LHC supersymmetry search results, and the early 2012 LHC Higgs search results.  We find that LEP and the LHC strongly shatter our trust in the CMSSM (with $M_0$ and $M_{1/2}$ below 2 TeV), reducing its posterior odds by approximately two orders of magnitude.  This reduction is largely due to substantial Occam factors induced by the LEP and LHC Higgs searches.
}
\authorrunning{C. Bal\'azs, A. Buckley, D. Carter, B. Farmer, M. White}
\titlerunning{Should we still believe in constrained supersymmetry?}
\maketitle
\tableofcontents
\section{Introduction}
\label{sec:intro}
%
%
Supersymmetry is an attractive and robust extension of the Standard Model of particle physics~\cite{Weinberg:2000cr}.  Weak scale supersymmetry resolves various shortcomings of the Standard Model, and explains several of its puzzling features~\cite{Kane:2000bp, Drees:2004jm, Baer:2006rs, Binetruy:2006ad, Terning:2006bq, Polonsky:2001pn}.  Coupled with high-scale unification, supersymmetry breaking radiatively induces the breakdown of the electroweak symmetry.  It also tames the quantum corrections to the Higgs mass, provides viable dark matter candidates, and is able to accommodate massive neutrinos and explain the cosmological matter-antimatter asymmetry~\cite{Pagels:1981ke, Goldberg:1983nd, Ramond:1999vh, Baer:2001vw, Balazs:2004ae}.  It is also an ideal framework to address cosmological inflation \cite{Nanopoulos:1982bv, Holman:1984yj}.


However, to date there is no experimental data providing direct evidence for supersymmetry in Nature.  The exclusion of supersymmetric models based on observation proves to be just as difficult as discovery, because the large number of parameters in the supersymmetry breaking sector makes supersymmetry (SUSY) sufficiently flexible to accommodate most experimental constraints.  The most predictive supersymmetric models are the constrained ones where theoretical assumptions about supersymmetry breaking are invoked, reducing the number of free parameters typically to a few.


The most studied SUSY theory is the constrained minimal supersymmetric standard model (CMSSM)~\cite{Dimopoulos:1981zb, Chamseddine:1982jx}.  Motivated by supergravity, in the CMSSM the spin-0 and spin-$1/2$ super-partners acquire common masses, $M_0$ and $M_{1/2}$, and trilinear couplings, $A_0$, at the unification scale.  The Higgs sector is parameterised by the ratio of the Higgs doublet vacuum expectation values (VEVs), $\tan\beta = v_u/v_d$, and the sign of the higgsino mass parameter, $\mathrm{sign}\,\mu$. 


Based on experimental data, an extensive literature delineates the regions of the CMSSM where its parameters can most probably fall.  After the early introduction of $\chi^2$ as a simple measure of parameter viability \cite{Baer:2003yh, Ellis:2003si} increasingly more sophisticated concepts were utilised, such as the profile likelihood and marginalised posterior probability and the corresponding confidence \cite{Bechtle:2009ty, Bechtle:2011it, Heinemeyer:2010de, Buchmueller:2010ai} or credible \cite{LopezFogliani:2009np, Cabrera:2009dm} regions.  The effect of the LHC data on the CMSSM has typically been presented in this general manner both in the frequentist \cite{Buchmueller:2011ki,Ellis:2012aa,Bechtle:2012zk} and the Bayesian \cite{Allanach:2011ut, Allanach:2011wi, Bertone:2011nj, Fowlie:2011mb} framework. To go beyond parameter estimation and obtain a measure of the viability of a model itself one has several options.  The most common frequentist measure is the p-value, the probability of obtaining more extreme data than the observed from the assumed theory\footnote{Here `more extreme' can be defined in numerous ways.} \cite{Buchmueller:2011sw, Buchmueller:2011ab}.  In the Bayesian approach model selection is based on the Bayes factor, and requires comparison to alternative hypotheses \cite{Starkman:2008py, Cabrera:2011ds, AbdusSalam:2011fc, Sekmen:2011cz, Strege:2011pk, Roszkowski:2012uf}.  



In the Bayesian framework the plausibility of the CMSSM can only be assessed when we consider it as one of a mutually exclusive and exhaustive set of hypotheses: $\text{CMSSM} \in \{H_i\}$.  The posterior probabilities of each of these hypotheses, in light of certain data, are given by Bayes' theorem
\begin{equation}
  P(H_i|data) = \frac{P(data|H_i) P(H_i)}{\sum_j P(data|H_j) P(H_j)} .
\label{eq:BayesTheorem2}
\end{equation}
Since the denominator in the right hand side is impossible to calculate, it is advantageous to compare the plausibility of the CMSSM to that of a reference model by forming the ratio 
\begin{equation}
  Odds(\frac{\text{CMSSM}}{\text{SM+DM}}|data) = 
  \frac{P(\text{CMSSM}|data)}{P(\text{SM+DM}|data)} .
\label{eq:Odds1}
\end{equation}
Here SM+DM denotes the Standard Model augmented with a simple dark matter candidate (which need not be specified explicitly so long as certain assumptions about its parameter space are satisfied; see section \ref{sec:Higgscons}), which we choose as our reference model.  Using eq. (\ref{eq:BayesTheorem2}) we can rewrite the odds in terms of ratios of marginalised likelihoods as
\begin{align}
  \hspace{2em}&\hspace{-2em}
  Odds(\frac{\text{CMSSM}}{\text{SM+DM}}|data) \label{eq:Odds2}\\ 
  &= \frac{P(data|\text{CMSSM})}{P(data|\text{SM+DM})} 
    \frac{P(\text{CMSSM})}{P(\text{SM+DM})} \nonumber\\ 
  &= B(data|\frac{\text{CMSSM}}{\text{SM+DM}}
) ~ Odds(\frac{\text{CMSSM}}{\text{SM+DM}}) .\nonumber
\end{align}
The second ratio on the right hand side is called the prior odds, and is incalculable within the Bayesian approach.  The first ratio, however, is calculable, and is commonly called the Bayes factor.  It gives the change of odds due to the newly acquired information.  

The Standard Model is the simplest choice for a reference model, given that it fits the bulk of the data and has been confirmed by experiments up to the electroweak scale. However, since it lacks a dark matter candidate and does not address the hierarchy problem, a straightforward comparison is not possible. Nevertheless, the SM can still be used as a reference if we factorise the Bayes factor into two pieces,
\begin{equation}
B(data)=B(d_2,d_1)=B_I(d_2|d_1)B_T(d_1) \hspace{5mm} 
\label{eqn:PBF0}
\end{equation}
where $B_T$ considers a ``baseline'' or ``training'' set of data $d_1$ including dark matter and electroweak const\-raints, and $B_I$ considers the subsequent impact of data of immediate interest $d_2$, which in this work we take to be a set of LEP and LHC searches (here, for simplicity, we have dropped the conditional on ``$\text{CMSSM}/\text{SM+DM}$'', which is shared by all terms).
Neither $B_T$ nor $B_I$ individually consider the full impact of all the available data, but each considers part of it in turn; as such they have been coined ``partial'' Bayes factors, or PBFs, in the statistics literature \cite{OHagen1995,berger1996intrinsic,berger1999default}. $B_I$ may be further split, allowing one to focus on the contributions of various new data in turn. We discuss the computation of PBFs more fully in section~\ref{sec:framework}.

The SM provides a good reference model for $B_I$, even though it cannot fully explain the ``baseline'' data, because any penalty for failing to explain part of the ``baseline'' data is shifted into $B_T$, which we do not compute. Our ``inference'' PBFs $B_I$ are thus constructed to extract only a comparison of how well the CMSSM explains the null LEP and LHC sparticle searches, 126 GeV Higgs hints, and direct dark matter searches, relative to the SM. It is for this reason that the details of the implicit dark matter sector are unimportant; the main requirement is that its parameters are constrained only by the ``baseline'' data, i.e. the ``inference'' data is assumed to have negligible impact (see section \ref{sec:Higgscons}).

An alternative perspective on $B_I$ is also possible. Since the difficulties in computing $B_T$ are of a similar nature to those involved in estimating the prior odds $Odds(\text{CMSSM}/\text{SM+DM})$ in the first place, it is useful to apply Bayes theorem using the training data $d_1$ to determine a new set of odds
\begin{align}
  \hspace{2em}&\hspace{-2em}
  Odds(\frac{\text{CMSSM}}{\text{SM+DM}}|d_1) \\ 
  &= B_T(d_1|\frac{\text{CMSSM}}{\text{SM+DM}}
) ~ Odds(\frac{\text{CMSSM}}{\text{SM+DM}}) .\nonumber
\end{align}
which are nevertheless still logically `prior' to the odds that are obtained after $d_2$ is considered. $B_I$ is then just the ordinary Bayes factor associated with updating from the `pre-$d_2$' to the `post-$d_2$' odds. The effect of the hierarchy and dark matter problems may thus be thought of in terms of their effect on the `pre-$d_2$' odds, as may a portion of the effect of changing parameter space priors. As far as our analysis is concerned the estimate of what these odds are is left to the readers subjective judgement, but since the same would be true if we started from `pre-$d_1$' odds we do not see this as a problem.

Given that the hierarchy and dark matter problems are important motivation for studying SUSY models it may not be clear what we hope to achieve by shifting them partially out of our considerations. This is discussed further in sections~\ref{sec:framework} and \ref{sec:bayesfactors}, but let us introduce the idea here. In studies of constraints on BSM physics, SUSY models in particular, statements along the lines of ``large parts of the parameter space are ruled out [by such-and-such a constraint]'' can often be found. It appears that such statements are made because there is an intuition that ruling out ``large'' parts of parameter space decreases the overall plausibility of a model. From a strict frequentist perspective such statements are nonsense, because the notion of parameter space volume makes no contribution to classical hypothesis tests, that is, there is no measure on the parameter space relevant to classical inference. On the other hand, to a Bayesian there is an extremely relevant measure on the parameter space: the probability measure defined by the prior. A primary motivation of this paper is thus to clarify how such statements can be defended, and quantified, from a Bayesian perspective and to highlight the caveats that must accompany them. The objects central to quantifying such statements are exactly the ``inference'' PBFs we compute, and by using the SM+DM as a reference we can say something about each candidate model in relative isolation and achieve inferences that we feel are closest to the spirit of these statements. 



%
%
To evaluate partial Bayes factors we will need to calculate marginalised likelihoods (or evidences) such as $P(data|H_i)$. 
These are calculated as integrals over the model parameters $\theta$,
\begin{equation}
  P(data|H_i) = \int \! P(data|H_i,\theta) P(\theta|H_i) \, \mathrm{d}\theta ,
\label{eq:likelihood}
\end{equation}
%
%
where the integral is over the set of $\theta$ values for which the prior $P(\theta|H_i)$ is non-zero. Here the notation $P(data|H_i,\theta)$ is understood as distinct from $P(data|H_i)$: the latter is the probability of observing $data$ averaged over the model parameters $\theta$ (computed by the marginalisation integral of eq. (\ref{eq:likelihood})), while $P(data|H_i,\theta)$ admits a standard frequentist interpretation as the probability of the data assuming the specific parameter space point $\theta$ to be generating it, i.e. as a likelihood function. While the likelihood function depends on the data in a straightforward manner, the choice of $P(\theta|H_i)$ describing the $a~priori$ distribution of the parameters is somewhat subjective.  We fix this initial prior (which, as we discuss further in sections~\ref{sec:term} and \ref{subsec:prior}, depends on certain ``training'' data, in this case the observed weak scale) based on naturalness arguments, following previous studies \cite{Allanach:2006jc, Allanach:2007qk, Cabrera:2008tj, Cabrera:2009dm, Cabrera:2010dh}.
The underlying idea is that some mechanism is required to protect the Higgs mass from quantum corrections~\cite{Hall:2011aa}; any new physics without such a mechanism must be fine tuned to a high degree in order for these (large) corrections to cancel each other.  If supersymmetry performs this task this then gaugino masses have to be light~\cite{Athron:2007ry, Cassel:2009cx, Horton:2009ed, Cabrera:2010dh, Cassel:2010px, Akula:2011jx, Arbey:2011un, Cassel:2011tg, Papucci:2011wy, Li:2011xg, Kang:2012tn}.  To investigate the dependence of our results on this natural prior we also calculate evidences using logarithmic priors. 


The remainder of this paper is structured as follows. In section~\ref{sec:framework} we briefly review the tools needed for performing sequential Bayesian updates and the computation of partial Bayes factors, and in section~\ref{sec:bayesfactors} we discuss in detail the computation of PBFs for the CMSSM vs SM+DM case along with some comments on their properties. In section~\ref{sec:datachanges} we outline the information changes occurring in each of our Bayesian updates and explain the terminology used to refer to these, while section~\ref{sec:term} contains important notes on the terminology needed to describe priors and posteriors in sequential analyses. Section~\ref{sec:Higgscons} details the computation of the evidences needed for the `Standard Model plus dark matter' half of our PBFs, including the details of the corresponding priors, followed by section~\ref{sec:method} which details the same for the CMSSM. In section~\ref{sec:likefunc} we present the details of our likelihood function and its components and in section~\ref{sec:results} we present and discuss our central results, the PBFs due to each of our updates. Conclusions follow in section~\ref{sec:concl}.


Note added: Due to the lengthy publishing process, this paper uses LHC Higgs and super-particle search constraints that are considerably earlier than its date of appearance.  Most significantly, when calculating PBFs we have used February 2012 ATLAS 4.9\,$\text{fb}^{-1}$ Higgs search data (in which the since discovered resonance at 126 GeV had a local significance of $3.5\sigma$), ATLAS 1\,$\text{fb}^{-1}$ direct sparticle search limits, as well as XENON100 direct dark matter search limits from 100 live days, all of which have since become stronger constraints. 

\section{Bayesian updating and partial Bayes factors}
\label{sec:framework}

%
%

The Bayesian framework describes how to update probabilities of competing propositions based on newly acquired information, where probability is interpreted as measuring a `degree of belief' in competing propositions \cite{jaynes2003probability}. This probability is subjective insofar as it depends upon a subjective `starting point', i.e. an initial set of prior odds and parameter prior distributions, but the updating procedure is completely objective. As eq. (\ref{eq:Odds2}) shows, the prior odds are updated to better reflect reality by multiplying them by Bayes factors to form posterior odds.  These Bayes factors therefore quantify the effect of the new information on the odds.  

It is easy to prove that once further information is available we can consider the earlier posterior odds as prior and fold in the new information by just multiplying these odds with a new Bayes factor.  To show this, we assume that there exist two sets of data, $d_1$ and $d_2$, and we examine their effect on the prior odds.  Using eq. (\ref{eq:BayesTheorem2}) the posterior odds considering both $d_1$ and $d_2$ can be written as
\begin{equation}
  Odds(\frac{H_i}{H_j}|d_1,d_2)
  = \frac{P(d_2|d_1,H_i)}{P(d_2|d_1,H_j)} 
    \frac{P(H_i|d_1)}{P(H_j|d_1)} .
\label{eq:Odds3}
\end{equation}
Comparing the first term on the right hand side to eq. (\ref{eq:Odds2}) we identify the Bayes factor induced by $d_2$.  Making this explicit we obtain 
\begin{equation}
  Odds(\frac{H_i}{H_j}|d_1,d_2) = 
  B(d_2|d_1,\frac{H_i}{H_j}) 
  \frac{P(H_i|d_1)}{P(H_j|d_1)} .
\label{eq:Odds4}
\end{equation}
Applying eq. (\ref{eq:BayesTheorem2}) again on the last term above, this transforms into 
\begin{equation}
  Odds(\frac{H_i}{H_j}|d_1,d_2) = 
  B(d_2|d_1,\frac{H_i}{H_j}) B(d_1|\frac{H_i}{H_j})
  \frac{P(H_i)}{P(H_j)} .
\label{eq:Odds5}
\end{equation}
By induction the above holds for any set of data $\{d_1, d_2, ..., d_n\}$.  That is Bayes factors factorise and update the odds multiplicatively,
\begin{align}
  \hspace{2em}&\hspace{-2em}
  Odds(\frac{H_i}{H_j}|d_1,d_2, ..., d_n) \label{eq:Odds6}\\
  &= \left( \prod_{k=1}^n B(d_k|d_{k-1},...,\frac{H_i}{H_j}) \right)
     \frac{P(H_i)}{P(H_j)} . \nonumber
\end{align}
In the language introduced in section~\ref{sec:intro}, we call each of the terms in the product of eq. (\ref{eq:Odds6}) a ``partial'' Bayes factor (PBF), though they are still just ordinary Bayes factors. The distinction lies only with the way data is grouped in the analysis; that is, whether certain information is incorporated into the prior odds or the likelihood function, and whether subsequent Bayesian updates occur or not. As a result, a useful perspective is that every Bayes factor is really a partial Bayes factor. This is essentially our view, and given our explicit separation of data into `training' and `inference' sets it is particularly useful to use the term PBF, as a constant reminder that our method shifts some of the impact of the training data into the prior odds.
 
Crucially, the size of a PBF induced by a certain set of data depends on what other data is already known and folded into the odds.  This can be understood by considering the following example.  Assume that data set $d_1$ excludes a certain portion of (say) the CMSSM parameter space, and $d_2$ excludes another portion that is fully contained within the portion \emph{already} excluded by $d_1$ (for simplicity assume that $d_1$ and $d_2$ have no effect on the alternate model, i.e. the SM+DM).  If we learn $d_1$ first, its PBF updates the prior odds by $B(d_1|\text{CMSSM}/\text{SM+DM})$.  Learning $d_2$ after this changes nothing so its induced PBF must be unity, i.e. $B(d_2|d_1,\text{CMSSM}/\text{SM+DM}) = 1$.  In contrast, when learning $d_1$ first and then $d_2$ their partial Bayes factors, $B(d_2|\text{CMSSM}/\text{SM+DM})$ and $B(d_1|d_2,\text{CMSSM}/\text{SM+DM})$, both have to be less than one, while their product must equal $B(d_1,d_2|\text{CMSSM}/\text{SM+DM})$. This final product is independent of the data ordering, but as we see the individual PBFs are not.


Since partial Bayes factors do not ``commute'' it is important that we define the order in which the data is learned.  To assess the role of LEP and the LHC in constraining the CMSSM we deviate slightly from the historic order in which data appeared.  
We assume that the initial odds contains information from various LEP direct sparticle search limits, the neutralino relic abundance, muon anomalous magnetic moment, precision electroweak measurements and various flavour physics observables.  This set of data forms our baseline.  
We then compute the partial Bayes factors induced by folding in the LEP Higgs search and XENON100 dark matter search limits, LHC 1\,$\text{fb}^{-1}$ direct sparticle search limits and February 2012 LHC Higgs search results. These PBFs are then an efficient summary of how much damage has been done to the plausibility of the CMSSM by this new data.

\section{Computing CMSSM vs SM+DM partial Bayes factors}
\label{sec:bayesfactors}

The marginalised likelihoods, or evidences, which appear in the Bayes factor of eq. (\ref{eq:Odds3}) contain a subtle difference from the general form described in eq. (\ref{eq:likelihood}), this being that they are conditional on {\em data} $d_1$:
\begin{equation}
P(d_2|d_1,H_i) = \int \! P(d_2|d_1,H_i,\theta) P(\theta|d_1,H_i) \, \mathrm{d}\theta . 
\label{eq:marginallikelihood}
\end{equation}
If $d_1$ and $d_2$ are statistically independent then the conditioning on $d_1$ drops out of the likelihood function, but it remains in the prior function $P(\theta|d_1,H_i)$. This prior may thus be called `informative' because it incorporates information from the likelihood $P(d_1|H_i,\theta)$, which has been folded in to an initial ``pre-$d_1$'' prior $P(\theta|H_i)$, in general resulting in an extremely complicated distribution which makes the integral difficult to evaluate.

Fortunately, there exists an alternative to directly evaluating the integral. From the definition of conditional probability we may write
\begin{equation}
P(d_2|d_1,H_i) = \frac{P(d_2,d_1|H_i)}{P(d_1|H_i)},
\label{eq:globev}
\end{equation}
where the numerator and denominator may be referred to as ``global'' evidences, since they are computed by integrating the global likelihood function over the parameter space, with the parameter space measure defined by the ``pre-$d_1$'' distribution for the model parameters, as is done in more conventional model comparisons~\cite{Feroz:2008wr,AbdusSalam:2009tr,Feroz:2009dv,Cabrera:2010xx,Pierini:2011yf}. We discuss the numerical details of the global evidence evaluation and priors in section~\ref{sec:method}, and the details of the global likelihood function in section~\ref{sec:likefunc}.

Bayes factors are only defined for a {\em pair} of hypotheses which are being compared, however it is useful to break them up into pieces which tell us something about what is happening in each hypothesis individually, so that we may more easily speculate about what effect variations in one hypothesis or the other might have. While the evidences themselves suit this purpose it can be more illuminating to break them up further, into a contribution from the maximum of the likelihood function of the new data, and an Occam factor.  The latter is defined only through its relationship to the evidence; it is what remains when the maximum value of the likelihood function is divided out:
\begin{equation}
\mathscr{O}(d_2|d_1;H_i) \equiv \frac{P(d_2|d_1,H_i)}{P(d_2|H_i,\hat{\theta})} .
\label{eq:OccamFact}
\end{equation}
Here $P(d_2|d_1,H_i)$ is the evidence associated with learning $d_2$ when $d_1$ is already known, as computed in eq. (\ref{eq:marginallikelihood}) and eq. (\ref{eq:globev}), and $P(d_2|H_i,\hat{\theta})$ is the maximum value of the likelihood function for $d_2$ that is achieved in the model $H_i$ (and $\hat\theta$ is the parameter space point in $H_i$ which achieves this maximum). $P(d_2|H_i,\hat\theta)$, coming as it does from the likelihood, does not depend on the prior\footnote{Strictly, some prior dependence remains due to the choice of parameter values considered possible by the prior, most often arising from the choice of scan range, however this is the same kind of dependence that exists in a frequentist analysis. As well as this there exists the possibility that $d_1$ strictly forbids certain values of $\theta$, and these too should be excluded from the computation of $P(d_2|H_i,\hat\theta)$.}: this dependence is entirely captured by the Occam factor. $P(d_2|H_i,\hat\theta)$ also has no dependence on $d_1$, with this dependence again contained in the Occam factor.

These two components of the evidence give us different information about the model. A Bayes factor (or PBF) is a ratio of evidences, so by decomposing evidences in this manner we will obtain in the PBF a product of ratios, one of which is a standard frequentist maximum likelihood ratio (considering just the new data $d_2$), and the other of which is a ratio of Occam factors. The maximum likelihood ratio tells us which model has the better fitting point with respect to $d_2$, but ignores all other aspects of the model and all other data. Complementing this the Occam factor tells us something about the relative {\em volume} of previously viable parameter space which is compatible with the new data $d_2$ in each model, where the measure of volume is defined by the informative prior $P(\theta|d_1,H_i)$, which has resulted from a previous Bayesian update and so ``knows'' about previous data $d_1$. The Occam factor can be roughly interpreted as the amount by which the new data $d_2$ collapses the parameter space when it arrives\footnote{The full volume of parameter space viable at this inference step, $V_\text{total}$, is defined by the informative prior.  If the likelihood function for the new data was constant in a region $V$ and zero outside of it, then the fraction $f=V/V_\text{total}$ would be the Occam factor.}, and its logarithm as a measure of the information gained about the model parameters~\cite{mackay2003information}

The impact of Occam factors on the model comparison can be seen by explicitly writing out the PBFs in terms of them:
\begin{align}
 B(d_2|d_1) &= \frac{P(d_2|d_1,H_i)}{P(d_2|d_1,H_j)} \label{eq:BayesFact2}\\
 &= \frac{P(d_2|H_i,\hat{\theta})}{P(d_2|H_j,\hat{\phi})}
 \frac{\mathscr{O}(d_2|d_1,H_i)}{\mathscr{O}(d_2|d_1,H_j)} \nonumber ,
\end{align}
where $\theta$ and $\phi$ parameterise $H_i$ and $H_j$ respectively (and $\hat{\phi}$ the analogue of $\hat{\theta}$). Schematically
\begin{equation}
 B = LR \times \frac{\mathscr{O}_{H_i}}{\mathscr{O}_{H_j}} ,
\label{eq:BayesFactSimple}
\end{equation}
where $LR$ denotes the maximum likelihood ratio for the new data $d_2$, and the rest of the abbreviated terms correspond directly to their partners in the more formal expression. We thus see two competing factors: a model is favoured if it achieves a high likelihood value for the new data somewhere in its parameter space, but disfavoured if the good-fitting region is not very compatible with the informed prior (i.e. if a good fit is achieved in only a small region, with `small' defined according to the probability measure of the informed prior). These effects are also relative; i.e. no objectively ``good'' likelihood value is needed, just one which is better than that achieved in alternate models, and likewise for the volume effects. 

Because the best fit point is only with respect to the new data it could be very different to the best fit point of the global likelihood function, and so may not appear to be a useful object to frequentist thinkers. However, in the Bayesian framework it is acknowledged that not all data relevant to inference can be expressed in the likelihood function, that is, the prior may contain real information. In our case the prior for each iteration (except the first) contains very concrete information; that coming from the rest of the likelihood. The best fit point with respect to the new data is thus indeed not so useful on its own (although it tells us something about the maximum goodness of fit possible in the model for that data), but extracting it from the evidence allows one to capture tension between the new and old data in a different way, i.e. in the Occam factor.

Eq. (\ref{eq:BayesFact2}) is completely general, except that the data must be independent. To gain some intuition about how PBFs select models we may now make some assumptions about how the global evidence for each model behaves under certain kinds of data changes. To begin with, in the case of adding new exclusion limits, the best-fit likelihood value of the new data is often very similar in large classes of models; specifically, it will be close in value to that for the SM, assuming no significant deviations from the SM predictions are observed. An interesting situation to consider is thus that in which we set the maximum likelihood value for new data to be equal in both models\footnote{I.e. in a generic event counting experiment we assume the expected number of signal events at the best fit point to be close to zero.}. Applying this assumption to the PBF gives us (for example):
\begin{equation}
 B(d_2|d_1) =
 \frac{\mathscr{O}(d_2|d_1,\text{CMSSM})}{\mathscr{O}(d_2|d_1,\text{SM+DM})} .
\label{eq:BayesFact4}
\end{equation}
If the CMSSM and SM+DM best fit values for the new data are similar then the Occam factors dominate our reasoning process. Models suffering large cuts to the parameter space become less believable, while those less damaged by the new limits become relatively more believable, as one intuitively expects.

Since this work is devoted to quantifying changes in odds, not odds themselves, we evaluate only the partial Bayes factor $B(d_2|d_1)$ for various data sets $d_2$; the calculation of the prior odds in eq. (\ref{eq:Odds6}) is not attempted and is impossible unless one is prepared to explore principles for defining measures on the global space of hypotheses --perhaps based on algorithmic probability (as advocated by Solomonoff\cite{Solomonoff19641} and others)-- or otherwise justify an `objective' origin for priors. From a purely subjective Bayesian perspective the prior odds can instead be allocated to the reader to estimate from their own knowledge base and philosophical preferences, to be modified by the PBFs we compute.

To close this section we wish to make an additional observation about our choice of the SM+DM as our reference model. It was recently shown in ref. \cite{Fichet:2012sn} that a model in which observable quantities enter directly as input parameters can be considered a ``puzzle'' from the perspective of naturalness considerations. Such a model lies at a natural boundary between a fully predictive (or ``natural'') model (which in effect has no free parameters, and for which the evidence collapses to a simple likelihood), and a fine-tuned model (in which `small' changes to parameters - where again `small' is defined relative to the measure set by the prior - produce large changes in predicted observations and for which the evidence due to learning the fine-tuning inducing data will be incredibly small, since only a tiny portion of parameter space predicts it correctly\footnote{Note that a very {\em small} value for the evidence from learning some data implies a very {\em large} amount of information was gained about the model. This may sound like a good thing, however it means that little was known about the model before this data arrived and so the model was not very useful for predicting what that data would be. PBFs penalise this failure, however if the information gain was sufficiently large then the model may in fact become highly predictive about future data, and may thus fare much better in future PBF tests.}). It is argued that such a ``puzzle'' model represents the only sensible reference point against which to measure naturalness. The changes in evidence in such a model can be easily computed, if one has enough data to define a prior for the observables, and it is argued that these be compared to the evidence changes that occur in a model of interest using a Bayes factor exactly as we compute; if the Bayes factor favours the ``puzzle'' model this is an indication that the model of interest is not a very natural explanation for the data and drives us to believe that a better model should exist. 

There is no reason to restrict this reasoning to only that data usually associated with fine-tuning, and as we have defined it our comparison SM+DM is just such a ``puzzle'' model\footnote{The reader may protest that the SM+DM is not just a fine-tuning ``puzzle'', it is a very extreme example of fine-tuning! However, this is only true if one considers it from a pre-`electroweak data' perspective. The SM+DM presumably suffers a very large PBF penalty for failing to predict the electroweak scale (and for this scale being observed very far from, say, the Planck scale, where {\em a priori} arguments based on the hierarchy problem may place it), however these considerations enter before the `baseline' data we choose for our inference sequence and so do not directly enter our PBFs. The complete assessment of which model best reflects reality should of course take these matters into account.}. Thus, if the reader prefers, they may interpret our computed PBFs not as tests of the CMSSM against any specific model, but as measures of how much better or worse than the ``puzzle'' model it predicts the new data (when constrained by the baseline data).

\subsection{Bayesian updates}
\label{sec:datachanges}
Here we outline the changes of information that we consider in this paper, and for which we compute the corresponding partial Bayes factors for the CMSSM vs SM+DM hypothesis test. We take as our initial information a conventional set of experimental data, including dark matter relic density constraints, muon anomalous magnetic moment measurements, LEP2 direct sparticle mass lower bounds, and various flavour observables. The full list and details of the likelihood function can be found in table~\ref{tab:likefunc} of section~\ref{sec:likefunc}. Notably, we do not include the LEP2 Higgs mass and cross section limits, nor any results from dark matter direct detection experiments or the LHC\footnote{Except for an early LHCb lower bound on $BR(B_s\to\mu\mu)$.}, because these are precisely the pieces of data whose impact on the CMSSM we wish to assess. To improve the brevity of later references, we name this initial data set the ``pre-LEP'' state of knowledge, to emphasise that the LEP Higgs bounds have been removed.

The shrewd reader will notice that we include many pieces of data in this initial set that were not yet measured when the LEP2 Higgs constraints began to exclude much of the low-mass CMSSM regions (most notably the WMAP measurements constraining the dark matter relic density), and that we neglect previous Higgs constraints, so our `initial' knowledge state is not truly representative of the experimental situation that existed around say 1998 (when the LEP bound was $m_h < 77.5$ GeV\cite{LEPHWG:1998} and would not have noticeably constrained our ``pre-LEP'' CMSSM parameter space had we included it). However there is no requirement that the analysis be chronologically accurate for meaningful results to be obtained. We maintain the rough correspondence simply to ease the interpretation of the results. In addition, most extra constraints in the initial set (aside from the WMAP data) tend to exclude parts of the CMSSM that the new data would also exclude, thus reducing the apparent strength of the latter.

From this initial data set we add in sequence the LEP2 Higgs constraints and XENON100 limits on the neutralino-nucleon elastic scattering cross section to form the ``LEP+XENON'' data set.  Next we add the 2011 1 $\text{fb}^{-1}$ LHC SUSY search results to form the ``ATLAS-sparticle'' data set.  Finally, we add the February 2012 LHC Higgs search results to form the ``ATLAS-Higgs'' data set.  The details of the likelihood functions for these new pieces of data are described in section~\ref{sec:likefunc}. This gives us four data sets and three sequential Bayesian updates, each of which is characterised by a partial Bayes factor. In addition, we compute results using two different ``pre-LEP'' distributions (i.e. priors) for the CMSSM parameter space (the description of which we leave to section~\ref{subsec:prior}, with some preliminary comments in section~\ref{sec:term}), giving two perspectives on each update and thus doubling the number of data sets and PBFs we obtain.

\subsection{A note on priors, posteriors, and terminology}
\label{sec:term}
Since we consider a sequence of Bayesian updates in this work, the conventional terminology used in more straightforward analyses becomes somewhat awkward; in particular, the usage of the words ``prior'' and ``posterior'' become more context-sensitive than usual. For any given Bayesian update, there are always probabilities that represent states of knowledge ``prior'' to the update, and corresponding probabilities that are logically ``posterior'' to the update; however, in a sequential analysis the posterior from one update acts as prior to the next, meaning that a single set of probabilities may be described as both ``prior'' \emph{and} ``posterior'' depending on the particular update being referenced, implicitly or explicitly, at the time.

Confusing the issue further is the technique we use to compute our PBFs, best illustrated by the structure of eq. (\ref{eq:globev}). Here we compute the evidence we are interested in, $P(d_2|d_1,H_i)$ (due to updating from data $d_1$ to $\{d_1,d_2\}$) by taking the ratio of the two ``global'' evidences $P(d_2,d_1|H_i)$ and ${P(d_1|H_i)}$ (due to updating from an implicit ``pre-$d_1$'' state of knowledge to $\{d_1,d_2\}$ and $d_1$ respectively), which are more straightforward to implement computationally. However this structure means we now have to be careful to be clear about the difference between the prior for the $d_1$ to $\{d_1,d_2\}$ update, $P(\theta|d_1,H_i)$, and the prior for the ``pre-$d_1$'' to $\{d_1,d_2\}$ or $d_1$ updates, $P(\theta|H_i)$. To aid in this distinction we refer to $P(\theta|H_i)$ as the ``pre-LEP'' prior, since the ``pre-LEP'' data set is the first we consider, and $P(\theta|D,H_i)$ as an ``informative'' prior, or where possible by a more explicit reference to the update to which it is prior, e.g. the ``LEP+XENON'' prior for the update from the ``pre-LEP'' to the ``LEP+XENON'' datasets (with the updates in our sequence occurring as described in section~\ref{sec:datachanges})

In the case of $H_i = \text{CMSSM}$, we do not ever explicitly compute the ``informative'' priors $P(\theta|D,H_i)$\footnote{This is a small lie; we do compute marginalised posteriors for each update, which indeed correspond to the ``informative'' priors for the subsequent update. Nevertheless we do not explicitly use them in this fashion.}, since we compute the required evidences using eq. (\ref{eq:globev}). On the other hand, in section~\ref{sec:Higgscons}, where $H_i = \text{SM}$, we \emph{do} explicitly compute and make use of these priors, so the terminology is particularly important there.

There is a final important note to be made on this topic, which is deeply connected to naturalness and the hierarchy problem. When we construct the ``pre-LEP'' priors  $P(\theta|H_i)$ for both the CMSSM and the SM+DM, it must be noted that large amounts of experimental data are taken into consideration when constructing them, so in no sense should they be though of as ``fundamental'' or ``data-free'' priors. This is true for all Bayesian global fits of such models of which we are aware.

The so-called ``natural'' (``pre-LEP'') prior we use for the CMSSM demonstrates this most explicitly. When scanning the CMSSM in the conventional parameter set $\{M_0,M_{1/2},A_0,\tan\beta,\text{sign}\mu\}$ one must remember that the codes generating the CMSSM spectrum make \emph{explicit} use of the observed $Z$ mass in order to reduce the dimensionality of the scan\footnote{The rest of the Standard Model parameters of course also enter explicitly, but we may reasonably consider priors over those to be statistically independent of the CMSSM parameters, such that measuring the values of these parameters results in PBFs of 1.}, which means that the ``pre-LEP'' prior $P(\theta|\text{CMSSM})$ should more correctly be written as $P(\theta|m_Z,\text{CMSSM})$, as should all priors set directly on these ``phenomenological'' CMSSM parameters. The ``natural'' prior explicitly acknowledges this fact and so begins from a ``pre-$m_Z$'' prior $P(\theta|\text{CMSSM})$ (which since no weak scale information is available must be formulated in terms of the more ``fundamental'' parameters {$\{M'_0,M'_{1/2},A'_0,B',\mu'\}$}, where the dashes acknowledge that the conventional set $\{M_0,M_{1/2},A_0,\tan\beta,\text{sign}\mu\}$ parameterises a (multi-branch) 4D hypersurface of the ``fundamental'' parameter space) which then effectively undergoes a Bayesian update, as features so prominently in our analysis, to the ``pre-LEP'' prior $P(\theta|m_Z,\text{CMSSM})$ by folding in the known Z boson mass. This update of course is accompanied by a PBF, and it is this PBF which penalises any tuning required to obtain the correct weak scale from a model, and which may be expected to extremely heavily prefer the CMSSM over the SM+DM no matter how large tuning becomes in the CMSSM. 

As mentioned in section~\ref{sec:intro} we do not compute the PBFs for this particular update, since it is difficult to do so rigorously and the focus of our paper is the CMSSM, rather than the SM+DM. Nevertheless we feel that this series of arguments is excellent motivation for so-called ``naturalness'' priors, and casts serious doubt on the logical validity of more conventional CMSSM priors, such as the log prior we use for comparison, which can in this light be understood to express some extremely odd beliefs about the ``fundamental'' parameters {$\{M_0,M_{1/2},A_0,B,\mu\}$}.  More to the point of this section, it is an excellent example of the type of ``background'' information on which many priors in the literature are implicitly conditional.

\section{Evidences for the Standard Model}
\label{sec:Higgscons}

For our purposes, we can consider all the parameters of the SM to be fixed by our initial experimental data or otherwise unaffected by the new data, with only the Higgs mass $m_h$ undetermined. The new data is also assumed to minimally affect any additional dark matter sector. The evidences for the combined SM plus dark matter (SM+DM) for each data transition can thus be computed entirely by considering the one-dimensional Higgs mass parameter space. This can be shown as follows.

The above assumptions allow us to separate the parameters and available data into three groups: (1) initial data $d_0$ which highly constrains the Standard Model parameters $\phi$ but less strongly constrains $m_h$; (2) data $d_\Omega$ which constrains only the dark sector parameters $\omega$, and whose effect on $\phi$ is negligible relative to $d_0$; and (3) `new' data $d_{new}$ constraining only the Higgs mass, with negligible effect on $\phi$ relative to $d_0$, and no impact on the dark sector parameters $\omega$. If we assume that the initial prior (or `{``pre-$\{d_0,d_\Omega\}$'' prior}' in the terminology introduced in section~\ref{sec:term}) for $\omega$ is independent of that for $m_h$ and $\phi$, i.e. $P(m_h,\phi,\omega|\text{SM+DM}) = $ $P(m_h,\phi|\text{SM+DM}) P(\omega|\text{SM+DM})$, and that the three sets of data are statistically independent, then the evidence associated with the new data $d_{new}$ can be written as
\begin{flalign}
 P(d_{new}|d_0,d_\Omega) &= \dfrac{P(d_{new},d_0,d_\Omega)}{P(d_0,d_\Omega)} &\big|\text{SM+DM},
\label{eq:SMseperable}
\end{flalign}
with
\begin{multline}
  P(d_{new},d_0,d_\Omega) = \int \! \mathrm{d}m_h\mathrm{d}\phi\,\mathrm{d}\omega \, P(d_{new}|m_h,\phi) \\
 \times P(d_0|m_h,\phi) P(d_\Omega|\phi,\omega) P(m_h,\phi,\omega) \text{~~~}\big|\text{SM+DM},
\end{multline}
and
\begin{multline}
  P(d_0,d_\Omega) = \int \! \mathrm{d}m_h\mathrm{d}\phi\,\mathrm{d}\omega \\
  \times P(d_0|m_h,\phi) P(d_\Omega|\phi,\omega) P(m_h,\phi,\omega) \text{~~~}\big|\text{SM+DM},
\end{multline}
where the ``$\big|\text{SM+DM}$'' notation indicates that all probabilities in the expression are conditional on ``SM+DM'', i.e. the combined model.
If $d_0$ sufficiently strongly constrains the SM parameters (except $m_h$) to $\phi'$ then to a good approximation $P(d_0|\text{SM+DM},m_h,\phi) \propto \delta(\phi' - \phi)f(m_h)$, where $f(m_h)$ describes the variation of the $d_0$ likelihood in the $m_h$ direction, and the proportionality constant divides out in the ratio. The $\phi$ integral is thus removed and the remaining integrals are separable. The integral over the dark sector parameters $\omega$ is identical in the numerator and denominator and thus vanishes, as does the prior density $P(\phi'|\text{SM+DM})$ (resulting from expanding $P(m_h,\phi'|\text{SM+DM})$ as $P(m_h|\text{SM+DM},\phi')P(\phi'|\text{SM+DM})$, evaluated at $\phi'$ due to the delta function), leaving us with
\begin{multline}
P(d_{new}|d_0,d_\Omega) = \\
	\frac{\int \! \mathrm{d}m_h \, 
		P(d_{new}|m_h,\phi') 
		f(m_h)
		P(m_h|\phi')}
	{\int \! \mathrm{d}m_h \,
		f(m_h)
		P(m_h|\phi')} \text{~~~}\big|\text{SM+DM}.
\label{eq:SMseperable2}
\end{multline}
We are free to choose the normalisation of $f(m_h)$, and it is convenient to choose it such that $\int \! \mathrm{d}m_h f(m_h) P(m_h|\text{SM+DM},\phi') = 1$, so that $f(m_h) P(m_h|\text{SM+DM},\phi')$ corresponds to the posterior probability density for $m_h$ once $d_0$ is considered, i.e. $P(m_h|d_0,\text{SM+DM},\phi')$. This density becomes the prior for the consideration of $d_{new}$. The evidence associated with learning $d_{new}$, starting from $d_0$ and $d_\Omega$, is thus shown to be the relatively straightforward integral
\begin{multline}
P(d_{new}|d_0,d_\Omega) = \\
\int \! \mathrm{d} m_h \, P(d_{new}|m_h,\phi')P(m_h|d_0,\phi') \text{~~~}\big|\text{SM+DM}.
\label{eq:SMHiggsOccamFact}
\end{multline}
as we intuitively expect. Importantly, this evidence is independent of the details of both the dark sector theory and the constraints $d_\Omega$, so long as the theory meets our criteria of not significantly affecting the predictions for $d_{new}$, nor is affected by the value of $m_h$\footnote{Within the range of $m_h$ values compatible with $d_{new}$, i.e the dark sector theory is permitted to exclude values of $m_h$ which are also well excluded by $d_{new}$.}. Any sufficiently decoupled dark sector satisfies this requirement.

We now evaluate eq. (\ref{eq:SMHiggsOccamFact}). The $d_0$ relevant for constraining $m_h$ are electroweak precision measurements, so we may build our ``pre-LEP'' prior $P(m_h|d_0, \text{SM+DM},\phi') = f(m_h) P(m_h|\text{SM+DM},\phi')$ based on these. Taking the most conservative $\Delta\chi^2$ curves from figure 5 of ref. \cite{EWWG:2010vi} as our electroweak constraints we reconstruct the corresponding likelihood function $f(m_h)$, and multiply this by an initial (i.e. ``pre-$\{d_0,d_\Omega\}$'') prior $P(m_h|\text{SM+DM},\phi')$ flat in $\log m_h$\footnote{For a scale parameter this is the Jeffreys prior.}.  Although this is done numerically it yields a prior close\footnote{These $\Delta\chi^2$ curves are almost quadratic in $\log m_h$, implying a close to Gaussian likelihood function, however we have digitised the most loose boundaries of the displayed curves to be conservative. As a result the likelihood function we reconstruct has a flat maximum from $\sim\!80$\,GeV to $\sim\!100$\,GeV.} to a broad Gaussian (in $\log m_h$ space) centred on $m_h=90$\,GeV with a $\log_{10}$ width of about $0.15$, i.e. $m_h = 90^{+35}_{-26}$\,GeV.

If the new data $d_{new}$ is the LEP2 $m_h$ likelihood function described in table~\ref{tab:likefunc} (let us call this $d_{LEP}$), then eq. (\ref{eq:SMHiggsOccamFact}) is now straightforward to evaluate numerically. Its value alone is not meaningful because the likelihood function is only defined up to a constant (which divides out in the PBF), however if we divide out the maximum likelihood value we recover the corresponding Occam factor, which we find to be 0.284, or about $1/3.5$. We have checked that choosing a flat initial prior for $m_h$ makes little difference to this result\footnote{If, due to tuning arguments, we except $m_h$ to adopt a value on the largest allowed scale, rather than all scales being equally likely, then a flat prior cut off at this scale may indeed better represent this belief. The lack of sensitivity of the informative ``pre-LEP'' prior to this choice reflects the fact that before the ``pre-LEP'' update the Standard Model prediction for the Higgs mass is already quite well constrained.}. We consider the corresponding effects on the CMSSM in section~\ref{sec:results}, however it is useful to mention here that the maximum likelihood values for both the SM+DM and CMSSM for this data are equal (since our simple model of the limit assumes the likelihood to be maximised for the background-only hypothesis), so the Occam factors themselves contain all the information about which model the PBF prefers. A more careful analysis of the LEP data would allow the CMSSM to receive a slight likelihood preference since it is has more parameters than the SM+DM and can in principle achieve a better fit to any observed deviation from the expected background, however since no significant excess was seen at LEP this effect will be small.

In addition to the evidence $P(d_{LEP}|d_0,\text{SM+DM})$, the computation of eq. (\ref{eq:SMHiggsOccamFact}) also produces for us (via Bayes' theorem) a new posterior distribution over $m_h$, which incorporates both $d_0$ and $d_{LEP}$ (with $d_\Omega$ having had no impact):
\begin{equation}
P(m_h|d_{LEP},d_0) = \frac{P(d_{LEP}|m_h)P(m_h|d_0)}{P(d_{LEP}|d_0)} \text{~~~}\big|\text{SM+DM},
\label{eq:SMpost}
\end{equation}
(for brevity we drop the conditionals on $\phi'$, as it is fixed from here on, and on $d_\Omega$, because our results were shown to be independent of it). This is the prior for the second iteration of our learning sequence, in which we consider the addition of the ATLAS sparticle search results, so we may call it the ``ATLAS-sparticle'' prior. These searches of course do not affect the Standard Model parameters, and our assumptions about the nature of the dark sector demand that it be similarly unaffected.  So the SM+DM evidence due to this update can be safely set to 1.

Finally, we consider the addition of the recent LHC Higgs search results. Since the sparticle searches had no impact the prior for this update is unchanged in form from eq. (\ref{eq:SMpost}), that is eq. (\ref{eq:SMpost}) also describes the ``ATLAS-Higgs'' prior. As we shall discuss further in section~\ref{sec:LHCHiggsL}, we constrain the CMSSM using only the results from the ATLAS $h\rightarrow\gamma\gamma$, $h\rightarrow ZZ\rightarrow 4l$ and $h\rightarrow WW\rightarrow 2l2\nu$ search channels\cite{ATLASaa:2012sk, ATLAS2l2v:2011aa, ATLAS4l:2012sm}, as these channels both dominate the constraints on the lightest CMSSM Higgs and are the only ones for which ATLAS provide signal best fit plots, which we require to perform our likelihood extraction.  CMS do not provide such plots for all channels so we are unable to incorporate the CMS results at this stage.  We constrain the cross sections for each of these channels separately in the CMSSM likelihood function since the factor by which they differ from the Standard Model prediction is not uniform across all channels, as is assumed in the ATLAS and CMS combinations. For the SM+DM evidence computation it would be optimal to include extra channels which can more powerfully exclude higher Higgs masses, however the strength of the 125 GeV excess in our chosen three channels is already sufficient to very strongly disfavour such Higgs masses, such that including these extra channels would negligibly improve our analysis.

\begin{figure}
\centerline{
\includegraphics[width=\columnwidth]{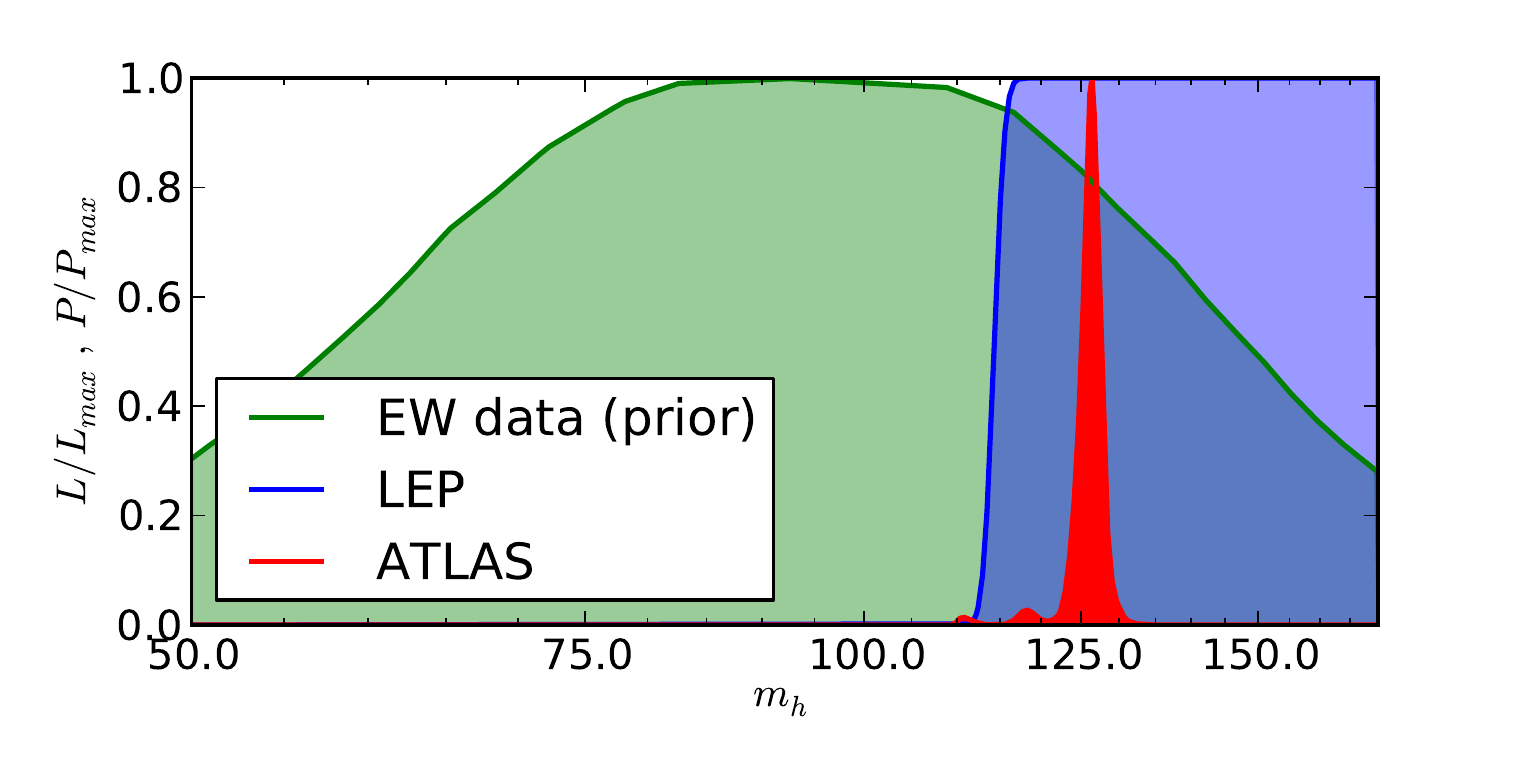}
}
\caption{The prior over the SM Higgs mass parameter derived from electroweak precision measurements (green), with LEP (blue) and ATLAS (red) Higgs search likelihood functions overlaid. The prior and likelihood functions are scaled against their maximum values.}
\label{fig:SMatlas}
\end{figure}

In figure \ref{fig:SMatlas} we show the ``pre-LEP'' prior for the SM Higgs parameter, derived from electroweak precision measurements, with the LEP and ATLAS Higgs search likelihood functions overlaid.  The LEP likelihood function is simply taken as a hard lower limit at 114.4\,GeV, convolved with a 1\,GeV Gaussian experimental uncertainty (as described in table~\ref{tab:likefunc}).  The ATLAS Higgs search likelihood function is reconstructed from the February 2012 combined Higgs search results~\cite{ATLAS:2012si} using the method described in section~\ref{sec:LHCHiggsL}.  Performing Bayesian updates with each of these likelihood functions in sequence we compute Occam factors of 0.284 and 0.02 respectively.

We note again that we have not folded in earlier LEP Higgs limits into the ``pre-LEP'' prior for the SM Higgs mass; for example the upper limits of around 80 GeV that existed in 1998. We have done this to avoid an arbitrary decision about exactly which limits to include, and because neglecting them only weakens the apparent damage that LEP did to the CMSSM. This occurs because CMSSM model points predicting Higgs masses of below 80 GeV are not common nor have particularly high likelihood in our ``pre-LEP'' CMSSM data set and so do not occur in most of the effective prior which arises from that data set, while a very sizable portion of the SM Higgs mass prior we have just constructed {\em is} below 80 GeV. Therefore, were we to include such a limit, it would increase the apparent damage done by the 114.4 GeV LEP limit to the CMSSM (relative to the SM), so leaving it out is a conservative choice. The impact on the corresponding Bayes factor can be fairly easily estimated anyway by considering eq. (\ref{eq:BayesFactSimple}), as follows. The amount of ``pre-LEP'' CMSSM posterior that would be affected is fairly negligible so we can ignore it in a rough estimate, while the amount of SM Higgs prior that would be cut off can be seen from figure~\ref{fig:SMatlas} to be about 1/3. The $114.4$ GeV limit would thus have its SM Occam factor increased from about 0.3 to about 0.5 (weakening it), and since the other components of the Bayes factor remain unchanged the effect would be about a $5/3$ boost in odds towards the SM, which, as we shall see in section~\ref{sec:results}, is of negligible importance. 

\section{Evidences for the CMSSM}
\label{sec:method}

To determine the CMSSM half of our partial Bayes factors we compute the CMSSM global evidence under each of the data sets described in section~\ref{sec:datachanges}, using two contrasting ``pre-LEP'' prior distributions for the parameters (see section~\ref{subsec:prior} for details). This requires the numerical mapping of the CMSSM global likelihood function for each data set (the details of which we discuss in section~\ref{sec:likefunc}).  To perform this mapping we use the public code \texttt{MultiNest~v2.12}~\cite{Feroz:2008xx,Feroz:2007kg}, which implements Skilling's nested sampling algorithm~\cite{skilling:395}\footnote{Aside from nested sampling and several variants of Markov Chain Monte Carlo methods, the list of techniques used to scan the CMSSM has expanded in recent years to also include genetic algorithms and neural networks \cite{Akrami:2009hp, Bridges:2010de}.}.  To compute the CMSSM predictions at each parameter space point we first generate the particle mass spectrum using \texttt{ISAJET~v7.81}~\cite{Paige:2003mg}.  We then pass the spectrum to \texttt{micrOmegas~v2.4.Q} \cite{Belanger:2010pz, Belanger:2008sj, Belanger:2006is} to compute the neutralino relic abundance, muon anomalous magnetic moment, spin-independent proton-neutralino elastic scattering cross section, and precision electroweak variable $\Delta\rho$.  We use \texttt{SuperISO~v3.1} \cite{Mahmoudi:2007vz, Mahmoudi:2008tp} to compute a number of flavour observables (the full set of likelihood constraints imposed is listed in table~\ref{tab:likefunc}).  We also estimate on a yes/no basis whether a given point can be considered excluded by ATLAS direct sparticle searches, using a machine learning technique which we describe in section~\ref{sec:LHCsparticle}.  Finally, the spectrum is passed to \texttt{HDECAY~v4.43}~\cite{Djouadi:1997yw}, which we use to compute the cross section ratio $\sigma_\text{CMSSM}/\sigma_\text{SM}$ for the following processes:
\begin{align}
gg&\to h\to \gamma\gamma,
\nonumber \\ gg&\to h\to ZZ\to 4l, \, \text{and}
\nonumber \\ gg&\to h\to WW\to 2l+2\nu.
\end{align}
We constrain these cross sections separately using the December 2011 - February 2012 ATLAS Higgs search results~\cite{ATLAS2l2v:2011aa,ATLASaa:2012sk,ATLAS4l:2012sm}.  \texttt{MultiNest} then guides the scan through a large number of sample points, returning a chain of posterior samples and the global evidence.

To ensure that the likelihood function is sampled densely enough to guarantee highly accurate evidence values, we run \texttt{MultiNest} with parameters guided by the recommendations of ref. \cite{Feroz:2011bj}, in which \texttt{MultiNest} is configured to sample the likelihood function to the accuracy required for frequentist analyses. Since our analysis is Bayesian we do not require as detailed information as frequentist scans in the vicinity of the maximum likelihood points, so we drop the recommended number of live points from $20k$ to $15k$ and relax the convergence criterion from $\texttt{tol}=10^{-4}$ to $\texttt{tol}=0.01$ to reduce the computational demand.  Additionally, we cluster in three dimensions ($M_0$, $M_{1/2}$ and $\tan \beta$) and set the efficiency parameter \texttt{efr} to 0.3.  Finally, we treat the top quark mass $m_t$ as a nuisance parameter (with the rest of the Standard Model parameters set to their central experimental values), so the dimensionality of the scanned parameter space is five.
The above \texttt{MultiNest} settings result in about $10^7$ evaluations of our likelihood function per run and posterior chains of about $2.5\times10^5$ good model points. The total number of likelihood evaluations over the whole project exceeded $10^8$.

\subsection{Priors and ranges}
\label{subsec:prior}

The shape of the ``pre-LEP'' prior $P(\theta|\,\text{CMSSM})$ reflects our relative belief in different parts of the parameter space before learning the any of the experiment information in our ``pre-LEP'', ``ATLAS-sparticle'' or ``ATLAS-Higgs'' likelihood functions (though as discussed in section~\ref{sec:term} they \emph{are} conditional on other `background' experimental knowledge, such as Standard Model parameter values, particularly $m_Z$).  By considering multiple of these priors we can analyse how a representative set of subjective beliefs about the CMSSM should be modified by new data.  We now describe the priors we use and explain our choices.


We allow the top quark mass to vary since, of the Standard Model parameters, its experimental uncertainty allows the largest variation in the CMSSM predictions. Since its value is to a large degree fixed by experiment we are able to set its ``pre-LEP'' prior to be a Gaussian with the experimental central value and width of $172.9 \pm 1.1$\,GeV~\cite{PDG:2010}.


To reduce the computational complexity of the problem we have only scanned the $\mu>0$ branch of the CMSSM. This is not optimal, however it almost certainly does not greatly affect our inferences, because the $\mu<0$ branch is already strongly disfavoured by the data in our ``LEP+Xenon'' set by a PBF of around 20-60 \cite{Feroz:2008wr}\footnote{The authors estimate these Bayes factors using both flat and log priors; here we refer to the log prior results only since we do not use flat priors. In addition, the $\delta a_{\mu}$ constraint is shown to strongly drive the preference of $\mu>0$ so if the validity of this constraint is questioned (we consider the effects of removing it in section~\ref{sec:results}) then the impact of ignoring the $\mu<0$ branch may also merit revisitation.}. The $\mu<0$ branch of the ``pre-LEP'' dataset is therefore less disfavoured than this, and so the fraction of parameter space disfavoured by this first update must be larger than we compute, making our estimated PBFs for it conservative. In subsequent updates the volume of parameter space left viable in the $\mu<0$ branch would be this factor of 20-60 smaller, and so changes to it would contribute by the same factor less to the corresponding PBFs, rendering it quite unimportant for those updates.

\subsubsection{Logarithmic prior}

Based on naturalness arguments there is a strong belief that all CMSSM parameters with mass dimensions, $\{M_0, M_{1/2}, A_0\}$, should be low. A flat ``pre-LEP'' prior distribution for these parameters would strongly conflict with this belief; with a flat prior a mass parameter would be considered 10 times more likely to be between 1 TeV and 10 TeV than between 100 GeV and 1 TeV, increasing 10 fold again each order of magnitude, which we consider undesirable. In contrast logarithmic priors favour neither low nor high scales, and so may be argued to represent a `neutral' position on the issue of naturalness. Such a prior is flat in $\log(\theta)$, resulting in $P(\theta|\,\text{CMSSM}) \propto 1/\theta$.  The log prior has the additional mathematical attraction of being the Jeffreys prior~\cite{jeffreys1961theory} for a scale parameter, i.e. it is minimally informative in the sense that it maximises the difference between the prior and the posterior for such parameters.  In our case $M_0$ and $M_{1/2}$ are assigned independent log priors, while $A_0$, and $\tan\beta$ are left with flat priors.  We make the latter choices because $A_0$ ranges over positive and negative values and so resists a log prior (due to the divergence that would occur at $|A_0|=0$) and because $\tan\beta$ varies over only one order of magnitude.  Each of these beliefs about the parameters are considered to be statistically independent, so the full prior is obtained simply by multiplying them all together:
\begin{equation}
 P(M_0,M_{1/2},A_0,\tan\beta|\,\text{CMSSM})
 \propto\frac{1}{M_0 M_{1/2}} .
\label{eq:logprior}
\end{equation}
Numerous studies of the CMSSM have already been performed using this prior~\cite{AbdusSalam:2009tr,Bertone:2011nj,Feroz:2011bj,Allanach:2011ut}, making it a good standard prior to consider. Such studies often also employ a flat prior in the mass parameters however we do not, for the reasons explained above, preferring to save our CPU time for the natural priors discussed below.

\subsubsection{Natural prior}

Naturalness is a theoretical consideration which can be used to set the shape of the ``pre-LEP'' prior distributions within the CMSSM, and which can be quantified in terms of fine-tuning.  Several measures of the degree of fine-tuning in a model exist\cite{Athron:2007ry}, but probably the most well known is the Barbieri-Guidice measure~\cite{barbieri1988upper}
\begin{equation}
 \Delta=\left|\frac{\partial\ln m_Z^2}{\partial\ln \theta}\right| ,
\label{eq:ftmeas}
\end{equation}
which quantifies the sensitivity of the $Z$ boson mass to the variation of the parameter $\theta$. In ref. \cite{Allanach:2007qk}, and later \cite{Cabrera:2008tj, Cabrera:2009dm, Cabrera:2010dh}, it was shown that a prior incorporating this measure to penalise high fine-tuning can be constructed from purely Bayesian arguments. This prior has the additional benefit of explicitly acknowledging the experimental data available prior to the ``pre-LEP'' update, specifically the $Z$ mass, on which \emph{all} priors for this update (and subsequent updates) are conditional (a discussion of the importance of this notion can be found in section~\ref{sec:term}).

The key idea is relaxing the usual requirement of the CMSSM that the $\mu$ parameter is fixed by the experimental value of $m_Z$ through the Higgs potential minimisation conditions and instead incorporating $m_Z$ into a likelihood function. One then starts from flat priors over the ``natural'' parameter set $M_0$, $M_{1/2}$, $A_0$, $B$ and $\mu$. Next the observed $m_Z$ is used to perform a Bayesian update, $\mu$ is marginalised out, and a transformation to the usual CMSSM parameter set is performed, introducing a Jacobian term penalising high $\tan\beta$ and a fine-tuning coefficient penalising high $\mu$ values, giving us the natural (or ``CCR'') prior~\cite{Roszkowski:2009ye}
\begin{multline}
 P_{\text{eff}}(M_0, M_{1/2} ,A_0, \tan\beta|\,\text{CMSSM}) \\
 \propto\frac{\tan^2\beta-1}{\tan^2\beta(1+\tan^2\beta)}\frac{B_{\text{low}}}{\mu_Z} .
\label{eq:CCRprior}
\end{multline}
Here $B_{\text{low}}$ is the low energy value of the $B$ parameter and $\mu_Z$ is the $\mu$ value required to produce the correct Z mass.  Operationally, we implement this prior by scanning the conventional parameter set with a flat prior and multiplying the above expression into the likelihood function.\footnote{We do this because $P_{\text{eff}}$ requires renormalisation group running to be evaluated, i.e. our spectrum generator needs to be run before we can evaluate $P_{\text{eff}}$.}

The above prior does not fully implement the Barbieri-Guidice measure because it only considers the fine-tuning of the $\mu$ parameter. In ref. \cite{Cabrera:2008tj} an extended version of this prior is constructed which also considers the tuning of the Yukawa couplings, and in refs.~\cite{Ghilencea:2012gz,Ghilencea:2012qk} a generalisation to the full parameter set is considered, but we choose to focus only on the simpler version in this work, since it captures a large amount of the fine-tuning effect and can be computed analytically once the spectrum generator ({\tt ISAJET}) has run.

\subsection{Effect of the parameter ranges on partial Bayes factors}
\label{sec:scanrange}

In many recent studies of the CMSSM, only relatively low mass regions of the parameter space have been considered, generally regions not much larger than $0<M_0,M_{1/2}<1$ TeV.  This is in part motivated by naturalness arguments, in part by the generally lower likelihood outside this region (largely driven by a poorer fit to $\delta a_\mu$), and perhaps largely because the LHC SUSY search limits will not reach deeper into the parameter space than this for several years yet.  Ideally, since we would like to consider changes in the total evidence for the CMSSM, it is desirable to consider the entire viable parameter space, since the more viable space that exists outside the LHC reach, the less the CMSSM will appear to be harmed by it.  However, it is extremely difficult to thoroughly scan the CMSSM out to very large values of $M_0$ and $M_{1/2}$ due to the computational expense of obtaining reliable sampling statistics. In addition, our study is primarily concerned with obtaining Bayesian evidence values, which involve integrals over the parameter space and so require us to perform particularly thorough scans in order to acquire them to sufficient accuracy for our study. If one is only concerned with identifying the major features of the posterior then less thorough scans often suffice.

Due to this, we focus only on the low mass region of the CMSSM. The apparent impact of the LHC on the CMSSM will thus be increased, though it will be a faithful estimate of the damage done to the low mass region. Importantly, the change this restriction makes to the final partial Bayes factors will be determined by the volume of the ``pre-LEP'' posterior (i.e. ``LEP+Xenon'' prior) that we neglect, not the full change of volume of the ``pre-LEP'' scan priors, as occurs for the global evidence. This is because our incremental evidences are ratios of global evidences, so factors due to the ``pre-LEP'' scan prior volume divide out. Indeed, were our scans to contain $100\%$ of the ``pre-LEP'' posterior, then further increases in the scan prior volume would have no effect at all\footnote{In practice a larger scan volume will decrease the scan resolution and reduce the accuracy of results, so scan prior volume dependence would still exist in this indirect form.}.

Finally, we should consider the bias that exists in our assessment of the damage done to the CMSSM due to our choice of the SM+DM as the alternate model. There of course exist numerous models which may be of more direct interest as alternatives to the CMSSM, which suffer more damage than our SM-like alternate does due to their larger parameter spaces, and comparing the CMSSM to these we would conclude that the posterior odds for it were better than when compared to our SM-like model. This consideration forms part of our motivation to present our results in terms of both partial Bayes factors and the constituent likelihood ratios and Occam factors, as we hope this allows the reader to more easily understand how changes in alternate model would affect our inferences. We will return to this discussion when we present our results in section~\ref{sec:results}. 



A summary of the priors and ranges used for this study are presented in table~\ref{tab:ranges}.

\begin{table}\footnotesize
   \begin{center}
    \begin{threeparttable}
      \begin{tabular}{| l | r |}
         \multicolumn{2}{l}{\textbf{Priors}} \\
         \hline
         Name & PDF \\
         \hline
         Log & $1/(M_0 M_{1/2})$ \\
         Natural & $\displaystyle\frac{\tan^2\beta-1}{\tan^2\beta(1+\tan^2\beta)}\frac{B_{\text{low}}}{\mu_Z}$ \\
         \hline
         \noalign{\smallskip}
         \multicolumn{2}{l}{\textbf{Ranges}} \\
         \hline
         Parameter & Range \\
         \hline
         $M_0$ & 10\,GeV -- 2\,TeV \\
         $M_{1/2}$ & 10\,GeV -- 2\,TeV \\
         $A_0$ & $-$3\,TeV -- 4\,TeV \\
         $\tan\beta$ & 0 -- 62 \\
	 $\text{sign}(\mu)$ & $+1$	\\
         $m_{t}$ & $172.9 \pm 1.1$\,GeV \\
         \hline

      \end{tabular}
    \end{threeparttable}
    \caption{\label{tab:ranges}Summary of the priors and ranges used in this study. The displayed PDFs for both priors are multiplied by a Gaussian for $m_t$ with mean and width specified by the values adjacent to $m_t$ in the table.}
   \end{center}
\end{table}

\section{Likelihood function}
\label{sec:likefunc}

We now detail the experimental data which goes into our likelihood function. Primarily this is summarised in table~\ref{tab:likefunc}.  Each of these components is considered to be statistically independent, so that the global likelihood function is simply the product of them:
\begin{equation}
 \mathscr L_{\text{Global}} = \prod_i \mathscr L_i = \prod_i P(d_i|\theta,\text{CMSSM}) .
\end{equation}
The 2011 constraints from the XENON100 experiment and the LHC Higgs and sparticle searches cannot be implemented via likelihood functions simple enough to list in a table, so we explain our treatment of them in sections~\ref{sec:special1} through ~\ref{sec:LHCHiggsL}.

\begin{table*}			
\begin{minipage}{\linewidth}	
\footnotesize
\centering
    \begin{threeparttable}	
      \begin{tabular}{| l | l | l | l |} 
         \hline
         Observable & Measured value & Computed by & Sources \\
         \hline
         \noalign{\smallskip}
         \multicolumn{2}{c}{\textbf{Gaussian likelihoods}} \\
         \hline
         $\Delta \rho$ & $0.0008\pm0.0017$ & \texttt{micrOmegas~2.4.Q} & \cite{PDG:2010}\footnotemark[1] \\
         $\Omega_\chi h^2$ & $0.1123\pm0.0035\pm10\%$ & \texttt{micrOmegas~2.4.Q} & \cite{Komatsu:2010fb}\footnotemark[2] \\
         $\delta a_{\mu}$ & $3.353\pm8.24 \, [\times 10^{-9}]$ & \texttt{micrOmegas~2.4.Q} & \cite{Benayoun:2011mm}\footnotemark[3] \\
         $BR(b\rightarrow s \gamma)$ & $3.55\pm0.26\pm5\% \, [\times 10^{-4}]$ & \texttt{SuperISO~3.1} & \cite{HFAG:2010}\footnotemark[4] \\
         $BR(B\rightarrow \tau \nu)$ & $1.67\pm0.39 \, [\times 10^{-4}]$ & \texttt{SuperISO~3.1} & \cite{HFAG:2010}\footnotemark[5] \\
         $\Delta_{0-}(B\rightarrow K^* \gamma)$ & $0.416\pm0.128$ & \texttt{SuperISO~3.1} & \cite{BABAR:2008cy}\footnotemark[6] \\
         $BR\left(\frac{B^+\rightarrow D_0 \tau \nu}{B^+\rightarrow D_0 e \nu}\right)$ & $0.029\pm0.039$ & \texttt{SuperISO~3.1} & \cite{Aubert:2007dsa}\footnotemark[7] \\
         $R_{l23}$ & $1.004\pm0.007$ & \texttt{SuperISO~3.1} & \cite{Antonelli:2008jg}\footnotemark[8] \\
         $BR(D_s\rightarrow \tau \nu)$ & $0.0538\pm0.0038$ & \texttt{SuperISO~3.1} & \cite{HFAG:2010}\footnotemark[9] \\
         $BR(D_s\rightarrow \mu \nu)$ & $5.81\pm0.47 \, [\times 10^{-3}]$ & \texttt{SuperISO~3.1} & \cite{HFAG:2010}\footnotemark[10] \\
	 \hline
         \noalign{\smallskip}
         \multicolumn{2}{c}{\textbf{Limits (erf)}} \\
         \hline
         $m_{\tilde g}$ & $>289 \pm 15$\,GeV (LEP2) & \texttt{ISAJET~7.81} & \cite{PDG:2006} \\
         $m_h$ & $>x\pm3$\,GeV\tnote{a} \, (LEP2) & \texttt{ISAJET~7.81} & \cite{Barger:2006dh}\footnotemark[11] \\
         \hline
         \noalign{\smallskip}
         \multicolumn{2}{c}{\textbf{Limits (hard cut)}} \\
         \hline
         \multicolumn{2}{|l|}{Other LEP2 direct sparticle mass 95\% C.L.'s} & \texttt{ISAJET~7.81} & \cite{PDG:2006} \\
         \hline
         \noalign{\smallskip}
         \multicolumn{2}{c}{\textbf{Special cases}} \\
         \hline
         $\sigma_{\tilde \chi_{0}-p}^{SI}$ & See text (XENON100) & \texttt{micrOmegas~2.4.Q} & \cite{Aprile:2011hi} \\
         $BR(B_s\rightarrow \mu^+ \mu^-)$ & $< 1.5 \times 10^{-8}$ (LHCb) & \texttt{SuperISO~3.1} & \cite{Bettler:2011rp}\footnotemark[12] \\
         $m_h$ & See text (LHC) & \texttt{ISAJET~7.81}, \texttt{HDECAY~4.43} & \cite{ATLAS:2012si, ATLASaa:2012sk, ATLAS2l2v:2011aa, ATLAS4l:2012sm} \\
         SUSY searches & See text (LHC) & \texttt{ISAJET~7.81}, \texttt{Herwig++~2.5.2}, & \cite{Aad:2011ib} \\
	   &  & \texttt{Delphes~1.9}, \texttt{PROSPINO~2.1} &  \\
         \hline
      \end{tabular}
      \begin{tablenotes}
         \item [a] $x$ determined from figure~3a of ref. \cite{Barger:2006dh} for each point. For nearly all CMSSM points $x = 114.4$\,GeV.
      \end{tablenotes}
      \end{threeparttable}
\caption{Summary of the likelihood functions and experimental data used in this analysis. Gaussian likelihoods: Likelihoods are modelled as Gaussians; where two uncertainties are stated the first arises from experimental/Standard Model sources, while the second is an estimate of the theoretical/computational uncertainty in the new physics contributions (and these are added in quadrature), otherwise the latter uncertainty is assumed to be small and treated as zero. Limits (erf): The listed central values are estimated 95\% C.L.'s, and are used to define a step function cut, which is convolved with the stated Gaussian estimate of the total (experimental and computation-based) uncertainty. Limits (hard cut): Step function likelihoods centred on the cited 95\% C.L.'s are used. Special cases: For details see sections~\ref{sec:special1} though \ref{sec:LHCHiggsL} (and footnote 12 for $B_s\rightarrow \mu^+ \mu^-$).}
\label{tab:likefunc}

\footnotetext[1]{Section `Electroweak model and constraints on new physics', p. 33 eq. (10.47). We take the larger of the 1 sigma confidence interval values. The full likelihood function is actually highly asymmetric and slightly disfavours values close to the Standard Model prediction, which we are effectively ignoring.}
\footnotetext[2]{Table 1 (WMAP + BAO + H0 mean). Theoretical uncertainties are not well know so we follow the estimates of ref. \cite{Bertone:2011nj}.}
\footnotetext[3]{Table 10 (Solution B).}
\footnotetext[4]{Table 129 (Average).}
\footnotetext[5]{Table 127.}
\footnotetext[6]{Page 17, uncertainties combined in quadrature.}
\footnotetext[7]{Table 1 (R value)}
\footnotetext[8]{Eq. (4.19).}
\footnotetext[9]{Figure 68, p. 225 (World average).}
\footnotetext[10]{Figure 67, p. 224 (World average).}
\footnotetext[11]{Figure 3a, p. 24.}
\footnotetext[12]{Figure 8. We use the full $CL_s$ curve rather than simply the 95\% confidence limit. Working backward from the $CL_s$ values given by the curve, assuming them to be instead $CL_{s+b}$ values, we determine the corresponding likelihood function which would generate these values (assuming a chi-square distributed test statistic). $CL_s$ intervals over-cover so this procedure is conservative.}
\end{minipage}
\end{table*}

\subsection{XENON100 limits}
\label{sec:special1}

The likelihood contribution from XENON does not yet have a significant impact on the CMSSM evidence so we have opted to simply model the likelihood as an error function of the WIMP-nucleon spin-independent elastic scattering cross section, which varies with the WIMP mass.  Our likelihood function for the cross section ($\sigma_{\tilde \chi_{0}-p}^{SI}$) is derived from the 90\% confidence limits published by the XENON100 experiment in figure~5 of ref. \cite{Aprile:2011hi}.  This limit is presented as a function of WIMP mass.  We fit the likelihood function with an error function such that it reproduces the correct 90\% C.L. and the correct apparent significance of the upper edge of the $1\sigma$ sensitivity band, based on the maximum likelihood ratio method, using a similar procedure to that used in ref. \cite{Buchmueller:2011aa} to estimate their likelihood function for a CMS multi-jet$+\cancel{E}_T$ search, which we summarise below.

XENON use the profile likelihood ratio test statistic
\begin{align}
 Q = -2\log(\lambda) &= -2\log\left(\frac{L_{s+b}(\sigma^\text{SI}_{pX};m_X)}{L_{s+b}(\hat{\sigma}^\text{SI}_{pX};m_X)}\right) \nonumber \\
&= -2\log\left(\frac{P(\text{data}|m_X,\sigma^\text{SI}_{pX})}{P(\text{data}|m_X,\hat\sigma^\text{SI}_{pX})}\right)
\end{align}
\label{eq:XENONQ}
to derive their exclusion limits, where $m_X$, $\sigma^\text{SI}_{pX}$ and $\hat{\sigma}^\text{SI}_{pX}$ are the hypothesised WIMP mass, spin-independent WIMP-proton scattering cross section, and best fit value of the latter for each mass slice, respectively. All nuisance variables are profiled over and limits are derived on the cross section for each fixed $m_X$, so the resulting profile likelihood ratio has one degree of freedom and $Q$ is asymptotically distributed as $f(Q|\sigma;m_X)=\chi^2_{k=1}(Q)$ (which XENON100 have confirmed is true to a good approximation via Monte Carlo \cite{Aprile:2011hx}). The cross section is proportional to the mean signal event rate $\mu$ for each $m_X$ slice,
so we may use the asymptotic expressions of \cite{cowan2011asymptotic} to express $Q$ in terms of $\mu$ as
\begin{equation}
Q = \frac{(\mu-\hat\mu)^2}{a^2},
\end{equation}
\label{eq:XENONQ2}
where $\hat\mu$ is the best fit signal event rate for some observed data, which is normally distributed with standard deviation $a$\footnote{We ignore the variation of $a$ with the predicted signal rate as it is small for small signal.}.

XENON report the observation of 3 events in their signal region, with an expected background of $1.8\pm0.6$ events, so $a=0.6$ and $\hat\mu=1.2$. The upper $90\%$ confidence limit is drawn on the contour on the $(m_X,\sigma^\text{SI}_{pX})$ plane on which the predicted mean event rate drops to the level producing $Q$ such that $p_s=\int \! \mathrm d Q \, f(Q|\sigma;m_X) = 0.1$\footnote{Actually the $CL_s$ method is used so the limit is drawn where $p=p_s/(1-p_b)=0.1$ \cite{Read:CLs}, but this correction weakens the limit so it is conservative to ignore it and in this case makes little difference anyway, given our other approximations.}, or $Q=2.71$. To fit our erf model likelihood we require a second contour of $Q$, and the expected$+1\sigma$ limit is a convenient choice. On this contour, a hypothetical observation of $1.8+0.6=2.4$ events is assumed, which produces a best fit signal mean of $\hat\mu=0.6$. Again the $90\%$ confidence limit is drawn where $Q$ drops to 2.71, which occurs at $\mu=\hat\mu+\sqrt{2.71}a \approx 1.59$. Knowing the predicted signal rate on this contour now allows us to infer the value of $Q$ on this contour given the actual observed data, i.e. from eq. (\ref{eq:XENONQ2}) $Q\approx(1.59-1.2)^2/0.6^2\approx0.417$. Our erf likelihood is fitted to reproduce these contours for each $m_X$ slice, thus producing an approximation of the full likelihood function.\footnote{In section~\ref{sec:LHCHiggsL} we construct the ATLAS Higgs search likelihood function using almost identical techniques, but argue that each fitted slice needs to be normalised relative to the others using the likelihood of the best fit point on each slice. This occurs because the best fit point of each slice lies a varying number of standard deviations from the zero signal point (zero cross section), which we know to have the same likelihood for every slice. A similar normalisation is in principle required to recover the true likelihood function computed by Xenon, however the variance of the limit appears to be approximately Gaussian in the logarithm of the cross section, making extrapolation of the likelihood to the zero cross section point extremely unreliable. In addition, the reconstruction method we use for the ATLAS Higgs search likelihood relies on plots of the signal best fit against $m_X$, whereas here we use a plot of the 90\% confidence limit. Performing the extraction using the limit curve requires more assumptions than a best fit curve, so combined with the logarithmic difficulty we judge that this technique would produce poor results, and so we prefer to stick with the simpler technique described. The Xenon limit turns out to be of very minor importance to our final inferences anyway so we are not concerned with small errors in our reconstructed likelihood. In hindsight we expect that even simply applying a hard cut at the observed Xenon limit would negligibly affect our inferences.} 

\subsubsection{Hadronic uncertainties}

The above procedure is simplistic, but it gives us a good enough estimate of the experimental uncertainty associated with $\sigma_{\tilde \chi_{0}-p}^{SI}$ for our purposes. In addition to this, we fold in an estimate of the associated theoretical uncertainties, assumed to be dominated by the uncertainties in the strange quark scalar density in the nucleon, in turn due mainly to the experimental uncertainty in the $\pi$-nucleon $\sigma$ term, $\Sigma_{\pi N}\equiv1/2(m_u+m_d)\langle N|\bar uu+\bar dd|N\rangle$. Numerous estimates of this quantity exist ($59\pm7$ \cite{Alarcon:2011zs}, $79\pm7$ \cite{Pavan:2001wz}, $\sim \!\! 45$ \cite{Gasser1991252}, $64\pm8$ \cite{Koch:1982pu} [MeV]) and it is not clear which are the most reliable so we opt to use a recent value on the low end of the spectrum based on lattice calculations, with a wide uncertainty ($39\pm14$ \cite{Giedt:2009mr}\footnote{From eq. (5), using the suggested $\sigma_s=50\pm8$ MeV and $\sigma_l=47\pm9$ MeV\cite{Young:2009zb}, with $m_s/m_l=m_s/(2(m_u+m_d))=26\pm4$\cite{PDG:2010}.}, with $\sigma_0\equiv1/2(m_u+m_d)\langle N|\bar uu+\bar dd-2\bar ss|N\rangle=36\pm7$\cite{Gasser:1982ap,Borasoy:1996bx,Sainio:2001bq,Knecht:1999dp}) as this produces low $\sigma_{\tilde \chi_{0}-p}^{SI}$ predictions and so a conservatively weak XENON100 constraint.

The computation of $\sigma_{\tilde \chi_{0}-p}^{SI}$ is performed by \texttt{micrOmegas~v2.4.Q}, and it accepts $\Sigma_{\pi N}$ as an input parameter, along with $\sigma_0$. To estimate the uncertainty in the computed cross section due to these quantities, they were first used to estimate the corresponding parameters $f_{T_q}^{(N)}\equiv\langle N|m_q\bar qq|N\rangle/m_N$ and their uncertainties (following~\cite{Ellis:2008hf}), which \texttt{micrOmegas} computes internally and uses in its computation of $\sigma_{\tilde \chi_{0}-p}^{SI}$. We find these to be $f_{T_u}^{(p)}=0.016\pm0.007$, $f_{T_d}^{(p)}=0.023\pm0.010$ and $f_{T_s}^{(p)}=0.039\pm0.026$, in close agreement with the values \texttt{micrOmegas} computes internally from our chosen $\Sigma_{\pi N}$ and $\sigma_0$. We have modified \texttt{micrOmegas} so that our computed uncertainties on the $f_{T_q}^{(N)}$ are then propagated alongside the $f_{T_q}^{(N)}$ themselves in the computation of $\sigma_{\tilde \chi_{0}-p}^{SI}$ and used to estimate the uncertainty on $\sigma_{\tilde \chi_{0}-p}^{SI}$ for each model point. This uncertainty is then added in quadrature to the width of the $\sigma_{\tilde \chi_{0}-p}^{SI}$ erf likelihood function (i.e. convoluted into it).

\subsubsection{Astrophysical uncertainties}

In our model of the $\sigma_{\tilde \chi_{0}-p}^{SI}$ likelihood function, we do not rigorously consider the effects of varying the astrophysical assumptions that XENON have made in their construction of their confidence limits. In their analysis XENON assume WIMPs to be distributed in an isothermal halo with $v_0=220$ km/s, galactic escape velocity $v_{esc}=544_{-46}^{+64}$ km/s, and a density of $\rho_{\chi}=0.3$\,GeV/cm$^3$, and we cannot change these without developing a model of the likelihood function based directly on the event rate observed by XENON100, as is done in ref. \cite{Buchmueller:2011ki} and~\cite{Bertone:2011nj}, for example. We have opted not to do this as~\cite{Bertone:2011nj} shows that marginalising over a range of plausible values near the nominal choice makes negligible difference to the impact the XENON100 experiment has on the CMSSM, and we prefer to avoid the additional increase in the dimensionality of the problem.

\subsection{1\,$\text{fb}^{-1}$ LHC sparticle searches}
\label{sec:LHCsparticle}

In late 2011 the ATLAS and CMS experiments~\cite{Aad:2008zzm,cms:2008zzk} updated their searches for supersymmetric particles using the 2011 1\,$\text{fb}^{-1}$ data set~\cite{Aad:2011ib,Aad:2011zj,Aad:2011cw,Aad:2011qa,ATLAS:2011ad,Chatrchyan:1381201,CMS-PAS-SUS-11-005,CMS-PAS-SUS-11-004}. Data collected from proton collisions at the Large Hadron Collider at $\sqrt{s}=7$\,TeV are analysed in a variety of final states, none of which show a significant excess over the expected Standard Model background. 
As the LHC is a proton-proton collider, one expects to dominantly produce coloured objects such as squarks and gluinos, whose inclusive leptonic branching ratios are relatively small, and hence the strongest CMSSM exclusions result from the ATLAS searches for events with no leptons and the CMS searches for sparticle production in hadronic final states. The ATLAS and CMS limits have a similar reach in the squark and gluino masses, and here we consider only the ATLAS zero lepton limits for simplicity. Recent interpretations of LHC limit results can be found in~\cite{Buchmueller:2011aa,Desai:2011th,Buchmueller:2011sw,Beskidt:2011qf,Buchmueller:2011ki,Allanach:2011qr}.

The ATLAS signal regions were each tuned to enhance sensitivity in a particular region of the $M_0$--$M_{1/2}$ plane. Events with an electron or muon with $p_T > 20$\,GeV were rejected. Table~\ref{tab:cuts} summarises the remaining selection cuts for each region, whilst table~\ref{tab:results} gives the observed and expected numbers of events. These numbers were used by the ATLAS collaboration to derive limits on $\sigma \times A \times \epsilon$, where $\sigma$ is the cross section for new physics processes for which the ATLAS detector has an acceptance $A$ and a detector efficiency of $\epsilon$. These results are also quoted in table~\ref{tab:results}.

\begin{table*}\footnotesize
\begin{center}
\begin{tabular}{|c|c|c|c|c|c|}
\hline
Region & R1 & R2 & R3 & R4 & RHM\\
\hline
Number of jets&$\ge 2$&$\ge 3$&$\ge 4$&$\ge 4$&$\ge 4$\\
$E_T^\text{miss}$ (GeV) & $>130$&$>130$ &$>130$ &$>130$ &$>130$ \\
Leading jet $p_{T}$ (GeV) & $>130$  &$>130$ &$>130$ &$>130$ &$>130$ \\
Second jet $p_{T}$ (GeV)& $>40$  & $>40$  &$>40$  & $>40$  & $>80$  \\
Third jet $p_{T}$ (GeV)& - & $>40$  &$>40$  & $>40$  & $>80$  \\
Fourth jet $p_{T}$ (GeV) & - & -  &$>40$  & $>40$  & $>80$  \\
$\Delta\phi($jet, $p_{T}^\text{miss})_\text{min}$ & $>0.3$ & $>0.25$ & $>0.25$ & $0.25$ & $>0.2$\\
$m_\text{eff} (GeV)$ & $>1000$ &  $>1000$  &  $>500$  &  $>1000$  &  $>1100$ \\
\hline
\end{tabular}
\end{center}
\caption{\label{tab:cuts} Selection cuts for the five ATLAS zero lepton signal regions. $\Delta\phi($jet, $p_{T}^\text{miss})_\text{min}$ is the smallest of the azimuthal separations between the missing momentum $p_{T}^\text{miss}$ and the momenta of jets with $p_T > 40$\,GeV (up to a maximum of three in descending $p_T$ order). The effective mass $m_\text{eff}$ is the scalar sum of $E_T^\text{miss}$ and the magnitudes of the transverse momenta of the two, three and four highest $p_T$ jets depending on the signal region. In the region RHM, all jets with $p_T>40$\,GeV are used to define $m_\text{eff}$.}
\end{table*}

\begin{table*}\footnotesize
\begin{center}
\begin{tabular}{|c|c|c|c|c|c|}
\hline
Region & R1 & R2 & R3 & R4 & RHM\\
\hline
Observed & 58 & 59 & 1118 & 40 & 18\\
Background & $62.4\pm 4.4\pm 9.3$&$54.9\pm 3.9\pm 7.1$&$1015 \pm 41\pm144$&$33.9\pm2.9\pm6.2$&$13.1\pm1.9\pm2.5$\\
$\sigma \times A \times \epsilon$ (fb)&22&25&429&27&17\\
\hline
\end{tabular}
\end{center}
\caption{\label{tab:results} Expected background yields and observed signal yields from the ATLAS zero lepton search using 1 fb$^{-1}$ of data~\cite{Aad:2011ib}. The final row shows the ATLAS limits on the product of the cross section, acceptance and efficiency for new physics processes.}
\end{table*}

The ATLAS collaboration have used the absence of evidence of sparticle production in 1\,fb$^{-1}$ of data to place an exclusion limit at the 95\% confidence level in the $M_0$--$M_{1/2}$ plane of the CMSSM for fixed $A_0$ and $\tan\beta$, and for $\mu > 0$, and all previous phenomenological interpretations of this limit in the literature have also ignored the $A_0$ and $\tan\beta$ dependence.  Ref. \cite{Allanach:2011ut}, for example, finds a negligible dependence of the limits on $A_0$ and $\tan \beta$. It is not guaranteed that this conclusion extends to the present limits, which are considerably stronger, so we reassess the $A_0$ $\tan \beta$ dependence of the new limits. To do this we simulate our own signal events for points in the full CMSSM using standard Monte Carlo tools coupled with machine learning techniques to reduce the total simulation time.

This section is structured as follows. Firstly, we explain and validate the tools we use to go from a set of CMSSM parameters to a signal expectation. We then examine why it can potentially be important not to neglect $A_0$ and $\tan\beta$ in LHC limits, by showing a class of model that fits the ATLAS data well but would be missed if one were to assert the limit as at $A_0 = 0$. Finally we address the fact that, when updating the posterior distributions obtained pre-LHC with the ATLAS results, it is not feasible to simulate every point in the posterior.  We therefore spend the remainder of this section developing a fast simulation technique derived by interpolating the output of a much smaller number of simulated points using a Bayesian Neural Network.

\subsubsection{Simulating the ATLAS results}

Given a signal expectation for a particular model, one can easily evaluate the likelihood of that model using the published ATLAS background expectation and observed event yield in each search channel. By simulating points in the full CMSSM parameter space, we can therefore investigate the LHC exclusion reach, provided that we can demonstrate that our simulation provides an adequate description of the ATLAS detector.

In this paper, we use \texttt{ISAJET~7.81}~\cite{Paige:2003mg} to produce SUSY mass and decay spectra then use \texttt{Herwig++~2.5.2}~\cite{Gieseke:2011na} to generate 15,000 Monte Carlo events. \texttt{Delphes~1.9}~\cite{Ovyn:2009tx} is subsequently used to provide a fast simulation of the ATLAS detector. The total SUSY production cross section is calculated at next-to-leading order using \texttt{PROSPINO~2.1}~\cite{Beenakker:1996ed}, where we include all processes except direct production of neutralinos, charginos and sleptons since the latter are sub-dominant. The ATLAS set-up differs from this only in the final step of detector simulation, where a full, \texttt{Geant\,4}-based simulation\cite{Agostinelli:2002hh} is used to provide a very detailed description of particle interactions in the ATLAS detector at vast computational expense.

It is clear that the \texttt{Delphes} simulation will not reproduce every result of the advanced simulation. Nevertheless, one can assess the adequacy of our approximate results by trying to reproduce the ATLAS CMSSM exclusion limits. We have generated a grid of points in the $M_0$--$M_{1/2}$ plane using the same fixed values of $\tan\beta=10$ and $A_0=0$ as the published ATLAS result. We must now choose a procedure to approximate the ATLAS limit setting procedure. ATLAS use both $CL_s$ and profile likelihood methods to obtain a 95\% confidence limit, using a full knowledge of the systematic errors on signal and background. Although the systematic error on the background is provided in the ATLAS paper, we do not have full knowledge of the systematics on the signal expectation, which may in general vary from point to point in the $M_0$--$M_{1/2}$ plane. Rather than implement these statistical techniques, we take a similar approach to that used in~\cite{Allanach:2011qr}, and use the published $\sigma \times A \times \epsilon$ limits to determine whether a given model point is excluded in a search channel. We use our simulation to obtain the $\sigma \times A \times \epsilon$ value for a given model point, and consider the model to be excluded if the value lies above the limit given in table~\ref{tab:results}. This allows us to draw an exclusion contour in each search channel, and we estimate the combined limit by taking the union of the individual exclusion contours for each channel (i.e. the most stringent search channel for a given model is used to determine whether it is excluded).  This method is not statistically rigorous, but it is conservative in the asymptotic limit for observations close to the expected background, for small signal hypotheses, assuming only positive linear correlations between channels\footnote{We demonstrate this in appendix \ref{app:approxCLs}}, and our scenario does not significantly depart from these conditions. Furthermore, the channel combination performed by ATLAS is very similar to our method: ATLAS estimate the combined limit by taking the limit from the channel with the most powerful \emph{expected} limit at each model point, whereas we take the most powerful \emph{observed} limit. Some further discussion of this difference can be found in appendix \ref{app:approxCLs}, though we conclude that the impact on our analysis is negligible.

\begin{figure*}
\sixgraphs{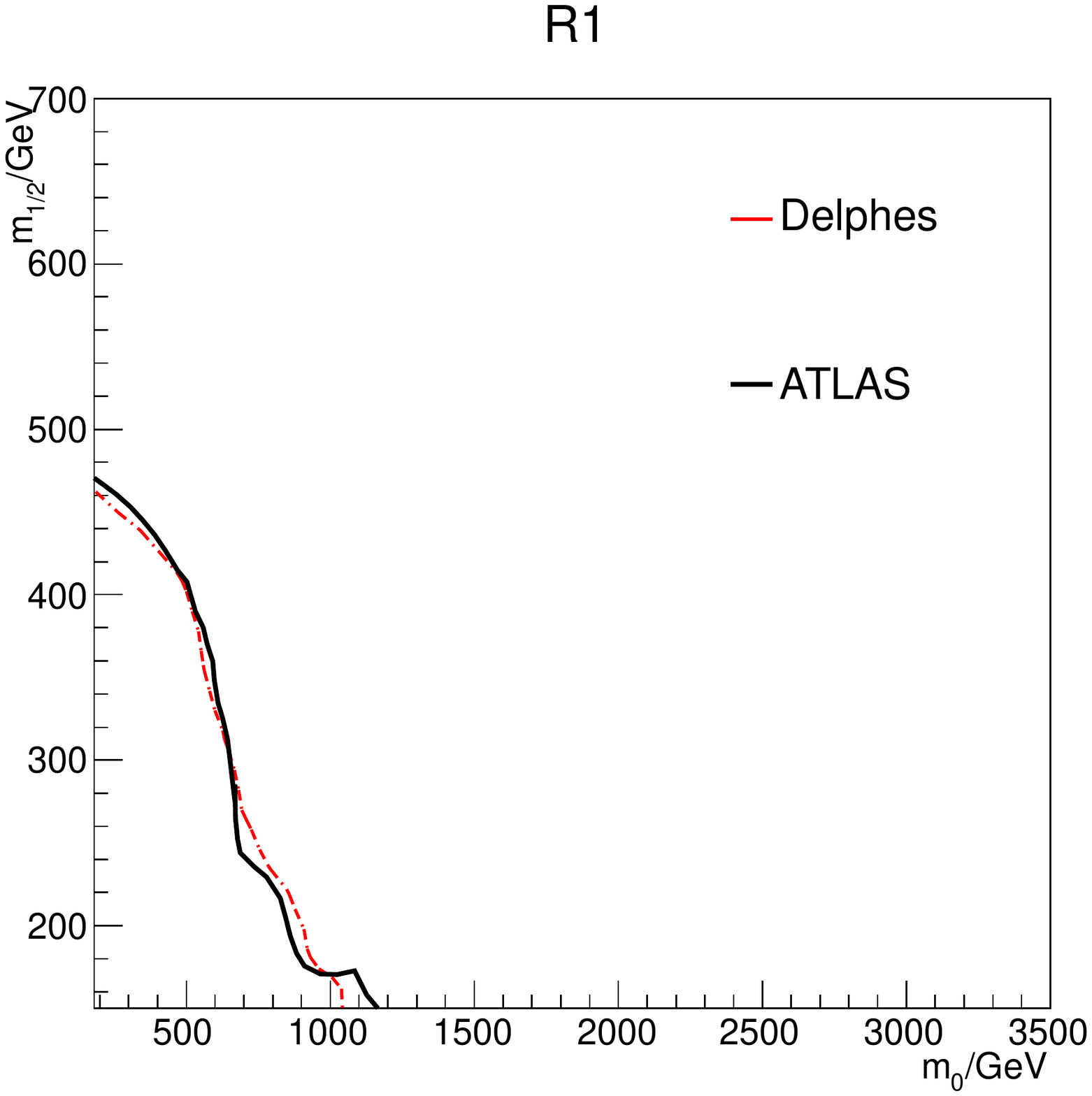}{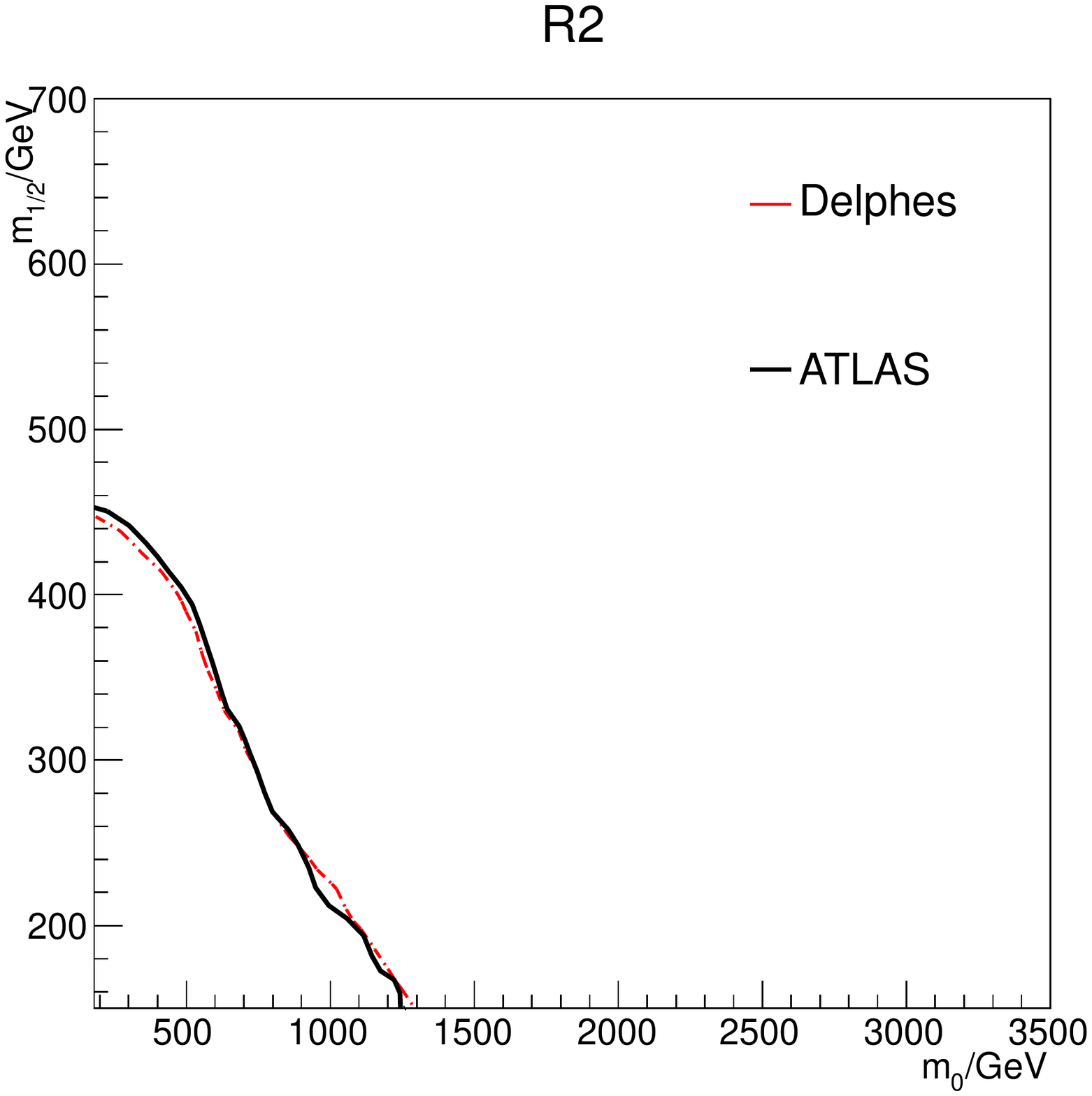}{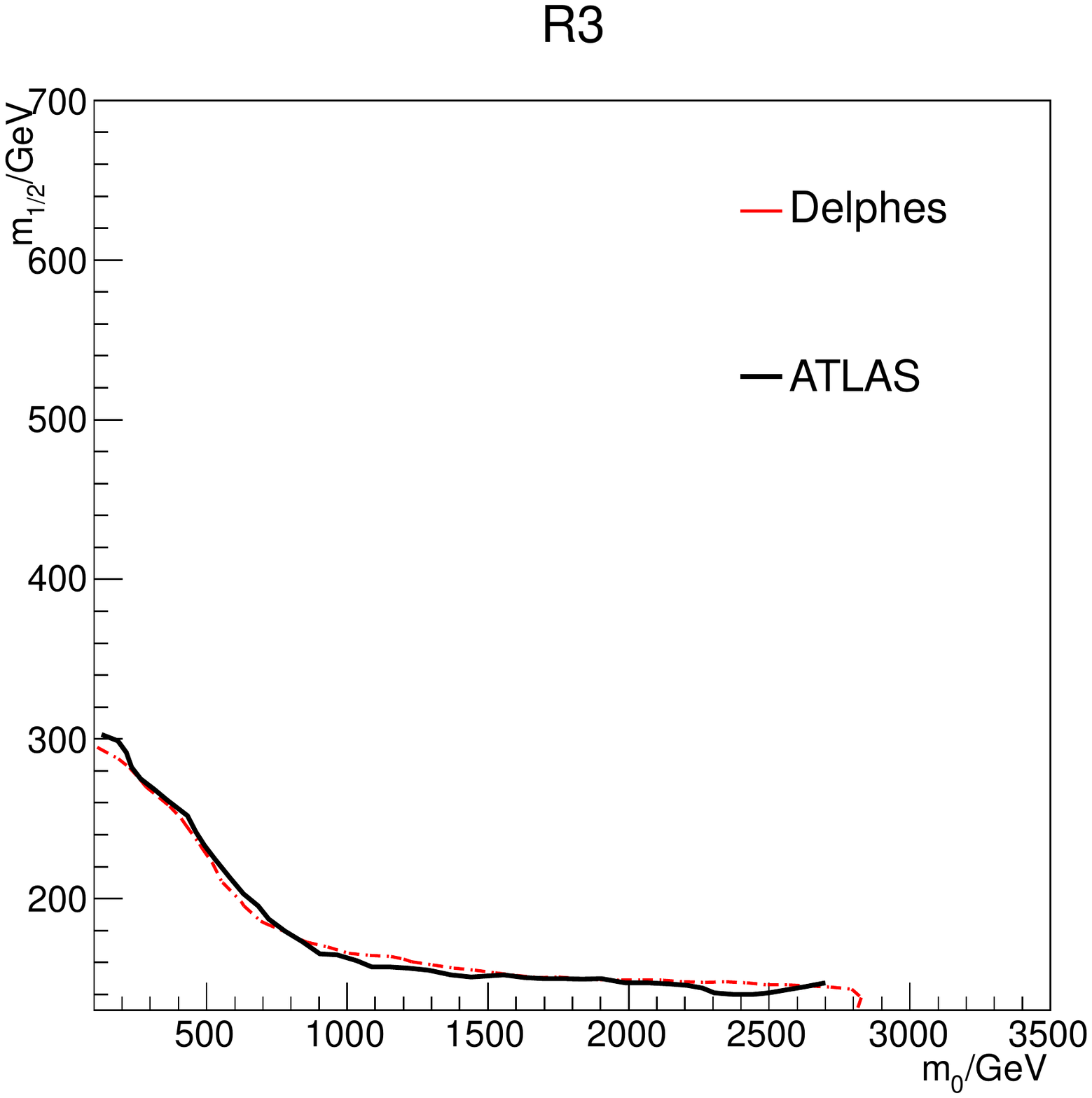}{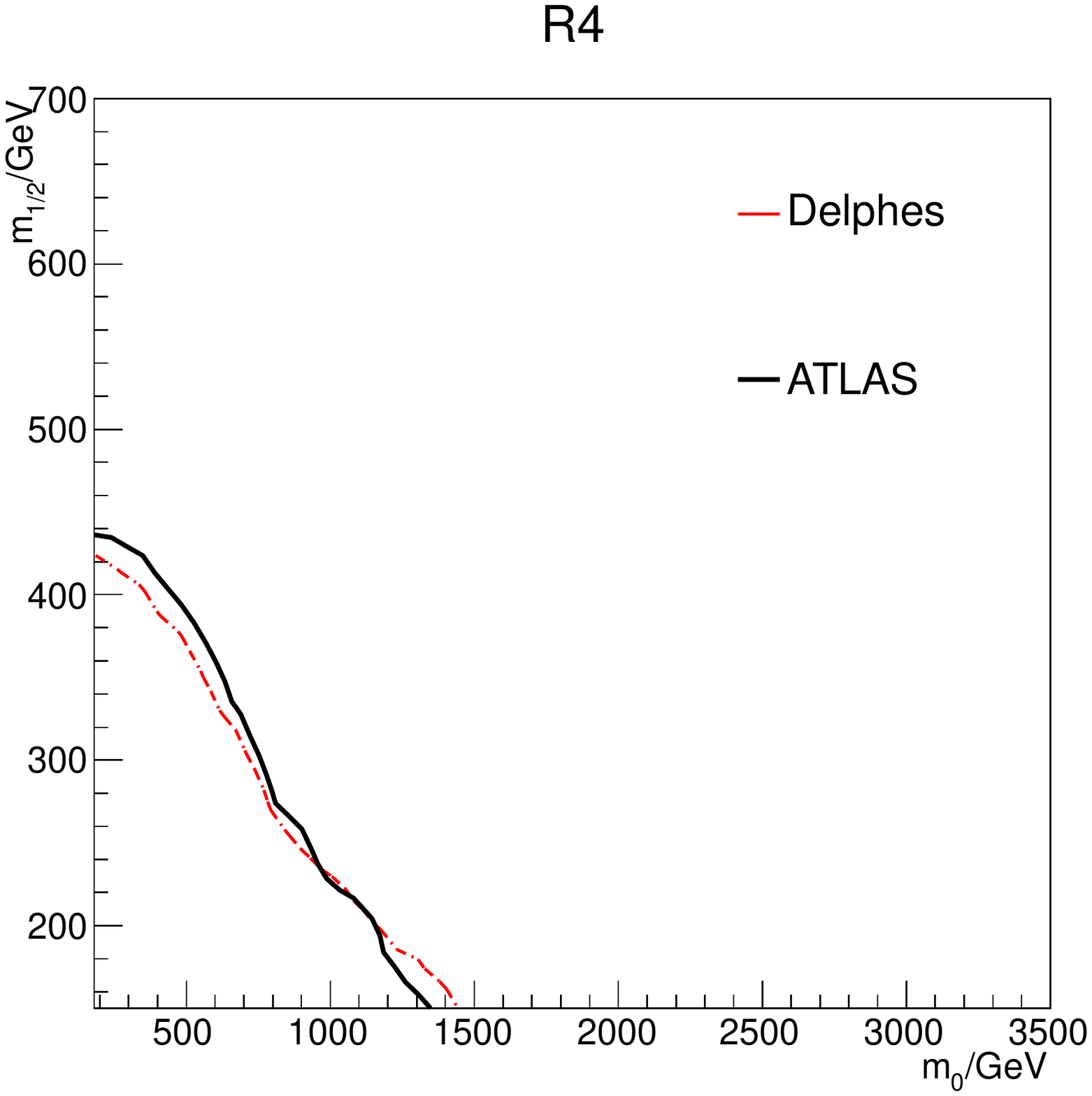}{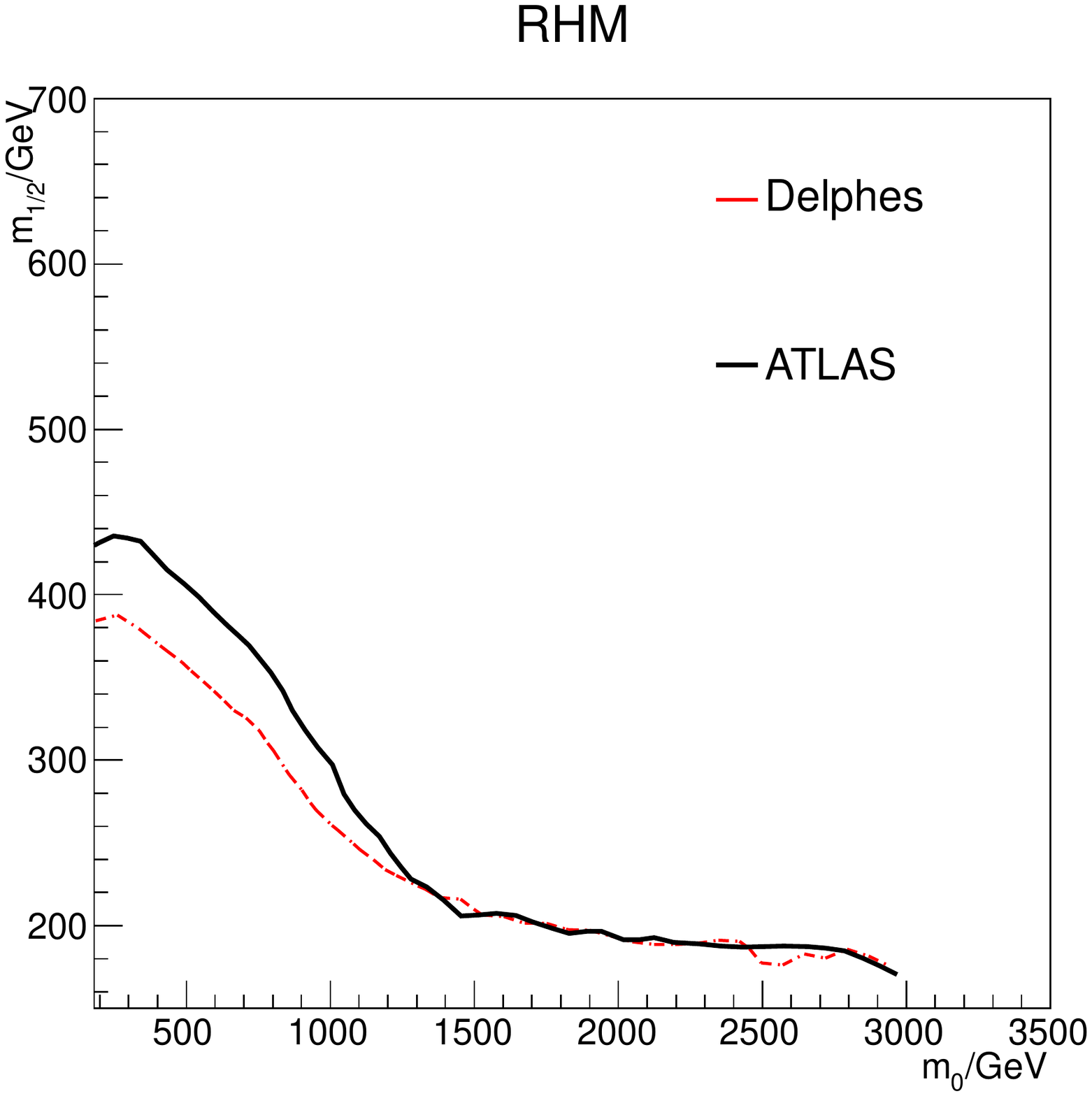}{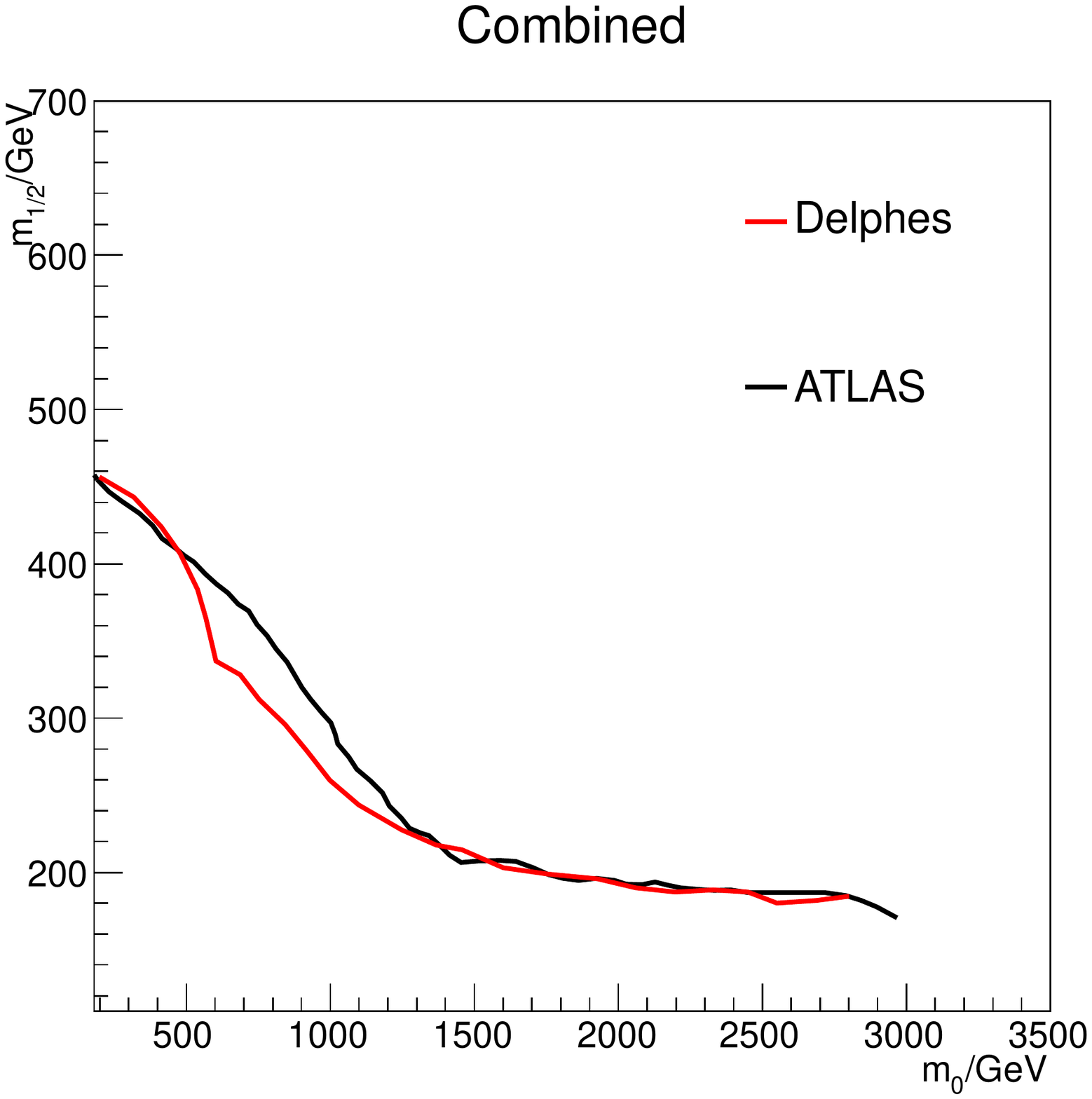}
\caption{Comparison between \texttt{Delphes} and ATLAS 95\% exclusion limits in the $M_0$--$M_{1/2}$ plane, for the signal regions R1, R2, R3, R4 and RHM defined in table~\ref{tab:cuts}. In the combined limit plot, the ATLAS limit is obtained using the ATLAS statistical combination, whilst the \texttt{Delphes} limit is obtained by taking the union of the \texttt{Delphes} limits for each signal region.}
\label{fig:limits1}
\end{figure*}

The procedure defined above neglects systematic errors on the signal and background yields and, as noted in~\cite{Allanach:2011qr}, this leads to a discrepancy between the \texttt{Delphes} results and the ATLAS limits in each channel. We follow~\cite{Allanach:2011qr} in using a channel dependent scaling to tune the \texttt{Delphes} output so that the limits in each channel match as closely as possible ``by eye''. We obtain factors of 0.82, 0.85, 1.25, 1.0 and 0.70 for the R1, R2, R3, R4 and RHM regions respectively. Comparisons between the resulting \texttt{Delphes} exclusion limit and the ATLAS limit are shown in figure~\ref{fig:limits1}, where we observe generally good agreement in all channels. The largest discrepancy is observed in the RHM channel, where we find that one cannot get the tail of the limit at large $M_0$ to agree with the ATLAS limit whilst simultaneously guaranteeing good agreement at low $M_0$. This is likely to be due to the fact that we have effectively assumed a flat systematic error over the $M_0$--$M_{1/2}$ plane. whereas the ATLAS results use a full calculation of the systematic errors for each signal point. It is important to notice however that the \emph{combined} limit will be dominated by regions R1 and R2 at low $M_0$, and thus by choosing to tune the RHM results in order to reproduce the large $M_0$ tail, one can ensure reasonable agreement of the combined limit over the entire range. Where disagreement remains, the \texttt{Delphes} limit is less stringent than the ATLAS limit, and hence using it gives us a conservative estimate of the ATLAS exclusion reach.

\subsubsection{The importance of $A_0$ and $\tan\beta$}

The ATLAS results in table~\ref{tab:results} demonstrate a small excess in the central value of the observed yield in the high multiplicity channel, RHM. Although one should assert that this has an innocent explanation (mostly likely an underestimate of the number of high multiplicity events in the SM due to a deficient Monte Carlo generator), it provides motivation to consider SUSY models in which there is a smaller amount of coloured sparticle production than in the bulk of the low mass CMSSM parameter space.

\begin{figure}
\centerline{
\includegraphics[width=\columnwidth]{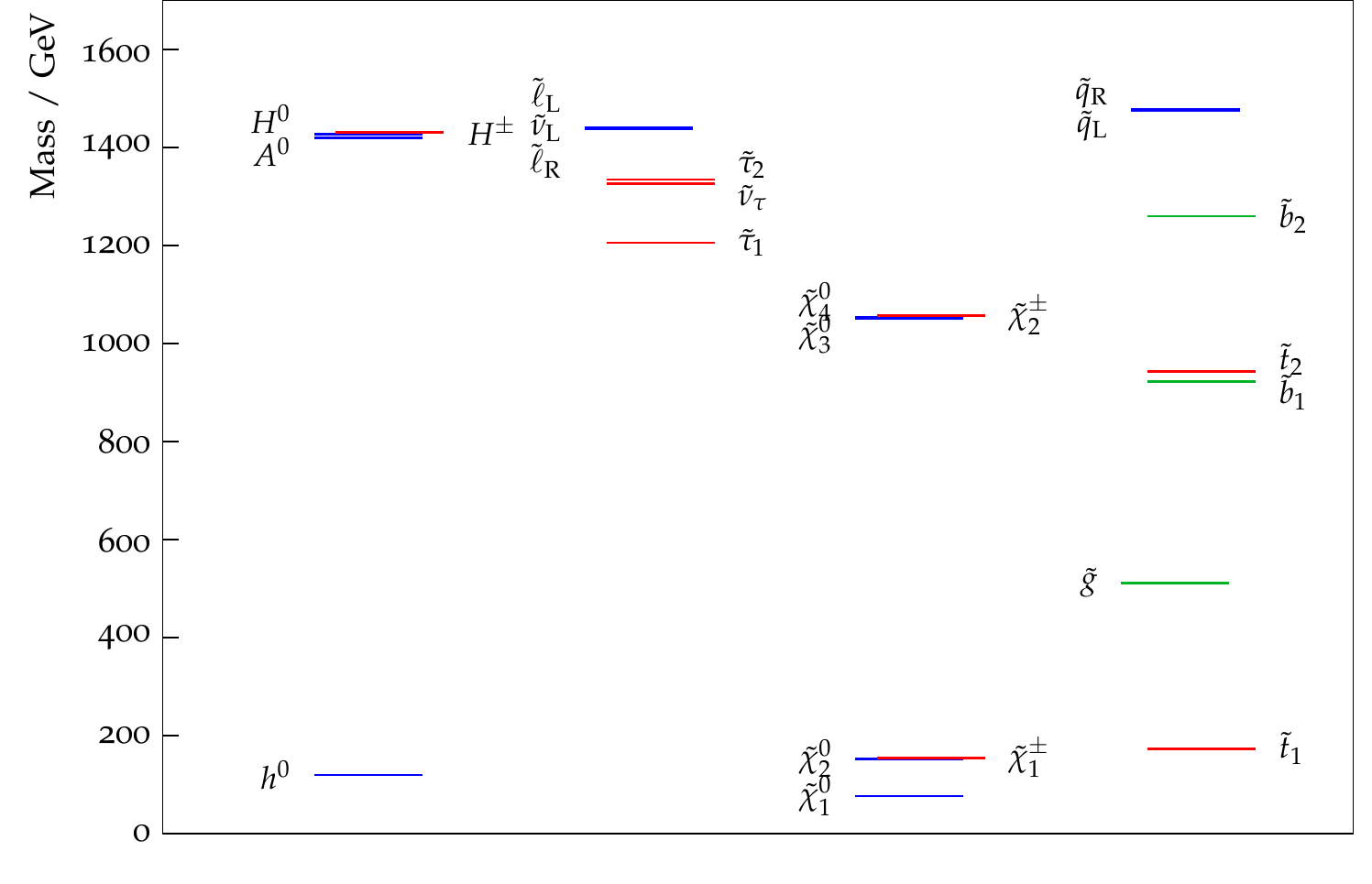}
}
\caption{The sparticle mass spectrum for a point with large $|A_0|$ capable of generating a small excess of high multiplicity events at the LHC. For details, see main text.}
\label{fig:mass}
\end{figure}

Such model points exist in the CMSSM at high-$M_0$ and high-$|A_0|$, in which most of the squarks are heavy except one stop quark whose mass gets pushed to lower values due to a large splitting between the $\tilde{t}_1$ and $\tilde{t}_2$ masses. These models furthermore exhibit low fine tuning, and would be capable of generating slightly higher masses for the lightest SUSY Higgs particle.  The mass spectrum of one such point is shown in figure~\ref{fig:mass}, with $M_0=1440$\,GeV, $M_{1/2}=177$\,GeV, $\tan\beta=27$, $A_0=-2950$\,GeV and $\mu > 0$~\footnote{The point was found during a wide ranging scan of the CMSSM parameter space, hence the esoteric choice of parameters.}. As ATLAS and CMS tighten the exclusion of SUSY models with several light squarks, models such as these are becoming much more important in the search for weak scale supersymmetry, and we therefore consider it important to add the effects of $A_0$ and $\tan\beta$ to our handling of LHC SUSY constraints.

The dependence on $\tan\beta$ is much weaker than that on $A_0$, as the ratio of Higgs doublet VEVs has a much greater impact on the Higgs sector of the CMSSM than on squark masses. However, large $\tan\beta$ values can reduce the stop and sbottom splitting induced by large values of $A_0$ as mentioned in the previous paragraph, and potentially swap the mass ordering of the $\tilde{t}_1$ and $\tilde{g}$ with corresponding effects on the phenomenology. As inclusion of all four continuous CMSSM parameters is technically possible, and $\tan\beta$ may influence the phenomenology of zero-lepton channels in certain regions of the $\{ M_0, M_{1/2}, A_0 \}$ parameter space, we hence include it in this study. We assess the value of having gone to this effort in section~\ref{sec:reassestanba0}, once the method itself has been described.

\subsubsection{Fast simulation using machine learning techniques}

Running the entire chain of \texttt{ISAJET}, \texttt{Herwig++}, \texttt{Delphes} and \texttt{PROSPINO} for a given model point takes $\sim 1$~hour in total on a typical CPU. Although one can trivially parallelise the simulation of different model points, it is still infeasible to simulate all of the $2 \times 10^6$ posterior samples required to reweight an existing data set, let alone the $10^7$ or more required if this was to be incorporated into the primary MultiNest run sequence for a full scan. If one were reweighting points after a scan had been completed, one could restrict the simulation to points that are reasonably probable, but even this still requires a very large number of CPU hours. It is this restriction that has prevented previous studies from considering the effects of $\tan\beta$ and $A_0$.

An obvious solution is to try and interpolate between a smaller grid of simulated values, such that one obtains a function that can give the signal expectation for \emph{any} CMSSM point within fractions of a second. This is a standard regression problem, and there is an extensive collection of efficient techniques in the literature for performing the interpolation. White, Buckley and Shilton have previously demonstrated good results in the CMSSM using machine learning algorithms~\cite{Buckley:2011kc} including both a Bayesian Neural Network (BNN) and a Support Vector Machine (SVM), with a zero lepton signal region from the ATLAS 2010 analysis as the test case. Here we go much further by interpolating all of the ATLAS zero lepton search channel results (from the 2011 analysis) and by combining the search channels to reproduce the ATLAS combined exclusion result.

Combining the search channels is non-trivial since ATLAS have not published enough information to determine the correlations between channels. We therefore continue to perform the approximate procedure outlined above. For any given model point, we can determine if it is excluded or still viable by choosing the most stringent limit on $\sigma \times A \times \epsilon$ for that point. Whilst this unfortunately only allows us to attach a discrete LHC-based likelihood to the points in the above posterior distributions, it is the most rigorous procedure that can be applied in the circumstances. We expect that this will slightly lower the apparent damaged done to the CMSSM by the ATLAS limits, as measured by the associated PBF, from what one would obtain with the full 4D likelihood. This is because we are effectively adding a significant amount of extra likelihood to all points which are ``not excluded'' (particularly those which are close to the limit), while removing likelihood from all ``excluded'' points. We expect the procedure to be adding more likelihood overall than is lost since points near the 95\% confidence limit have quite low likelihood to begin with. Since it is an integral over the likelihood function which leads to the evidence values used in the Bayes factors, an overall increase in likelihood will increase the CMSSM evidence and thus lower the apparent damage to the CMSSM.  This argument is valid unless the low likelihood points encompass a large prior volume, in which case their contribution to the evidence can be significant.
Furthermore, we expect the `true' likelihood map to quite sharply transition from strong to very weak exclusion of model points in the vicinity of the limit; for example the approximate 2D likelihood map computed in \cite{Bechtle:2012zk} shows this transition occurring over a range of around 50 GeV in $M_{1/2}$. We thus expect any errors introduced into our analysis due to the step-function approximation to the limits and approximate combination procedure to be small.

\begin{figure}
  \begin{center}
    \includegraphics[width=\columnwidth]{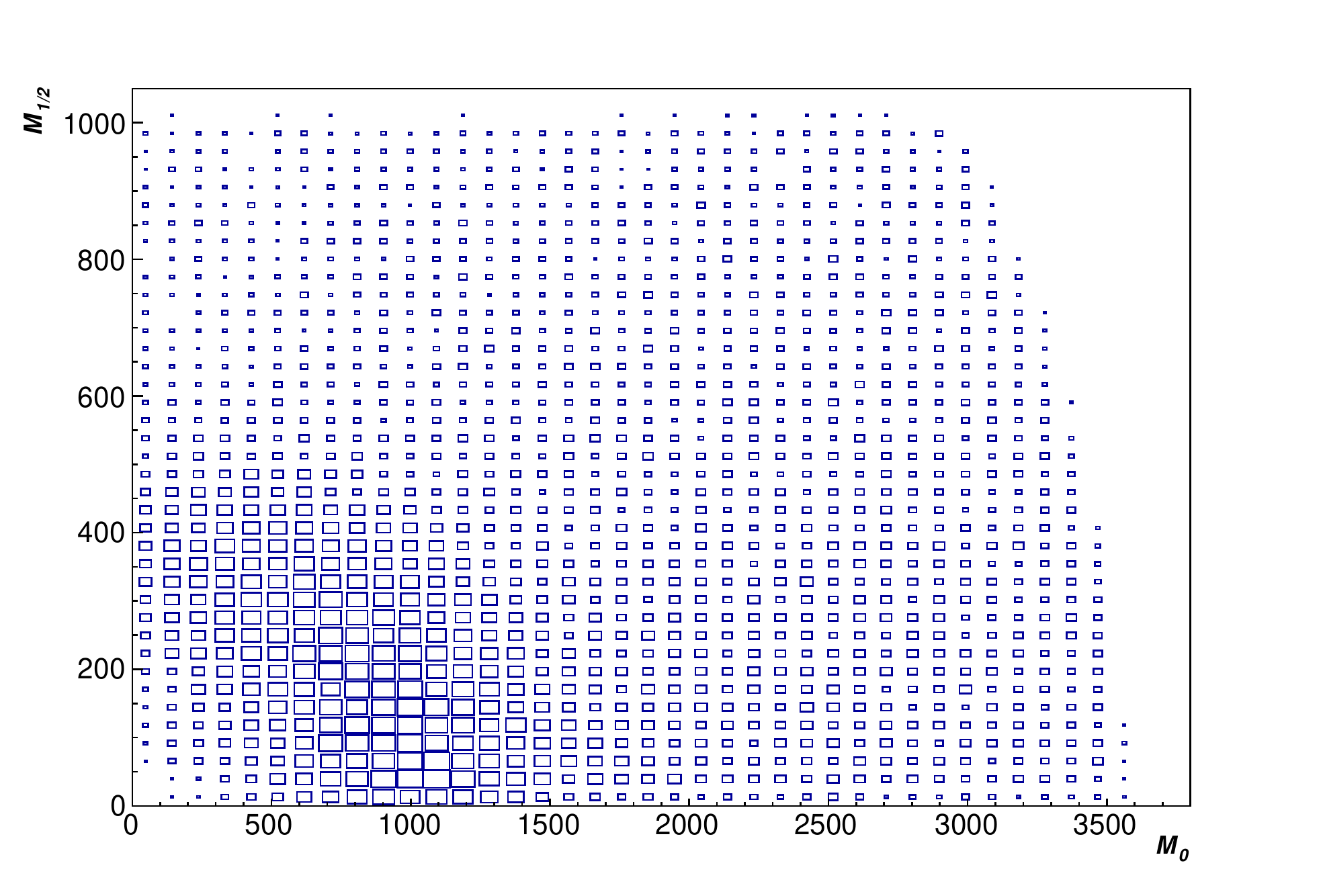}
  \end{center}
\caption{The sampled distribution in the $M_0$--$M_{1/2}$ plane for BNN training, shown after removal of failed \texttt{ISAJET} and \texttt{PROSPINO} points.}
\label{fig:m0m12sampling}
\end{figure}

Our study in~\cite{Buckley:2011kc} demonstrated successful interpolation of the signal expectation itself. Given that we here want to apply only a discrete likelihood based on whether a point is excluded or not excluded, one can use a Bayesian neural net (BNN) as a classifier rather than a regressor (the former being the discrete case of the latter). For each channel in table~\ref{tab:results} we have used the BNN implementation in the \texttt{TMVA} package~\cite{Hocker:2007ht,Zhong:2011xm} to classify SUSY parameter points into two classes:
\begin{enumerate} \itemsep-5pt
\item \textbf{Excluded: }$(\sigma \times A \times \epsilon \times f)>l$ \\
\item \textbf{Not excluded: } $(\sigma \times A \times \epsilon \times f)<l$ \\
\end{enumerate} \vspace{-10pt}
where $l$ is the limit for that channel given in table~\ref{tab:results} and $f$ is the scaling factor applied to the channel to obtain a close match with the ATLAS results. The success of the classification depends critically on the quality of the training data, and it is particularly essential to ensure that the training data adequately cover the limit $(\sigma \times A \times \epsilon \times f)=l$. In the $M_0$--$M_{1/2}$ plane, this limit is traced by the exclusion limits in figure~\ref{fig:limits1}. To maximise the accuracy of the BNN training in the region of maximum analysis sensitivity, while still achieving sufficiently comprehensive coverage of the $M_0$--$M_{1/2}$ plane, we hence sample training data using a hybrid distribution composed of distinct two functions in $M_0$--$M_{1/2}$:
\begin{itemize}
\item uniform sampling in $M_0 \in [10, 4000]$\,GeV, and a falling exponential distribution with width 500\,GeV for $M_{1/2} \in [10, 1000]$\,GeV;
\item sampling from a ellipse with Gaussian profile, constructed such that it intersects the $M_0$ axis at 1\,TeV with width 300\,GeV, and intersects the $M_{1/2}$ axis at 350\,GeV with width 105\,GeV.
\end{itemize}
Sampling weight was distributed equally between these two distribution components, the resulting sampling density being shown in figure~\ref{fig:m0m12sampling}. $A_0$ and $\tan\beta$ were sampled uniformly from $A_0 \in [-3000, 4000]$\,GeV and $\tan\beta \in [0, 62]$ regardless of the distribution type being used in $M_0$--$M_{1/2}$.

We generated two sets of training data of 25,000 points for the $\mu>0$ branch, and a further 5,000 points on which to validate the classification performance.

\begin{figure}
  \centering
  \includegraphics[width=\columnwidth]{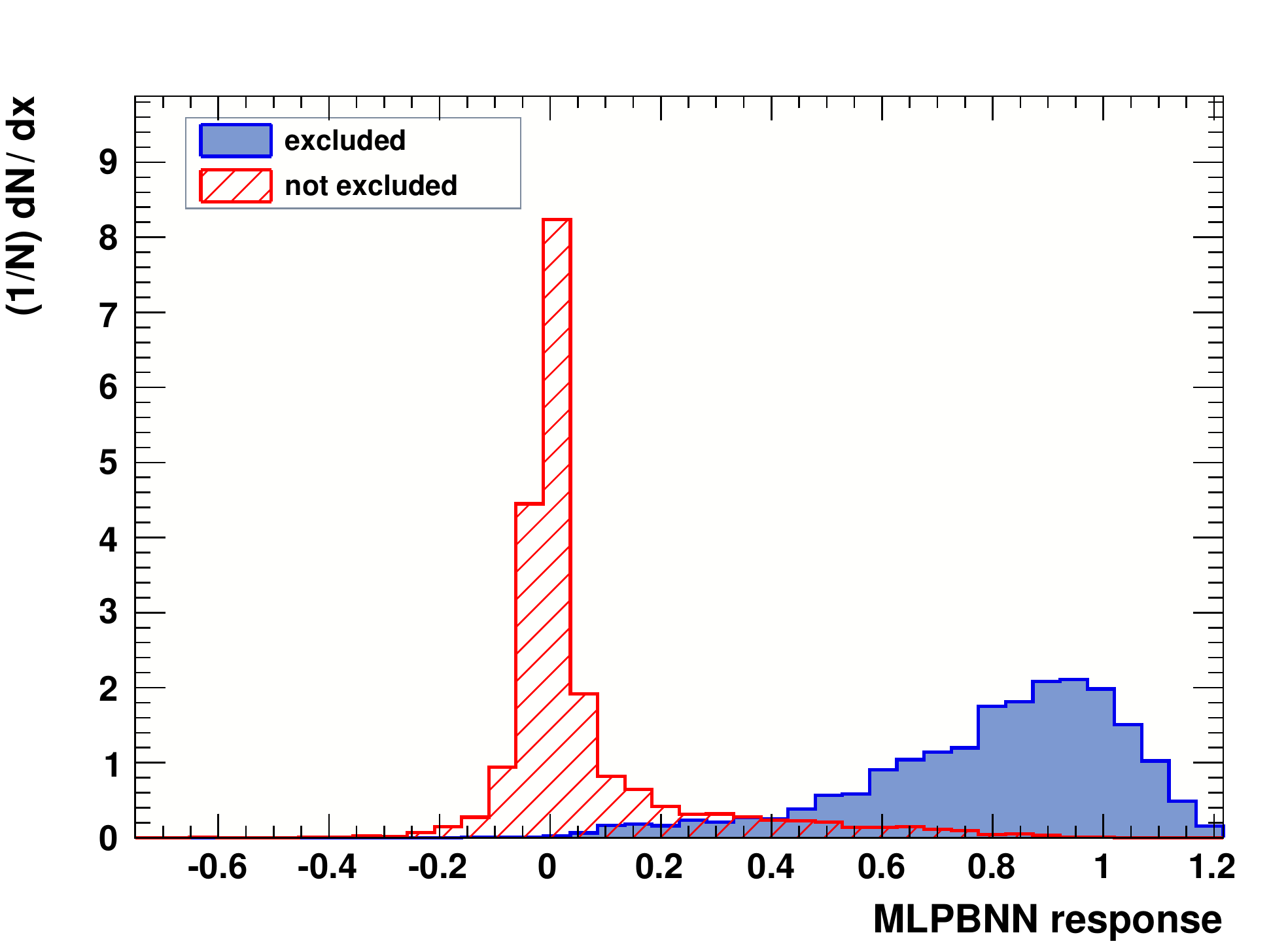}
 \caption{\label{fig:dists}Distributions of the BNN response for ``not excluded'' points and ``excluded'' points, for the ATLAS R1 search channel. The MLPBNN response variable offers good discrimination between the two classes of SUSY model.}
\end{figure}

The output of the BNN classification is a mapping between the CMSSM input parameters $M_0$, $M_{1/2}$, $A_{0}$, $\tan\beta$, $\mathrm{sign}(\mu)$ and a continuous variable that offers good discrimination between the ``excluded'' and ``not excluded'' points. Sample distributions of this variable (the ``MLP Response'') for ``excluded'' and ``not excluded'' points are shown in figure~\ref{fig:dists} for the ATLAS R1 search channel. By choosing a suitable cut on this value, one can determine whether a given point is excluded given the input parameters. The cut value must be chosen to provide a familiar compromise between \emph{efficiency} and \emph{purity}. A cut that is too low will lead to large numbers of points that are ``not excluded'' being classified as ``excluded''. On the contrary, a cut that is too high will lead to large numbers of points that are ``excluded'' being classified as ``not excluded''. We select our cut to minimise the former outcome- it is much worse to claim points are excluded when they are not excluded than to miss excluded points that should be excluded, since in the latter case one can present conservative results that, nevertheless, are not false.

\begin{figure}
 \centering
 \subfloat{
   \includegraphics[width=\columnwidth,viewport=0 0 550 380,clip]{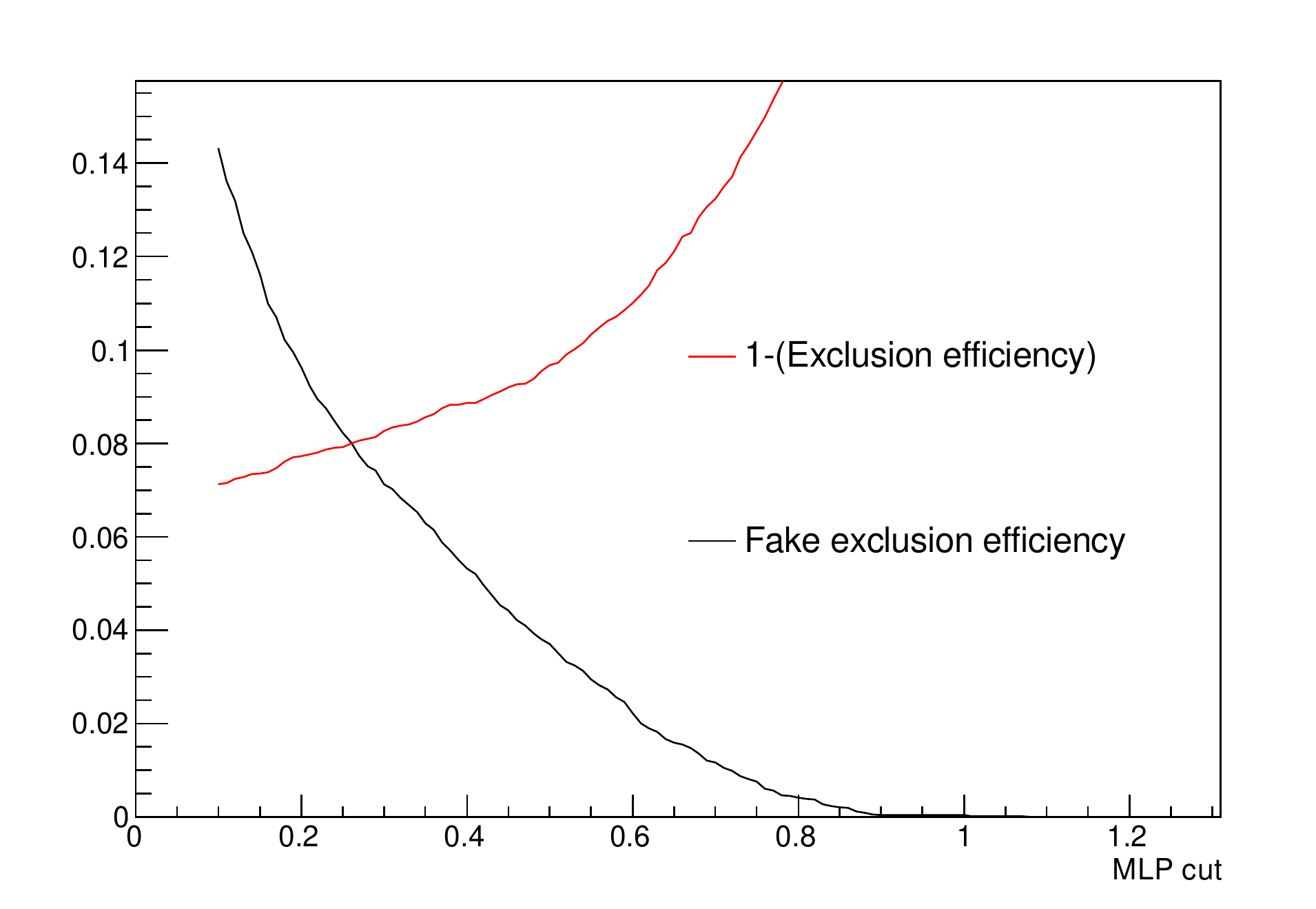}
 } \\
 \subfloat{
   \includegraphics[width=\columnwidth]{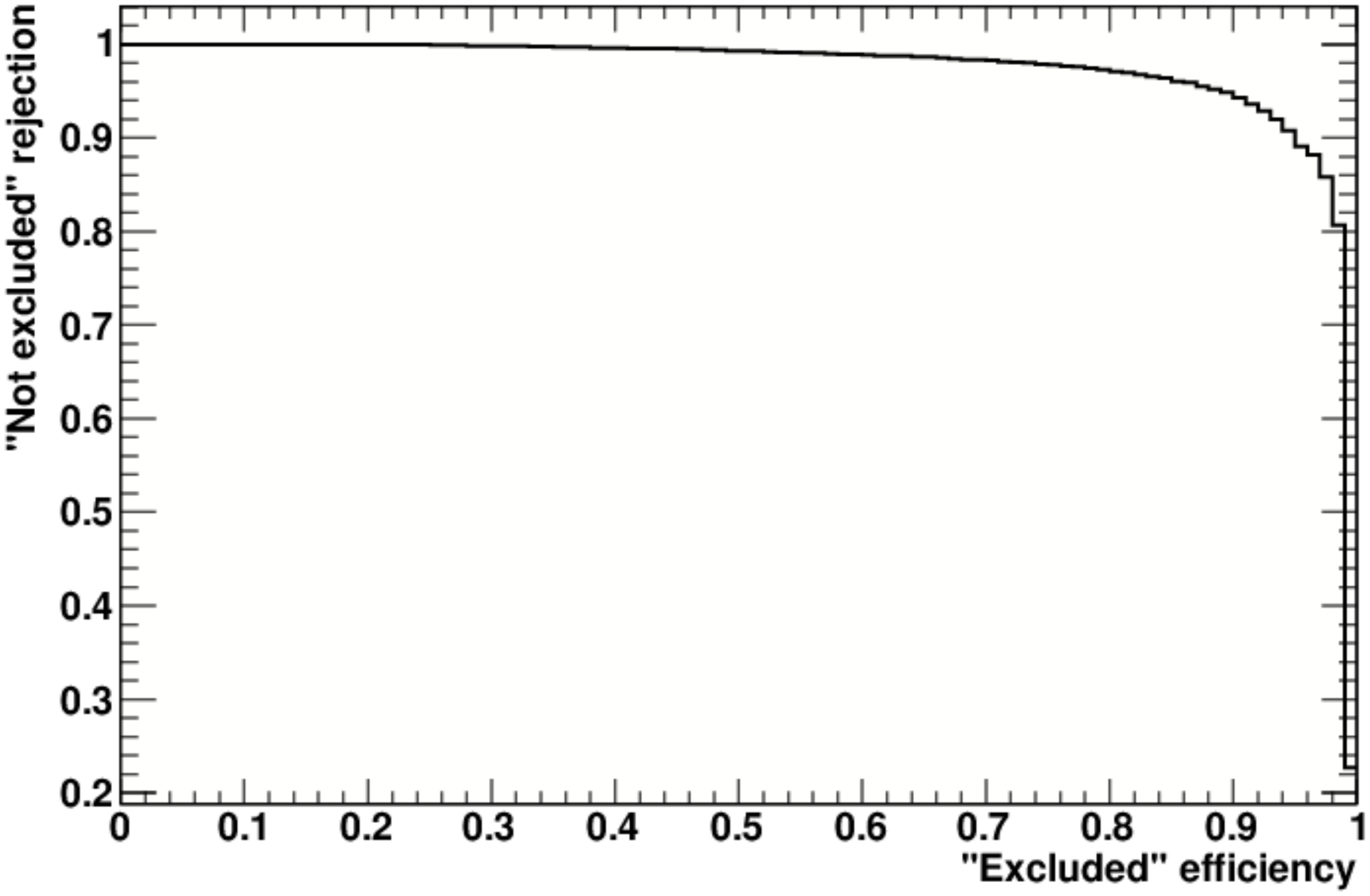}
 }
 \caption{\label{fig:curve}Fake exclusion rate and missed exclusion rate for the ATLAS R1 search channel vs the cut on the MLPBNN response.  The lower figure shows the equivalent Receiver Operating Characteristics (ROC) curve, demonstrating that for an exclusion efficiency of 90\%, models that are not excluded are rejected 95\% of the time (giving a fake exclusion rate of 5\%). Note: a ``rejected'' response from the classifier signals that it is assigning the point to the ``not-excluded'' category.}
\end{figure}

Figure~\ref{fig:curve} shows an example of this optimisation for the ATLAS R1 search channel. The black line shows the fraction of SUSY points in our test sample of 5,000 points that are labelled ``excluded'' when they should be ``not excluded'' vs the cut value on the MLP response. The red line shows the fraction of SUSY points that are labelled ``not excluded'' when the should be labelled ``excluded''\footnote{All other points in the test sample are ``not excluded'' and labelled as such, or ``excluded'' and labelled as such.}. By choosing an MLP cut of 0.5, one can keep the fake exclusion rate below 5\% whilst only missing 10\% of points which should be excluded. This is a very good performance considering that we now have the ability to apply results to the full parameter space of the CMSSM. A summary of the performance for each channel after choosing suitable MLP response cut values is provided in table~\ref{tab:bnnperf}. There is an element of subjectivity in choosing suitable cut values. We do not allow the efficiency for excluding points to drop below 90\%, but for channels where one can obtain a higher efficiency whilst keep the false exclusion rate below $\sim 4\%$ we choose the cut appropriately.

\begin{table*}\footnotesize
\begin{center}
\begin{tabular}{|c|c|c|c|c|c|}
\hline
Region & R1 & R2 & R3 & R4 & RHM\\
\hline
MLPBNN response cut value & 0.53 & 0.51 & 0.2 & 0.45 & 0.46\\
Fraction labelled ``Excluded'' when ``Not Excluded'' & 3.2$\%$ & 3.5$\%$ & 3.2$\%$ & 4.2$\%$ & 3.5$\%$ \\
Fraction labelled ``Not Excluded'' when ``Excluded'' & 10.0$\%$ & 10.0$\%$&6.8$\%$ & 10.0$\%$ & 8.1$\%$\\
\hline
\end{tabular}
\end{center}
\caption{\label{tab:bnnperf} MLPBNN response cut values for each ATLAS search channel, with performance statistics for the chosen cut value. We choose to accept a lower rate of labelling excluded points as ``not excluded'' (and thus missing excluded points) to keep the rate of false exclusion low.}
\end{table*}
Table~\ref{tab:bnnperf} demonstrates that the false exclusion rate remains at the few per cent level in each search channel whilst we can exclude 90$\%$ of the points that should be excluded. We have succeeded in obtaining an efficient and robust classifier for SUSY model points. For the ``ATLAS-sparticle'' and ``ATLAS-Higgs'' data sets this classifier was incorporated into the full MultiNest run sequence, and thus used to concentrate the scans on regions considered ``not excluded'' by the classifier.

\subsubsection{Variation of exclusion limits with $A_0$ and tan$\beta$}
\label{sec:reassestanba0}

\begin{figure*}
    \centering
    \includegraphics[width=\textwidth]{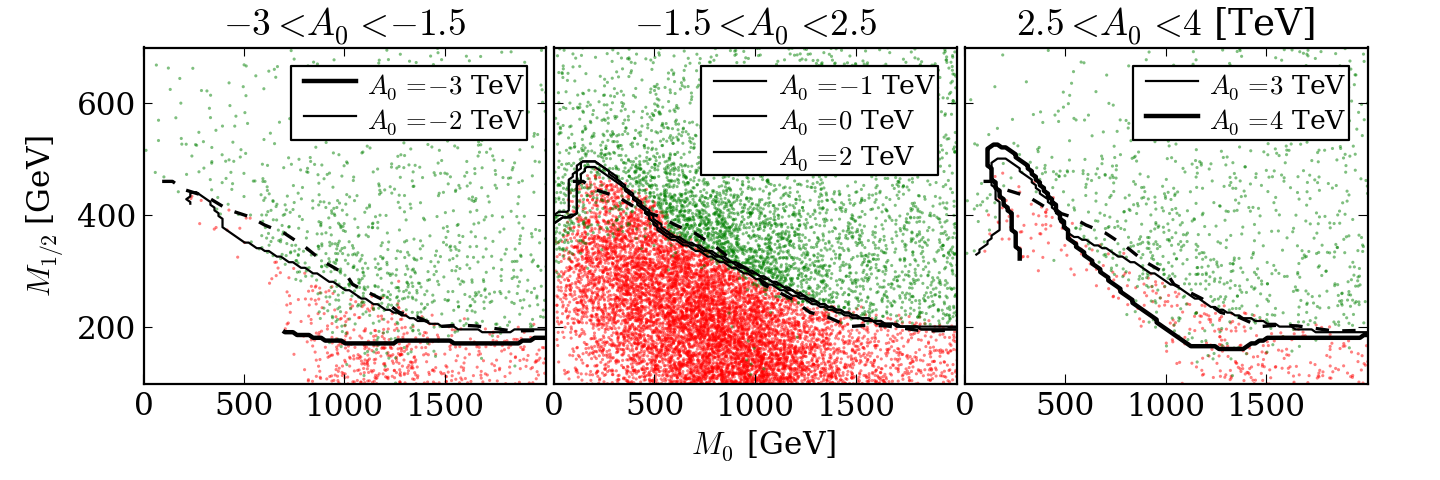}
    \caption{ATLAS 1 fb$^{-1}$ sparticle search 95\% confidence limits as estimated by the BNN classifier, displayed in the $(M_0,M_{1/2})$ plane for various values of $A_0$ (specified in the legends), with $\tan\beta=10$ (though little variation occurs with $\tan\beta$). The $A_0$ range displayed above each plot indicates the cut made on the training data in each plot, where the red points are excluded and the green not excluded as determined from simulated events. The official ATLAS limit (dashed), determined for $A_0=0$ and tan$\beta=10$, is also displayed for comparison. The cuts made on the neural net response are tuned on the conservative side, so the increased contamination of the generically excluded region with not-excluded models, which occurs for large values of $A_0$, causes the classifier to weaken the limit in these regions to avoid false exclusions. The empty regions at high $A_0$ and low $(M_0,M_{1/2})$ are excluded on physical grounds. Note: in the center plot no effort is made to distinguish the different neural net limits since they are extremely similar.}
    \label{fig:A0tanblimits}
\end{figure*}

With the classifier trained we now reassess how worthwhile it was to estimate the full 4D limit. To do this we examine the position of the limit, as estimated by the classifier, in the ($M_0$, $M_{1/2}$) plane for a range of $A_0$ and tan$\beta$ values and compare these to the official ATLAS limit. A representative set of these limits is shown in figure \ref{fig:A0tanblimits}.

The classifier limit is observed to be largely unchanged from the ATLAS limit for $A_0$ values between $-2$ and $3$ TeV, for all tan$\beta$, except for the stau-neutralino coannihilation region at very low $M_0$, which ATLAS miss due to the coarseness of their grid (and which we suspect escapes detection due to the combination of increased slepton pair production and a compressed mass spectrum rendering coloured production with several hard jets less visible), however for $A_0$ outside this range the classifier limit is seen to weaken in $M_{1/2}$ above $M_0\sim500$ GeV, quickly dropping to around 200 GeV. Comparing these limits to the training data, it appears that this occurs because the boundary of the excluded/not-excluded regions becomes less well defined. The increased `contamination' of the generically excluded region with not-excluded points causes the classifier, with the response cuts we have chosen, to ``play it safe'' and avoid the false exclusions by weakening the limit. From this we conclude that $A_0$ variation in particular is of importance to interpretations of LHC limits if regions with very large $|A_0|$ are of interest, but otherwise may be fairly safely neglected.

We pre-empt our results to say that we find that there is not much posterior probability located outside $-2\text{ TeV}<A_0<3\text{ TeV}$ in any of our scans when using a log prior, and so our 4D treatment of the limit will have had little impact on our log prior results. However, when using the `natural' prior we indeed find a significant amount of probability below $A_0=-2$ TeV in all scans except the baseline scan, part of which would have been excluded had we not allowed the limit to vary with $A_0$. These posteriors are a by-product of our central results but we include them in Appendix \ref{app:plots} in part to illustrate this point.

We next compare our findings on the $A_0$-tan$\beta$ dependence of the sparticle search limits to those of other groups. In ref. \cite{Bechtle:2012zk} more recent 4.7 fb$^{-1}$ ATLAS and CMS limits\cite{ATLAS-CONF-2012-033,CMS-PAS-SUS-12-005} were studied and no $A_0$-tan$\beta$ dependence was observed within systematic uncertainties. To support this assertion the signal yield for a handful of points is presented, and shown to remain within systematics, however these all have $A_0$ equal to either 0 or 1 TeV, and our study agrees that little $A_0$-tan$\beta$ variation should be seen in this range. The total $A_0$ range scanned exceeded $|5|$ TeV, so according to our findings some dependence may be expected, however since these are different limits to those we have used (as they were not yet available at the time our computations were performed), it is plausible that their dependence on $A_0$ is indeed weaker. This may occur because the 1 fb$^{-1}$ ATLAS search we study contains a modest excess, which increases the likelihood for high $A_0$ points, and thus increases the $A_0$ dependence of the limit. In the 4.7 fb$^{-1}$ search no excess as large as this was observed, so this extra source of $A_0$ dependence may be absent.

Older studies exist which also contribute to this picture. In ref. \cite{Allanach:2011ut} a 35 pb$^{-1}$ limit was studied and it was concluded that assuming it to be $A_0$ and tan$\beta$ independent was a reasonable approximation, however it was also observed that points with high $|A_0|$ exhibited the largest disagreements with the $A_0=0$ limit, as we observe. Furthermore, only one point outside $-1.5\text{ TeV}<A_0<2.5\text{ TeV}$ was studied (and this excluded for other reasons), well within the range our results indicate to be `safe'. 

To conclude, the overall impact on our study of using the 4D limit appears to be minimal, however as limits increase to higher $M_0$ and $M_{1/2}$ and larger $|A_0|$ values become more plausible (driven by a need to fit the $125$ GeV Higgs candidate discussed in the next section) then it will become more important to use the full limits, with this importance potentially increasing with the size of any excesses. The value of including $A_0$ and tan$\beta$ dependence in approximations to LHC search likelihood functions will thus need to be reassessed for each new limit which is produced.

\subsection{February 2012 ATLAS Higgs search results}
\label{sec:LHCHiggsL}

In December 2011 ATLAS and CMS released preliminary results for their combined Higgs searches~\cite{ATLAS-CONF-2011-163,CMS-PAS-HIG-11-032} showing strong hints for the presence of a SM-like Higgs boson in the vicinity of 125\,GeV. The official combinations were released in February 2012 \cite{ATLAS:2012si,Chatrchyan:2012tx} with little change. Others have considered the impact that the existence of such a Higgs, if confirmed, would have on the CMSSM parameter space~\cite{Buchmueller:2011ab, Akula:2011aa, Kadastik:2011aa, Roszkowski:2012uf, Baer:2012uy, Ellis:2012aa, Bechtle:2012zk}. We are interested in the state of knowledge as of February 2012, so we do not make such assumptions. Instead, we reconstruct the full likelihood based on the public results.  We use the February 2012 ATLAS 4.9\,$\text{fb}^{-1}$ Higgs search data in which the since discovered resonance at 126 GeV had a local significance of $3.5\sigma$.  Our method bears some similarity to that used by~\cite{Azatov:2012bz}, however we work from signal best fit plots, not $CL_S$ limit plots. Since CMS do not produce signal best fit plots for all the channels we require we use only the ATLAS results. We will now detail our method.

To construct signal best fit plots ATLAS and CMS use the log-likelihood ratio test statistic
\begin{align}
 Q = -2\log(\lambda) &= -2\log\left(\frac{L_{s+b}(\mu)}{L_{s+b}(\hat\mu)}\right) \\
 &= -2\log\left(\frac{P(\text{data}|m_h,\mu)}{P(\text{data}|m_h,\hat\mu)}\right) . \nonumber
\end{align}
\label{eq:CLsQ}
Here $m_h$ is the Higgs mass parameter, $\mu$ the cross section scaling parameter (the factor which multiplies the SM prediction for the Higgs cross section for a given channel to achieve the hypothesised value, i.e. $\mu = \sigma/\sigma_\text{SM}$) and $\hat\mu$ the value of $\mu$ which maximises the likelihood for a fixed $m_h$ value. Nuisance variables are profiled over. A $\pm1\sigma$ error band is also presented, the extents of which give the values of $\mu$ for which $Q$ rises to 1 for each value of $m_h$.  Examples of such plots are shown in figure 8 of ref. \cite{ATLAS-CONF-2011-163}.

Following ref. \cite{cowan2011asymptotic}, if one assumes Wald's asymptotic approximation to be valid (which ATLAS confirms to be true to good accuracy for the three individual channels we use \cite{ATLASaa:2012sk, ATLAS2l2v:2011aa, ATLAS4l:2012sm} as well as for the combination in ref. \cite{ATLAS:2012si}) then $Q$ can be written as
\begin{equation}
 Q = \frac{(\mu - \hat{\mu})^2}{a^2},
\end{equation}
\label{eq:CLsQ2}
where it is assumed that $\hat{\mu}$ is normally distributed with mean $\mu$ and standard deviation $a$ (when the data is generated by the signal plus background model with the parameters $m_h$ and $\mu$), and where both $\hat\mu$ and $a$ depend on the model parameters $m_h$ and $\mu$ we are testing. All the information regarding systematic and statistical uncertainties is carried by $a$. If $a$ did not vary with $\mu$ then we could immediately determine $Q$ from the best fit and $\pm1\sigma$ curves (taking the largest deviation from $\mu$ as $a$ to be conservative) of the published best-fit plots, and thus extract the likelihood ratio for all $\mu$ in each $m_h$ slice. In fact this is exactly what we do, and this is safe because signals are at this stage small, which implies that the distributions of $Q$ cannot be very different between $\mu=\hat\mu$ and $\mu=0$ (or else the establishment or exclusion of a signal at much higher significance would be possible). So assuming $a$ to remain constant for each $m_h$ is sufficient for our purposes. The reconstructed likelihood will be accurate near $\mu=\hat\mu$ and lose accuracy far from the best fit point, however the likelihood is low for such parameters so this mistake will have little impact on our results.

We now can obtain the likelihood ratio for every value of $\mu$ in each $m_h$ slice, however the slices are not scaled correctly relative to each other. We can fix this by noting that the points $\mu=0$ for each $m_h$ are degenerate (because $m_h$ makes no difference to predictions if $\mu=0$).  We can thus scale the likelihood of each $m_h$ slice relative to the likelihood at $\mu=0$, i.e. instead of $Q$ we can work with the test statistic $Q_{CL_s}$ (so called because of its use in constructing $CL_s$ limits):
\begin{align}
 Q_{CL_s} &= -2\log\left(\frac{L_{s+b}}{L_{b}}\right) \\
 &= -2\log\left(\frac{P(\text{data}|m_h,\mu)}{P(\text{data}|m_h,\mu=0)}\right) . \nonumber
\end{align}
\label{eq:CLsQ3}
Applying Wald's approximation again we obtain~\cite{cowan2011asymptotic}
\begin{align}
 \begin{split}
 Q_{CL_s} &= -2\log\left(\frac{P(\text{data}|m_h,\mu)}{P(\text{data}|m_h,\hat\mu)}\right) \\
&\phantom{=} + 2\log\left(\frac{P(\text{data}|m_h,\mu=0)}{P(\text{data}|m_h,\hat\mu)}\right)
 \end{split} \\
   &= \frac{(\mu - \hat{\mu})^2}{a^2} - \frac{\hat{\mu}^2}{a^2} . 
\end{align}
\label{eq:CLsQ4}
As above, $\hat\mu$ and $a$ can be extracted for every $m_h$ slice from the publicly available plots, but now the new term correctly normalises the slices relative to each other. The likelihood we extract contains an extra constant factor due to the $\mu=0$ contribution however this is of no importance for our analysis.

\begin{figure*}[htbp]
    \centering
    \includegraphics[height=2.5in]{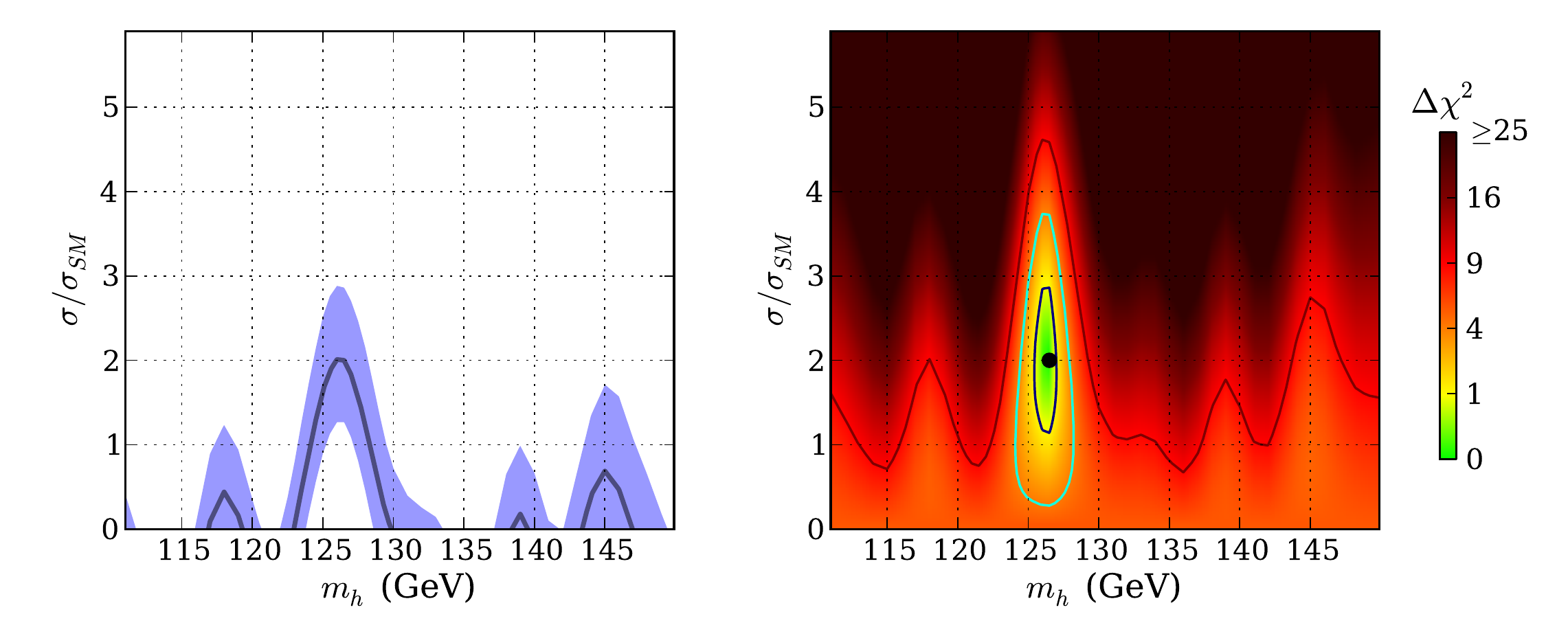}
    \caption{Reconstructed likelihood function for the diphoton channel, using asymptotic approximations for the test statistic distributions. In the left frame a reproduction of the ATLAS signal best fit plot from ref. \cite{ATLASaa:2012sk} (with $\pm1\sigma$ band) is shown, while on the right is the reconstructed $\Delta\chi^2$ map, with the  $\Delta\chi^2 = 1,4,9$ contours shown.  We ignore the negative $\sigma/\sigma_{SM}$ region as it is not relevant for the models we consider. In our scans this likelihood (and those for the other channels) is convolved with a further 1 GeV Gaussian on $m_h$ to account for theoretical/numerical uncertainty in the value of $m_h$ computed at each model point, and so the best fit region is extended in $m_h$ by an extra GeV or so.}
    \label{fig:Ldemo}
\end{figure*}

In figure~\ref{fig:Ldemo} we show the likelihood function reconstructed from the ATLAS diphoton channel results, as an example.  
We checked the consistency of our reconstruction by combining the three search channels we use and comparing the result to a reconstruction of the official ATLAS channel combination, finding good agreement.
In our scans this likelihood is further convolved with a 1 GeV width Gaussian uncertainty in the $m_h$ direction to account for theoretical uncertainty in the $m_h$ value computed at each model point.

\section{Results}
\label{sec:results}

\subsection{Profile likelihoods and marginalised posteriors}

Before presenting our main results (the partial Bayes factors) we show ancillary results from the datasets that we used to calculate them.  These are the profile likelihood functions and marginalised posterior PDFs over the CMSSM parameter space for each dataset, and may be found in figures \ref{fig:likelog}-\ref{fig:pos1CCR} in appendix \ref{app:plots}. These figures show the evolution of the profile likelihoods and posteriors from the ``pre-LEP'' situation (first row of each figure), to including the LEP and the XENON100 data (second row), to adding the LHC sparticle searches (third row), to folding in the 2012 February Higgs search results.  The figures reflect the well known effect: LEP has pushed the viable sparticle masses upwards substantially.  Specifically, LEP eliminated some of the lowest $M_{1/2}$ region, the region with the lowest fine-tuning, and created the small hierarchy problem.  The LHC sparticle searches directly lower the likelihood only in the lowest $M_0$-$M_{1/2}$ corner.  This leaves the bulk of the highest likelihood region toward slightly higher $M_0$ and $M_{1/2}$\footnote{The apparent 'thinning' of the likelihood toward higher $M_0$ and $M_{1/2}$ is a mere sampling artefact.}.  The 2012 February Higgs data seriously damages the high likelihood region at the lowest $M_0$-$M_{1/2}$, resulting in the relative enhancement of high negative $A_0$ regions with high Higgs masses, and the likelihood is pushed toward even higher $M_{1/2}$. Interestingly the highest likelihood region hardly moves, despite predicting a Higgs mass much below $125$ GeV (it is instead around $115$ GeV). This is because the ATLAS Higgs signal is not yet strong enough to conclusively outweigh the observables which strongly favour the low mass region, particularly $\delta a_{\mu}$, however extremely strong tension is created which causes the evidence to drop significantly and PBF to strongly disfavour the CMSSM. As can be seen in the profile likelihoods of figure~\ref{fig:likelogno(g-2)} and the PBFs of figure~\ref{fig:BayesFactors}, removing the $\delta a_{\mu}$ constraint indeed goes a significant way towards relieving this tension and reducing the total damage to the CMSSM.  At the same time the mid-$\tan\beta$ region emerges with the highest likelihood.

The evolution of the marginalised posteriors follows a similar pattern. The 68 and 95 percent credible regions follow the general trend of the highest likelihood, moving toward higher $M_0$ and $M_{1/2}$.  
Despite the inclusion of an increasing amount of data these credible regions are seen to ``spread out'' rather than ``shrink'' (as do the corresponding confidence regions) which is a signal that the global goodness of fit is worsening. The new data is excluding the part of the parameter space that was favoured by earlier data, causing tension among the likelihood components. The new best fit regions are not favoured with the same relative strength as the old ones, so globally poorer fitting points become less poor \emph{relative} to the new best fit, and so become included in both the confidence and credible regions, which are thus enlarged. The poorer (on average) likelihood values also feed into the evidence, causing it to lower accordingly. We remind the reader that lower evidence does not always signal a decrease in fit quality ---it can occur simply due to the reduction of viable parameter space--- however in this case fit quality is a significant factor.

A notable difference between the log and natural prior cases is that in the natural prior case the posterior exhibits a strong preference for $\tan\beta < 10$, where the $\mu$ fine tuning is generally low, which is decreased only a small amount by the new data, while in the log prior case there is clear movement of the preferred regions to higher $\tan\beta$. In conjunction, in order to maintain low $\tan\beta$, the natural prior scan is forced towards large negative $A_0$ values, while the log prior viable regions end up centred on $A_0=0$, although with sizable variance.


\subsection{Partial Bayes factors and their interpretation}

Based on the likelihood functions and posterior probabilities shown in the previous section, we have computed partial Bayes factors which update the odds of the CMSSM `existing' relative to our SM-like reference model for several data changes.  As described in section~\ref{sec:datachanges} the following data sets were utilised in our study:

\begin{description}
\item[\textbf{Pre-LEP}:] All constraints listed in table~\ref{tab:likefunc} are imposed except for the LEP Higgs lower bound, the XENON100 limit, the ATLAS direct sparticle search limits, and the ATLAS Higgs search results.
\item[\textbf{LEP+XENON100}:] As Pre-LEP, but including the LEP Higgs and XENON100 limits.
\item[\textbf{ATLAS-sparticle}:] As LEP+XENON100, but including the ATLAS direct sparticle search limits.
\item[\textbf{ATLAS-Higgs}:] As ATLAS-sparticle, but including the ATLAS Higgs search results.
\end{description}

\begin{figure*}
\begin{center}
\includegraphics[width=0.70\paperwidth]{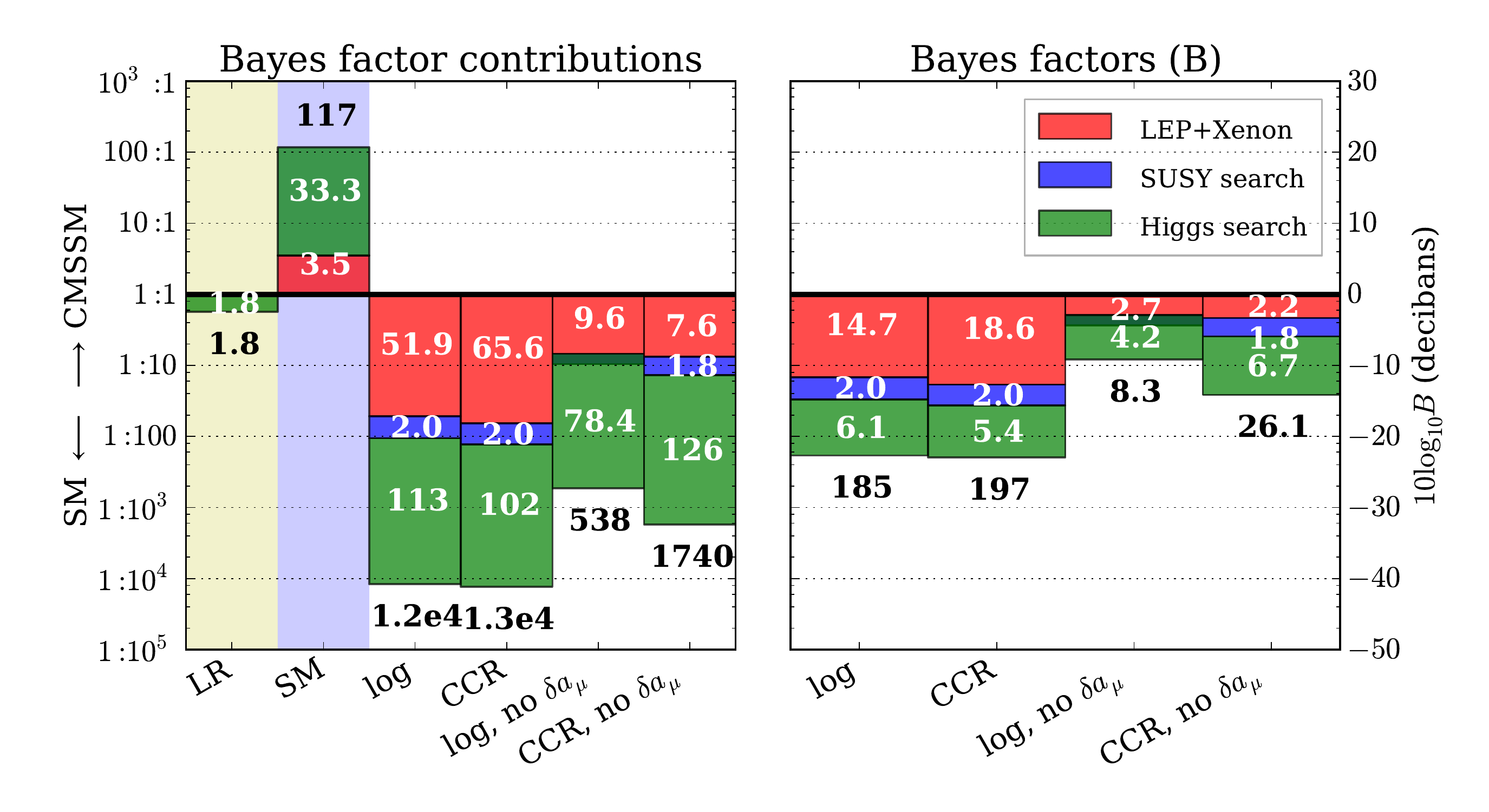}
\end{center}
\caption{(right) Partial Bayes factors for the three Bayesian updates we consider, for a hypothesis test of the CMSSM against the Standard Model (SM) augmented with a simple dark matter candidate, as computed using both `log' and `natural' (CCR) ``pre-LEP'' priors for the CMSSM and both with and without the $\delta a_\mu$ constraint imposed. We begin by updating from the ``pre-LEP'' situation to including the LEP Higgs search and the XENON100 data (red), to adding the ATLAS 1 $\text{fb}^{-1}$ sparticle searches (blue), to folding in the 2012 February ATLAS Higgs search results (green). (left) We also show the breakdown of each PBF into the maximum likelihood ratio of the data added in each transition (yellow highlight), and the ``Occam'' factors for each transition for both the SM (blue highlight) and the CMSSM (remainder). If one was willing to bet even odds on the CMSSM and SM at the ``pre-LEP'' stage, the product of these PBFs (as stacked) give the posterior odds with which one should now gamble on these models, given our ``pre-LEP'' parameter space priors and data assumptions. The cumulative effect of these PBFs is an almost 200 fold swing in the odds away from the CMSSM, reduced to a 10-30 fold swing if the $\delta a_\mu$ constraint is dropped. PBFs of the former strength represent significant experimental disfavouring of the low energy CMSSM and could only be outweighed by very strong prior odds (determined by considerations outside the scope of our analysis), while the latter values (with $\delta a_\mu$ removed) are of only moderate strength and are unlikely to dominate over prior considerations. Despite differences in the details of posteriors obtained under the two priors used, the Bayes factors themselves remain remarkably robust, although this robustness is partially compromised if $\delta a_\mu$ is ignored since it is a powerful constraint which helps the baseline (``pre-LEP'') data to dominate over differences between ``pre-LEP'' priors.
}
\label{fig:BayesFactors}
\end{figure*}

The global evidence for each dataset is computed in the $\mu > 0$ branch of the CMSSM, for both the log and natural prior as described in section~\ref{subsec:prior}, giving us a total of 8 data sets which have resulted from around 100 million likelihood evaluations in total.  From these global evidences we compute PBFs for the Bayesian updates
\begin{align*}
\text{Pre-LEP}&\to\text{LEP+XENON100}
\\ \text{LEP+XENON100}&\to\text{ATLAS-sparticle}
\\ \text{ATLAS-sparticle}&\to\text{ATLAS-Higgs}
\label{eq:transitions}
\end{align*}
according to the prescription of eq. (\ref{eq:globev}) and (\ref{eq:BayesFact2}), for each choice of ``pre-LEP'' prior.  We also compute a `cumulative' PBF by multiplying together the PBFs in the sequence of updates.  We present the results in table~\ref{tab:bayesfactors} and figure \ref{fig:BayesFactors}.

\begin{table*}\footnotesize
   \begin{center}
    \begin{threeparttable}
      \begin{tabular}{| l | l | l | l | l | l | l |}
         \hline
         Scenario & $\ln\mathcal Z$ & LR & $\mathcal O$ & $B$ & $B_\text{cumulative}$ & Strength of $B$ \\
         \hline
         \noalign{\smallskip}
         \multicolumn{2}{c}{\textbf{SM+DM}} \\
         \hline
         Pre-LEP & $0^*$ & - & - & - & - & - \\
         \hline
         LEP+XENON100 & $-1.26$ & - & $1:3.52$ & - & - & - \\
         \hline
         ATLAS-sparticle & $-1.26$ & - & $1:1$ & - & - & - \\
         \hline
	   ATLAS-Higgs& $-5.16$ & - & $1:33.3$ & - & - & - \\
	\hline
         \noalign{\smallskip}
         \multicolumn{2}{c}{\textbf{CMSSM (log priors)}} \\
         \hline
         Pre-LEP & $54.30(2)$ & - & - & - & - & - \\
         \hline
         LEP+XENON100 & $50.34(2)$ & $1:1$ & $1:51.9(1)$ & $1:14.7(4)$ & $1:14.7(4)$ & Strong \\
         \hline
         ATLAS-sparticle & $49.62(2)$ & $1:1$ & $1:2.04(5)$ & $1:2.04(5)$ & $1:30.1(8) $ & Barely worth mentioning \\
         \hline
	   ATLAS-Higgs& $43.91(2)$ & $1:1.8$ & $1:113(3)$ & $1:6.1(2)$ & $1:185(5)$& Substantial \\
	\hline
	\noalign{\smallskip}
         \multicolumn{2}{c}{\textbf{CMSSM (natural priors)}} \\
         \hline
         Pre-LEP & $44.73(2)$ & - & - & - & - & -\\
         \hline
         LEP+XENON100 & $40.54(2)$ & $1:1$ & $1:65.6(2)$ & $1:18.6(6)$ & $1:18.6(6)$ & Strong \\
         \hline
         ATLAS-sparticle & $39.87(2)$ & $1:1$ & $1:1.97(6)$ & $1:1.97(6)$ & $1:37(1)$ & Barely worth mentioning \\
         \hline
	   ATLAS-Higgs& $34.29(2)$ & $1:1.8$ & $1:102(3)$ & $1:5.4(2)$ & $1:197(6)$ & Substantial \\
	\hline
      \end{tabular}
    \begin{tablenotes}
         \item [*] We have computed $\Delta\ln\mathcal Z$ directly for the SM+DM so this zero is an arbitrary initial value, for illustrative purposes only.
    \end{tablenotes}
    \end{threeparttable}
    \caption{\label{tab:bayesfactors} Summary and interpretation of our results.  The global log evidence values $\ln\mathcal Z$ and statistical uncertainties are presented as computed by \texttt{MultiNest} for each scan, except in the case of the SM+DM for which we have computed $\Delta\ln\mathcal Z$ values directly.  The partial Bayes factor (PBF) $B$ is shown for the Bayesian update to the data set of each row from that of the previous row. The cumulative PBF $B_\text{cumulative}$ is the product of $B$ with all of the previous PBFs. The components of the PBFs are also shown: the LR column shows the maximum likelihood ratio between the SM+DM and CMSSM for the newly added data (which is only different from one for the ATLAS Higgs search data, where we see that the maximum likelihood is a factor of 1.8 higher in the SM+DM than the CMSSM), and the $\mathcal O$ column shows the Occam factors. The final column offers an interpretation of the strength of each PBF according to the Jeffreys scale as listed in table~\ref{tab:jeffreysscale}. The combined effect of all experiments on a `pre-LEP' odds ratio is seen to be a `Decisive' shift away from the low energy CMSSM when judged by the Jeffreys scale, using either prior.}
   \end{center}
\end{table*}

The first column of table~\ref{tab:bayesfactors} lists the datasets we have computed. The second shows the global log evidence value $\ln\mathcal Z$ and its statistical uncertainty as computed by \texttt{MultiNest}, using the method described in section~\ref{sec:method}, except in the case of the SM+DM evidences, where $\Delta\ln\mathcal Z$ values are computed as described in section~\ref{sec:Higgscons}.  In the fifth column the PBF $B$ is shown for the Bayesian update to the dataset of each row from that of the previous row, followed by the cumulative PBF $B_\text{cumulative}$, which is the product of $B$ with all of the previous PBFs, in column six. Columns three and four show the breakdown of each PBF into the maximum likelihood ratio for the new data and the respective Occam factors for each model, as defined in eq. (\ref{eq:OccamFact}).  The final column offers an interpretation of the strength of each PBF according to the Jeffreys scale as listed in table~\ref{tab:jeffreysscale}. A graphical representation of these results (and those of table~\ref{tab:amubayesfactors}) is presented in figure~\ref{fig:BayesFactors}.

Table~\ref{tab:bayesfactors} and figure~\ref{fig:BayesFactors} show that the LEP Higgs limit very strongly reduced our trust in the low mass CMSSM.  The LHC sparticle limits induced a much smaller and not very significant additional reduction, and finally the LHC Higgs signal hints cause a `substantial' additional swing against the CMSSM.  The combined effect of all experiments (aside from the LHC Higgs data) on a pre-LEP odds ratio is seen to be a shift against the low mass CMSSM of a strength above the level considered `Decisive' on the Jeffreys scale.  These findings are robust against the shapes of the prior probabilities of the CMSSM parameters that we have considered, although they would be weakened by priors which strongly favoured high $M_0$ and $M_{1/2}$ values.  Presently the impact of XENON100 is negligible, but we remind the reader that the apparent strength of each piece of data is dependent on the order in which it is added. The XENON100 results appear largely irrelevant because they exclude regions of the CMSSM parameter space already excluded by the LEP Higgs searches and have only a small impact on the surviving parameter space. 
In case of the Standard Model XENON100 is completely irrelevant and the 1:3.52 Occam factor comes from the LEP Higgs searches alone. 
One may expect that the reinforcement of previous exclusions by independent experiments should count for something in the Bayesian framework (i.e. as reassurance that no mistakes were made by either experiment), however in the current analysis all data in the likelihood function is assumed to be 100\% reliable and so we are not considered to learn anything new by ``doubling up''. In order to see such effects in an analysis a measure of doubt about the reliability of experimental data would need to be introduced.

It has been noted previously that the $\delta a_\mu$ constraint is in considerable tension with several other observables~\cite{Feroz:2009dv, Cabrera:2010xx} and indeed this tension plays a strong role in the damage to the CMSSM that we observe since $\delta a_\mu$ strongly favours the now excluded lowest mass regions. However, there remains some controversy over its value~\cite{Hoecker:2010qn, Goecke:2010if, Hagiwara:2011af, Bodenstein:2011qy, Benayoun:2011mm, Goecke:2011bm}, so we consider the impact on our inferences if we remove it from our likelihood function.  It is too computationally expensive to do this by completing a full set of new scans, so we subtract it from the likelihood function of our original data sets in a similar `afterburner' manner as is done in ref. \cite{Buchmueller:2011sw}. The accuracy of the results obtained this way is lower than those obtained from full scans, particularly because the higher $M_0$, $M_{1/2}$ regions are substantially under-sampled with the $\delta a_\mu$ removed. The resulting PBFs, which we present in table~\ref{tab:amubayesfactors}, are thus offered as rough estimate only, and can be expected to overestimate the damage done to the CMSSM.

Since the $\delta a_\mu$ constraint pushes the posteriors strongly down in $M_0$ and $M_{1/2}$, removing it makes us much less surprised that no direct evidence for the low-mass CMSSM was seen at LEP or in the LHC sparticle searches.  This is reflected in weaker partial Bayes factors than in table~\ref{tab:bayesfactors}.  The LEP results in particular are seen to cause much less damage. The combined effect of both colliders on a pre-LEP odds ratio is seen to be greatly reduced; for log and natural priors the final cumulative Bayes factors are weakened to `Substantial' and `Strong' shifts away from the CMSSM respectively, also demonstrating through the increased prior dependence that $\delta a_\mu$ plays an important role in constraining the initially viable parameter space, i.e. in building the informative priors from the ``pre-LEP'' dataset.

As mentioned in section~\ref{subsec:prior} we were driven to ignore the $\mu<0$ branch of the CMSSM parameter space by computational restrictions and due to its poorer fit to data, particularly $\delta a_\mu$. However, given the significant (relative) boost to confidence in the CMSSM that is gained by removing $\delta a_\mu$, and the potential for the boost to be even larger had the $\mu<0$ branch not been ignored, it would be interesting to take this branch into account in future work. Confidence in the $\delta a_\mu$ constraint is thus seen to remain an important issue.

We understand that some readers may remain confused as to how the removal of the $\delta a_\mu$ constraint can improve the performance of the CMSSM in our analysis. To understand this, it is important to remember that the $\delta a_\mu$ constraint entered into our analysis as part of our `baseline' dataset, which was used to effectively create an informative prior for the CMSSM (i.e. the posterior resulting from the inclusion of this baseline data). The only effect of $\delta a_\mu$ is thus to help determine the initially viable regions of parameter space in each model. Since the SM is highly constrained it cannot `tune' its prediction of $\delta a_\mu$ to match experiments, so there is no change to its parameter space whether $\delta a_\mu$ is included or not (and since the DM sector is assumed to be unaffected by all data after the baseline the parts of the PBFs originating from it remain 1 even if it is constrained by $\delta a_\mu$). On the other hand, the initially viable parameter space of the CMSSM is severely restricted --to low $M_0$ and $M_{1/2}$ values-- by the demand that it reproduce the observed $\delta a_\mu$ value. This leaves the CMSSM highly vulnerable to damage from the LEP2 Higgs limits, which is reflected in the large PBF against it when the LEP2 data is introduced. Removing $\delta a_\mu$ alleviates this extremely strong tension, greatly reducing the corresponding PBF. The relative maximum likelihood penalty against the SM+DM that one may expect to be present due to $\delta a_\mu$ is {\em not} present in our PBFs, because we have separated it into the {\em prior}, that is ``pre-LEP'', odds. Our analysis is not designed to assess these odds, however when interpreting our PBFs the reader must keep in mind which data we have dealt with in the PBFs and which they are left to consider in their personal ``pre-LEP'' prior odds.

To reiterate: the reader may feel that we are not considering the direct effects of the various observables that form our initial ``pre-LEP'' data set on the model comparison. This is completely true; all this data has contributed to previous `iterations' of the Bayesian update process, and must be considered in the prior (``pre-LEP'') odds for the current analysis. We have done this to distance our analysis as much as possible from the impact of the somewhat subjective ``pre-LEP'' parameter space priors, in order to more robustly isolate the impact of the experiments under study. 
Based on the results presented in tables \ref{tab:bayesfactors} it appears that only a mild ``pre-LEP'' prior dependence remains if the $\delta a_\mu$ constraint is removed, as the results of table \ref{tab:amubayesfactors} show. The cost of this robustness is that some of the burden of interpretation remains on the reader. For example, say after the ``pre-LEP'' update one believes the odds for the CMSSM vs the SM-like model to be 1:1, after considering all the data in our `baseline' (``pre-LEP'') dataset along with personal theoretical biases. Our PBFs then dictate how one is required to modify these beliefs in light of the data featured in the subsequent updates. As a more explicit demonstration of how this should be done we offer the following toy thought process, considering the PBFs of table \ref{tab:bayesfactors};  
``According to Bal\'azs et al. the total CMSSM:SM+DM Bayes factor for learning the LEP2 Higgs limits, Xenon100 limits, 1 fb$^{-1}$ sparticle search limits, and early 2012 Higgs search results, is about 1:200 (CMSSM:SM+DM). The ``pre-LEP'' parameter space priors they have used roughly correspond to my expectations about the CMSSM, so I accept this number. Multiplying this by my personal ``pre-LEP'' odds, which I estimate to be roughly 50:1 (favouring the CMSSM), I obtain posterior odds of about 1:4, now in favour of the Standard Model by a moderate amount. 

Although crude, and not rigorous as to the details of what it means to believe that the CMSSM will be discovered (which is a serious question in of itself, requiring that we be far more thorough with the definition of the propositions which we so simply represent by the symbols ``CMSSM'' and ``SM+DM'' in this work, remembering that the interpretation of Bayesian inference which we follow is primarily as a theory of reasoning about the truth or falsity of propositions in the face of uncertainty), we hope that this example helps to clarify the meaning of our results.


While this paper was under review the ATLAS and CMS Higgs searches made significant progress, with the local p-values for the signal ``hints'' utilised in this work increasing in significance to over ``5 sigma'', leading to the announcement of the discovery of a new integer-spin resonance \cite{:2012gk,:2012gu}. We expect this to increase the degree to which the CMSSM is disfavoured in our results, since the decrease in parameter space compatible with the stronger measurement will be larger in the CMSSM than the SM+DM, and potential discrepancies in the branching ratios from SM predictions are not significant enough to make much of an impact. In addition, ATLAS and CMS have released the results of numerous supersymmetry searches using up to 5 $\text{fb}^{-1}$ of data (for example \cite{ATLAS-CONF-2012-033,CMS-PAS-SUS-12-005}, utilising the full 2011 data set; results of similar strength utilising 2012 data also exist, but no combination of 2011 and 2012 data is yet available), a significant increase over the 1 $\text{fb}^{-1}$ results we have used. No hints of new physics have been seen, and though the improved limits do cut a small way into the posterior remaining in our final ``ATLAS-Higgs'' datasets the improvement is not sufficient to significantly alter the PBFs we have computed; we expect considerably less than an extra factor of two shift against the CMSSM \footnote{We note that the new limits cut off much less than half of the posterior remaining in the ``ATLAS-Higgs'' dataset (shown in the last frame of figure~\ref{fig:pos1CCR}), so the corresponding ``additional'' PBF is likewise much less than two}
.

\begin{table*}\footnotesize
   \begin{center}
    \begin{threeparttable}
      \begin{tabular}{| l | l | l | l | l | l | l |}
         \hline
         Scenario & $\ln\mathcal Z$ & LR & $\mathcal O$ & $B$ & $B_\text{cumulative}$ & Strength of $B$ \\
         \hline
         \noalign{\smallskip}
         \multicolumn{2}{c}{\textbf{CMSSM (log priors)}} \\
         \hline
         Pre-LEP & $36.69(2)$ & - & - & - & - & - \\
         \hline
         LEP+XENON100 & $34.43(2)$ & $1:1$ & $1:9.6(2)$ & $1:2.72(6)$ & $1:2.72(6)$ & Barely worth mentioning \\
         \hline
         ATLAS-sparticle & $34.77(2)$ & $1:1$ & $1:0.72(2)$ & $1:0.72(2)$ & $1:1.95(5) $ & Barely worth mentioning$^*$ \\
         \hline
	   ATLAS-Higgs& $29.42(2)$ & $1:1.8$ & $1:78(2)$ & $1:4.2(2)$ & $1:8.3(1)$& Substantial \\
	\hline
	\noalign{\smallskip}
         \multicolumn{2}{c}{\textbf{CMSSM (natural priors)}} \\
         \hline
         Pre-LEP & $29.29(2)$ & - & - & - & - & -\\
         \hline
         LEP+XENON100 & $27.27(2)$ & $1:1$ & $1:7.6(2)$ & $1:2.15(6)$ & $1:2.15(6)$ & Barely worth mentioning \\
         \hline
         ATLAS-sparticle & $26.67(2)$ & $1:1$ & $1:1.81(6)$ & $1:1.81(6)$ & $1:3.9(1)$ & Barely worth mentioning \\
         \hline
	   ATLAS-Higgs& $20.87(2)$ & $1:1.8$ & $1:126(4)$ & $1:6.7(2)$ & $1:26.1(8)$ & Substantial \\
	\hline
      \end{tabular}
    \footnotetext{$^*$ Here is the
footnote}
    \begin{tablenotes}
         \item [*] This Bayes factor appears to indicate a slight increase in the viable CMSSM parameter space, which is impossible. It is therefore certain to be an artefact of reweighting process.
    \end{tablenotes}
    \end{threeparttable}
    \caption{\label{tab:amubayesfactors} Summary and interpretation of our results, with the $\delta a_\mu$ constraint removed.  Columns as in table~\ref{tab:bayesfactors}. We have dropped the SM+DM rows because they are unchanged from table~\ref{tab:bayesfactors}. The $\delta a_\mu$ constraint pushes the posteriors strongly down in the mass parameters, so removing it makes us much less surprised that no direct evidence for the low-mass CMSSM was seen at LEP or in the LHC sparticle searches.  This is reflected in the weaker PBFs than in table~\ref{tab:bayesfactors}.  The LEP results in particular are seen to be much less surprising.  The combined effect of both colliders on a `pre-LEP' odds ratio is seen to be downgraded from a `Decisive' to `Substantial' (by the Jeffreys scale) shift away from the low energy CMSSM when using the log prior, and to be downgraded from `Decisive' to `Strong' when using the natural prior. Since these results were obtained by reweighting scan data whose sampling was optimised for likelihoods containing the $\delta a_\mu$ constraint, the reweighted posteriors are expected to be substantially under-sampled in the higher $\{M_0, M_{1/2}\}$ regions, causing the PBFs listed in this table to overestimate the penalty to the CMSSM, the correction of which would further weaken these PBFs relative to those in table~\ref{tab:bayesfactors}. Confidence in the $\delta a_\mu$ constraint is thus seen to have a very large impact, and is likely to remain an important issue in models beyond the CMSSM.}
   \end{center}
\end{table*}

\begin{table}\footnotesize
   \begin{center}
      \begin{tabular}{| r | l |}
	\hline
	$B$           & Strength of evidence \\
	\hline
	$< 1:1$       & Negative \\
	1:1 to 3:1    & Barely worth mentioning \\
	3:1 to 10:1   & Substantial \\
	10:1 to 30:1  & Strong \\
	30:1 to 100:1 & Very strong \\
	$> 100:1$	  & Decisive \\
	\hline
	\end{tabular}
    \caption{\label{tab:jeffreysscale} The Jeffreys scale for interpreting Bayes factors.  We use this scale to interpret our results for $B$ and $B_\text{cumulative}$.}
  \end{center}
\end{table}

\section{Conclusions}
\label{sec:concl}

We examined the viability of the low energy CMSSM, the corner of the parameter space with $M_0$ and $M_{1/2}$ restricted below 2 TeV, in the light of data from before LEP to the recent measurements of the LHC.  To quantify this viability we computed the partial Bayes factors associated with learning the LEP Higgs limits, XENON100 dark matter limits, LHC sparticle searches, and the 2012 LHC Higgs hint, in sequence, in a straightforward Bayesian hypothesis test of the CMSSM against a SM-like model.

Interpreting the relative change of belief in the CMSSM induced by these PBFs in terms of the Jeffreys scale we concluded the following.  The LEP Higgs limit strongly reduced our trust in the low energy CMSSM, as is well known. The LHC sparticle limits deal a much smaller and not yet very significant additional blow. Lastly, the LHC Higgs hints are already strong enough that they have a substantial impact (on the ``pre-LEP'' scenario) even if the previous damage is ignored.  
When considering the cumulative effect of all three data changes we found that support for the CMSSM, as measured by the posterior odds, is reduced relative to the SM-like alternative by a decisive 200 fold.  These findings are robust against the shape of prior probabilities of the CMSSM parameters we considered (and are expected to remain so under other reasonable choices for priors), however they are severely weakened if the sometimes contentious muon anomalous magnetic moment constraint is removed from consideration. Presently the impact of XENON100 is negligible, although in the near future dark matter direct detection is expected to further reduce our belief in the low energy corner of the CMSSM, unless they discover a positive signal soon. 

The strength of these results is largely due to the very small amount of CMSSM parameter space in the posterior of the initial (``pre-LEP'') data set, which forms the informative prior for the next update, which is capable of producing a lightest Higgs of around 125 GeV as is required to explain the LHC Higgs hints, and so is quite expected from that perspective.  The ease with which this can be accommodated in the SM-like model causes the more `wasteful' CMSSM to be strongly disfavoured. The CMSSM would not fare as poorly in a test against a perhaps more realistic model of similar parameter space complexity, unless that model naturally produces a compatible Higgs in a much more substantial portion of its otherwise viable parameter space. Likewise if there exists a good reason to restrict the parameter space prior for the CMSSM to those regions that produce a relatively viable Higgs, such as some motivation from a higher energy theory, then our large penalising Occam factors may be largely negated. This is essentially the Bayesian manifestation of a naturalness problem; the CMSSM is now a highly unnatural model (completely separately from the little hierarchy problem, which is associated with data in our baseline set) due to the small amount of parameter space capable of fitting both the Higgs observations and previous data, and this is strong motivation to search for a more complete theory (if not for a completely different theory) to explain why this small portion of parameter space should be chosen by Nature.

\section*{Acknowledgements}

The authors are indebted to Sudhir Gupta and Doyoon Kim for their assistance with the calculation of Higgs boson production cross sections and decays.
BF is thankful to Farhan Feroz for assistance with \texttt{MultiNest}.  MJW thanks Teng Jian Khoo and Ben Allanach for conversations regarding the calculation of ATLAS-based likelihoods for candidate SUSY models.
This research was funded in part by the ARC Centre of Excellence for Particle Physics at the Tera-scale, and in part by the Project of Knowledge Innovation Program (PKIP) of Chinese Academy of Sciences Grant No. KJCX2.YW.W10.  AB acknowledges the support of the Scottish Universities Physics Alliance.  The use of Monash Sun Grid (MSG) and Edinburgh ECDF high-performance computing facilities is also gratefully acknowledged.
Most numerical calculations were performed on the Australian National Computing Infrastructure (NCI) National Facility SGI XE cluster and Multi-modal Australian ScienceS Imaging and Visualisation Environment (MASSIVE) cluster.

\bibliographystyle{utphys} 
\bibliography{CMSSMev} 

\providecommand{\href}[2]{#2}\begingroup\raggedright\begin{thebibliography}{100}

\bibitem{Weinberg:2000cr}
S.~Weinberg,
``{The quantum theory of fields. Vol. 3: Supersymmetry},''.

\bibitem{Kane:2000bp}
G.~L. Kane,
``{Supersymmetry: Squarks, photinos, and the unveiling of the ultimate laws of
  nature},''.

\bibitem{Drees:2004jm}
M.~Drees, R.~Godbole, and P.~Roy,
``{Theory and phenomenology of sparticles: An account of four-dimensional N=1
  supersymmetry in high energy physics},''.

\bibitem{Baer:2006rs}
H.~Baer and X.~Tata,
``{Weak scale supersymmetry: From superfields to scattering events},''.

\bibitem{Binetruy:2006ad}
P.~Binetruy,
``{Supersymmetry: Theory, experiment and cosmology},''.

\bibitem{Terning:2006bq}
J.~Terning,
``{Modern supersymmetry: Dynamics and duality},''.

\bibitem{Polonsky:2001pn}
N.~Polonsky, ``{Supersymmetry: Structure and phenomena. Extensions of the
  standard model},'' {\em Lect. Notes Phys.} {\bfseries M68} (2001) 1--169,
\href{http://arxiv.org/abs/hep-ph/0108236}{{\ttfamily arXiv:hep-ph/0108236
  [hep-ph]}}.

\bibitem{Pagels:1981ke}
H.~Pagels and J.~R. Primack, ``{Supersymmetry, Cosmology and New TeV
  Physics},''
\href{http://dx.doi.org/10.1103/PhysRevLett.48.223}{{\em Phys. Rev. Lett.}
  {\bfseries 48} (1982) 223}.

\bibitem{Goldberg:1983nd}
H.~Goldberg, ``{Constraint on the Photino Mass from Cosmology},''
  \href{http://dx.doi.org/10.1103/PhysRevLett.50.1419,
  10.1103/PhysRevLett.50.1419}{{\em Phys.Rev.Lett.} {\bfseries 50} (1983)
  1419}.

\bibitem{Ramond:1999vh}
P.~Ramond,
``{Journeys beyond the standard model},''.

\bibitem{Baer:2001vw}
H.~Baer, C.~Balazs, M.~Brhlik, P.~Mercadante, X.~Tata, {\em et~al.}, ``{Aspects
  of supersymmetric models with a radiatively driven inverted mass
  hierarchy},'' \href{http://dx.doi.org/10.1103/PhysRevD.64.015002}{{\em
  Phys.Rev.} {\bfseries D64} (2001) 015002},
\href{http://arxiv.org/abs/hep-ph/0102156}{{\ttfamily arXiv:hep-ph/0102156
  [hep-ph]}}.

\bibitem{Balazs:2004ae}
C.~Balazs, M.~S. Carena, A.~Menon, D.~Morrissey, and C.~Wagner, ``{The
  Supersymmetric origin of matter},''
  \href{http://dx.doi.org/10.1103/PhysRevD.71.075002}{{\em Phys.Rev.}
  {\bfseries D71} (2005) 075002},
\href{http://arxiv.org/abs/hep-ph/0412264}{{\ttfamily arXiv:hep-ph/0412264
  [hep-ph]}}.

\bibitem{Nanopoulos:1982bv}
D.~V. Nanopoulos, K.~A. Olive, M.~Srednicki, and K.~Tamvakis, ``{Primordial
  Inflation in Simple Supergravity},''
\href{http://dx.doi.org/10.1016/0370-2693(83)90954-1}{{\em Phys. Lett.}
  {\bfseries B123} (1983) 41}.

\bibitem{Holman:1984yj}
R.~Holman, P.~Ramond, and G.~G. Ross, ``{Supersymmetric Inflationary
  Cosmology},''
\href{http://dx.doi.org/10.1016/0370-2693(84)91729-5}{{\em Phys. Lett.}
  {\bfseries B137} (1984) 343--347}.

\bibitem{Dimopoulos:1981zb}
S.~Dimopoulos and H.~Georgi, ``{Softly Broken Supersymmetry and SU(5)},''
  \href{http://dx.doi.org/10.1016/0550-3213(81)90522-8}{{\em Nucl.Phys.}
  {\bfseries B193} (1981) 150}.

\bibitem{Chamseddine:1982jx}
A.~H. Chamseddine, R.~L. Arnowitt, and P.~Nath, ``{Locally Supersymmetric Grand
  Unification},'' \href{http://dx.doi.org/10.1103/PhysRevLett.49.970}{{\em
  Phys.Rev.Lett.} {\bfseries 49} (1982) 970}.

\bibitem{Baer:2003yh}
H.~Baer and C.~Balazs, ``{Chi**2 analysis of the minimal supergravity model
  including WMAP, g(mu)-2 and b $\to$ s gamma constraints},''
  \href{http://dx.doi.org/10.1088/1475-7516/2003/05/006}{{\em JCAP} {\bfseries
  0305} (2003) 006},
\href{http://arxiv.org/abs/hep-ph/0303114}{{\ttfamily arXiv:hep-ph/0303114
  [hep-ph]}}.

\bibitem{Ellis:2003si}
J.~R. Ellis, K.~A. Olive, Y.~Santoso, and V.~C. Spanos, ``{Likelihood analysis
  of the CMSSM parameter space},''
  \href{http://dx.doi.org/10.1103/PhysRevD.69.095004}{{\em Phys.Rev.}
  {\bfseries D69} (2004) 095004},
\href{http://arxiv.org/abs/hep-ph/0310356}{{\ttfamily arXiv:hep-ph/0310356
  [hep-ph]}}.

\bibitem{Bechtle:2009ty}
P.~Bechtle, K.~Desch, M.~Uhlenbrock, and P.~Wienemann, ``{Constraining SUSY
  models with Fittino using measurements before, with and beyond the LHC},''
  \href{http://dx.doi.org/10.1140/epjc/s10052-009-1228-3}{{\em Eur.Phys.J.}
  {\bfseries C66} (2010) 215--259},
\href{http://arxiv.org/abs/0907.2589}{{\ttfamily arXiv:0907.2589 [hep-ph]}}.

\bibitem{Bechtle:2011it}
P.~Bechtle, K.~Desch, H.~Dreiner, M.~Kramer, B.~O'Leary, {\em et~al.},
  ``{Present and possible future implications for mSUGRA of the non-discovery
  of SUSY at the LHC},''
\href{http://arxiv.org/abs/1105.5398}{{\ttfamily arXiv:1105.5398 [hep-ph]}}.

\bibitem{Heinemeyer:2010de}
S.~Heinemeyer and G.~Weiglein, ``{Predicting Supersymmetry},''
  \href{http://dx.doi.org/10.1016/j.nuclphysbps.2010.09.007}{{\em
  Nucl.Phys.Proc.Suppl.} {\bfseries 205-206} (2010) 283--288},
\href{http://arxiv.org/abs/1007.0206}{{\ttfamily arXiv:1007.0206 [hep-ph]}}.

\bibitem{Buchmueller:2010ai}
O.~Buchmueller, R.~Cavanaugh, D.~Colling, A.~De~Roeck, M.~Dolan, {\em et~al.},
  ``{Frequentist Analysis of the Parameter Space of Minimal Supergravity},''
  \href{http://dx.doi.org/10.1140/epjc/s10052-011-1583-8}{{\em Eur.Phys.J.}
  {\bfseries C71} (2011) 1583},
\href{http://arxiv.org/abs/1011.6118}{{\ttfamily arXiv:1011.6118 [hep-ph]}}.

\bibitem{LopezFogliani:2009np}
D.~E. Lopez-Fogliani, L.~Roszkowski, R.~R. de~Austri, and T.~A. Varley, ``{A
  Bayesian Analysis of the Constrained NMSSM},''
  \href{http://dx.doi.org/10.1103/PhysRevD.80.095013}{{\em Phys.Rev.}
  {\bfseries D80} (2009) 095013},
\href{http://arxiv.org/abs/0906.4911}{{\ttfamily arXiv:0906.4911 [hep-ph]}}.

\bibitem{Cabrera:2009dm}
M.~E. Cabrera, J.~A. Casas, and R.~Ruiz~d Austri, ``{MSSM Forecast for the
  LHC},'' \href{http://dx.doi.org/10.1007/JHEP05(2010)043}{{\em JHEP}
  {\bfseries 1005} (2010) 043},
\href{http://arxiv.org/abs/0911.4686}{{\ttfamily arXiv:0911.4686 [hep-ph]}}.

\bibitem{Buchmueller:2011ki}
O.~Buchmueller, R.~Cavanaugh, D.~Colling, A.~De~Roeck, M.~Dolan, {\em et~al.},
  ``{Supersymmetry and Dark Matter in Light of LHC 2010 and Xenon100 Data},''
  \href{http://dx.doi.org/10.1140/epjc/s10052-011-1722-2}{{\em Eur.Phys.J.}
  {\bfseries C71} (2011) 1722},
\href{http://arxiv.org/abs/1106.2529}{{\ttfamily arXiv:1106.2529 [hep-ph]}}.

\bibitem{Ellis:2012aa}
J.~Ellis and K.~A. Olive, ``{Revisiting the Higgs Mass and Dark Matter in the
  CMSSM},''
\href{http://arxiv.org/abs/1202.3262}{{\ttfamily arXiv:1202.3262 [hep-ph]}}.

\bibitem{Bechtle:2012zk}
P.~Bechtle, T.~Bringmann, K.~Desch, H.~Dreiner, M.~Hamer, {\em et~al.},
  ``{Constrained Supersymmetry after two years of LHC data: a global view with
  Fittino},''
\href{http://arxiv.org/abs/1204.4199}{{\ttfamily arXiv:1204.4199 [hep-ph]}}.

\bibitem{Allanach:2011ut}
B.~Allanach, ``{Impact of CMS Multi-jets and Missing Energy Search on CMSSM
  Fits},'' \href{http://dx.doi.org/10.1103/PhysRevD.83.095019}{{\em Phys.Rev.}
  {\bfseries D83} (2011) 095019},
\href{http://arxiv.org/abs/1102.3149}{{\ttfamily arXiv:1102.3149 [hep-ph]}}.

\bibitem{Allanach:2011wi}
B.~Allanach, T.~Khoo, C.~Lester, and S.~Williams, ``{The impact of the ATLAS
  zero-lepton, jets and missing momentum search on a CMSSM fit},''
  \href{http://dx.doi.org/10.1007/JHEP06(2011)035}{{\em JHEP} {\bfseries 1106}
  (2011) 035},
\href{http://arxiv.org/abs/1103.0969}{{\ttfamily arXiv:1103.0969 [hep-ph]}}.

\bibitem{Bertone:2011nj}
G.~Bertone, D.~G. Cerdeno, M.~Fornasa, R.~Ruiz~de Austri, C.~Strege, {\em
  et~al.}, ``{Global fits of the cMSSM including the first LHC and XENON100
  data},'' \href{http://dx.doi.org/10.1088/1475-7516/2012/01/015}{{\em JCAP}
  {\bfseries 1201} (2012) 015},
\href{http://arxiv.org/abs/1107.1715}{{\ttfamily arXiv:1107.1715 [hep-ph]}}.

\bibitem{Fowlie:2011mb}
A.~Fowlie, A.~Kalinowski, M.~Kazana, L.~Roszkowski, and Y.~S. Tsai, ``{Bayesian
  Implications of Current LHC and XENON100 Search Limits for the Constrained
  MSSM},''
\href{http://arxiv.org/abs/1111.6098}{{\ttfamily arXiv:1111.6098 [hep-ph]}}.

\bibitem{Buchmueller:2011sw}
O.~Buchmueller, R.~Cavanaugh, A.~De~Roeck, M.~Dolan, J.~Ellis, {\em et~al.},
  ``{Supersymmetry in Light of 1/fb of LHC Data},''
\href{http://arxiv.org/abs/1110.3568}{{\ttfamily arXiv:1110.3568 [hep-ph]}}.

\bibitem{Buchmueller:2011ab}
O.~Buchmueller, R.~Cavanaugh, A.~De~Roeck, M.~Dolan, J.~Ellis, {\em et~al.},
  ``{Higgs and Supersymmetry},''
\href{http://arxiv.org/abs/1112.3564}{{\ttfamily arXiv:1112.3564 [hep-ph]}}.

\bibitem{Starkman:2008py}
G.~D. Starkman, R.~Trotta, and P.~M. Vaudrevange, ``{Introducing doubt in
  Bayesian model comparison},''
\href{http://arxiv.org/abs/0811.2415}{{\ttfamily arXiv:0811.2415
  [physics.data-an]}}.

\bibitem{Cabrera:2011ds}
M.~E. Cabrera, J.~Casas, V.~A. Mitsou, R.~Ruiz~de Austri, and J.~Terron,
  ``{Histogram comparison as a powerful tool for the search of new physics at
  LHC. Application to CMSSM},''
\href{http://arxiv.org/abs/1109.3759}{{\ttfamily arXiv:1109.3759 [hep-ph]}}.

\bibitem{AbdusSalam:2011fc}
S.~AbdusSalam, B.~Allanach, H.~Dreiner, J.~Ellis, U.~Ellwanger, {\em et~al.},
  ``{Benchmark Models, Planes, Lines and Points for Future SUSY Searches at the
  LHC},'' \href{http://dx.doi.org/10.1140/epjc/s10052-011-1835-7}{{\em
  Eur.Phys.J.} {\bfseries C71} (2011) 1835},
\href{http://arxiv.org/abs/1109.3859}{{\ttfamily arXiv:1109.3859 [hep-ph]}}.

\bibitem{Sekmen:2011cz}
S.~Sekmen, S.~Kraml, J.~Lykken, F.~Moortgat, S.~Padhi, {\em et~al.},
  ``{Interpreting LHC SUSY searches in the phenomenological MSSM},''
\href{http://arxiv.org/abs/1109.5119}{{\ttfamily arXiv:1109.5119 [hep-ph]}}.

\bibitem{Strege:2011pk}
C.~Strege, G.~Bertone, D.~Cerdeno, M.~Fornasa, R.~de~Austri, {\em et~al.},
  ``{Updated global fits of the cMSSM including the latest LHC SUSY and Higgs
  searches and XENON100 data},''
\href{http://arxiv.org/abs/1112.4192}{{\ttfamily arXiv:1112.4192 [hep-ph]}}.

\bibitem{Roszkowski:2012uf}
L.~Roszkowski, E.~M. Sessolo, and Y.-L.~S. Tsai, ``{Bayesian Implications of
  Current LHC Supersymmetry and Dark Matter Detection Searches for the
  Constrained MSSM},''
\href{http://arxiv.org/abs/1202.1503}{{\ttfamily arXiv:1202.1503 [hep-ph]}}.

\bibitem{OHagen1995}
A.~O'Hagan, ``Fractional bayes factors for model comparison,'' {\em Journal of
  the Royal Statistical Society. Series B (Methodological)} {\bfseries 57}
  no.~1, (1995) pp. 99--138.

\bibitem{berger1996intrinsic}
J.~Berger and L.~Pericchi, ``The intrinsic bayes factor for model selection and
  prediction,'' {\em Journal of the American Statistical Association}
  {\bfseries 91} no.~433, (1996) 109--122.

\bibitem{berger1999default}
J.~Berger and J.~Mortera, ``Default bayes factors for nonnested hypothesis
  testing,'' {\em Journal of the American Statistical Association} {\bfseries
  94} no.~446, (1999) 542--554.

\bibitem{Allanach:2006jc}
B.~Allanach, ``{Naturalness priors and fits to the constrained minimal
  supersymmetric standard model},''
  \href{http://dx.doi.org/10.1016/j.physletb.2006.02.052}{{\em Phys.Lett.}
  {\bfseries B635} (2006) 123--130},
\href{http://arxiv.org/abs/hep-ph/0601089}{{\ttfamily arXiv:hep-ph/0601089
  [hep-ph]}}.

\bibitem{Allanach:2007qk}
B.~C. Allanach, K.~Cranmer, C.~G. Lester, and A.~M. Weber, ``{Natural priors,
  CMSSM fits and LHC weather forecasts},''
  \href{http://dx.doi.org/10.1088/1126-6708/2007/08/023}{{\em JHEP} {\bfseries
  0708} (2007) 023},
\href{http://arxiv.org/abs/0705.0487}{{\ttfamily arXiv:0705.0487 [hep-ph]}}.

\bibitem{Cabrera:2008tj}
M.~E. Cabrera, J.~A. Casas, and R.~Ruiz~de Austri, ``{Bayesian approach and
  Naturalness in MSSM analyses for the LHC},''
  \href{http://dx.doi.org/10.1088/1126-6708/2009/03/075}{{\em JHEP} {\bfseries
  03} (2009) 075},
\href{http://arxiv.org/abs/0812.0536}{{\ttfamily arXiv:0812.0536 [hep-ph]}}.

\bibitem{Cabrera:2010dh}
M.~E. Cabrera, ``{Bayesian Study and Naturalness in MSSM Forecast for the
  LHC},''
\href{http://arxiv.org/abs/1005.2525}{{\ttfamily arXiv:1005.2525 [hep-ph]}}.

\bibitem{Hall:2011aa}
L.~J. Hall, D.~Pinner, and J.~T. Ruderman, ``{A Natural SUSY Higgs Near 126
  GeV},''
\href{http://arxiv.org/abs/1112.2703}{{\ttfamily arXiv:1112.2703 [hep-ph]}}.

\bibitem{Athron:2007ry}
P.~Athron and .~Miller, D.J., ``{A New Measure of Fine Tuning},''
  \href{http://dx.doi.org/10.1103/PhysRevD.76.075010}{{\em Phys.Rev.}
  {\bfseries D76} (2007) 075010},
\href{http://arxiv.org/abs/0705.2241}{{\ttfamily arXiv:0705.2241 [hep-ph]}}.

\bibitem{Cassel:2009cx}
S.~Cassel, D.~Ghilencea, and G.~Ross, ``{Testing SUSY},''
  \href{http://dx.doi.org/10.1016/j.physletb.2010.03.032}{{\em Phys.Lett.}
  {\bfseries B687} (2010) 214--218},
\href{http://arxiv.org/abs/0911.1134}{{\ttfamily arXiv:0911.1134 [hep-ph]}}.

\bibitem{Horton:2009ed}
D.~Horton and G.~Ross, ``{Naturalness and Focus Points with Non-Universal
  Gaugino Masses},''
  \href{http://dx.doi.org/10.1016/j.nuclphysb.2009.12.031}{{\em Nucl.Phys.}
  {\bfseries B830} (2010) 221--247},
\href{http://arxiv.org/abs/0908.0857}{{\ttfamily arXiv:0908.0857 [hep-ph]}}.

\bibitem{Cassel:2010px}
S.~Cassel, D.~Ghilencea, and G.~Ross, ``{Testing SUSY at the LHC: Electroweak
  and Dark matter fine tuning at two-loop order},''
  \href{http://dx.doi.org/10.1016/j.nuclphysb.2010.03.031}{{\em Nucl.Phys.}
  {\bfseries B835} (2010) 110--134},
\href{http://arxiv.org/abs/1001.3884}{{\ttfamily arXiv:1001.3884 [hep-ph]}}.

\bibitem{Akula:2011jx}
S.~Akula, M.~Liu, P.~Nath, and G.~Peim, ``{Naturalness, Supersymmetry and
  Implications for LHC and Dark Matter},''
\href{http://arxiv.org/abs/1111.4589}{{\ttfamily arXiv:1111.4589 [hep-ph]}}.

\bibitem{Arbey:2011un}
A.~Arbey, M.~Battaglia, and F.~Mahmoudi, ``{Implications of LHC Searches on
  SUSY Particle Spectra: The pMSSM Parameter Space with Neutralino Dark
  Matter},'' \href{http://dx.doi.org/10.1140/epjc/s10052-011-1847-3}{{\em
  Eur.Phys.J.} {\bfseries C72} (2012) 1847},
\href{http://arxiv.org/abs/1110.3726}{{\ttfamily arXiv:1110.3726 [hep-ph]}}.

\bibitem{Cassel:2011tg}
S.~Cassel, D.~Ghilencea, S.~Kraml, A.~Lessa, and G.~Ross, ``{Fine-tuning
  implications for complementary dark matter and LHC SUSY searches},''
  \href{http://dx.doi.org/10.1007/JHEP05(2011)120}{{\em JHEP} {\bfseries 1105}
  (2011) 120},
\href{http://arxiv.org/abs/1101.4664}{{\ttfamily arXiv:1101.4664 [hep-ph]}}.

\bibitem{Papucci:2011wy}
M.~Papucci, J.~T. Ruderman, and A.~Weiler, ``{Natural SUSY Endures},''
\href{http://arxiv.org/abs/1110.6926}{{\ttfamily arXiv:1110.6926 [hep-ph]}}.

\bibitem{Li:2011xg}
T.~Li, J.~A. Maxin, D.~V. Nanopoulos, and J.~W. Walker, ``{Natural Predictions
  for the Higgs Boson Mass and Supersymmetric Contributions to Rare
  Processes},'' {\em Phys.Lett.} {\bfseries B708} (2012) 93--99,
\href{http://arxiv.org/abs/1109.2110}{{\ttfamily arXiv:1109.2110 [hep-ph]}}.

\bibitem{Kang:2012tn}
Z.~Kang, J.~Li, and T.~Li, ``{On the Naturalness of the (N)MSSM},''
\href{http://arxiv.org/abs/1201.5305}{{\ttfamily arXiv:1201.5305 [hep-ph]}}.

\bibitem{jaynes2003probability}
E.~Jaynes and G.~Bretthorst, {\em Probability theory: the logic of science}.
\newblock Cambridge Univ Pr, 2003.

\bibitem{Feroz:2008wr}
F.~Feroz, B.~C. Allanach, M.~Hobson, S.~S. AbdusSalam, R.~Trotta, {\em et~al.},
  ``{Bayesian Selection of sign(mu) within mSUGRA in Global Fits Including
  WMAP5 Results},'' \href{http://dx.doi.org/10.1088/1126-6708/2008/10/064}{{\em
  JHEP} {\bfseries 0810} (2008) 064},
\href{http://arxiv.org/abs/0807.4512}{{\ttfamily arXiv:0807.4512 [hep-ph]}}.

\bibitem{AbdusSalam:2009tr}
S.~S. AbdusSalam, B.~C. Allanach, M.~J. Dolan, F.~Feroz, and M.~P. Hobson,
  ``{Selecting a Model of Supersymmetry Breaking Mediation},''
  \href{http://dx.doi.org/10.1103/PhysRevD.80.035017}{{\em Phys. Rev.}
  {\bfseries D80} (2009) 035017},
\href{http://arxiv.org/abs/0906.0957}{{\ttfamily arXiv:0906.0957 [hep-ph]}}.

\bibitem{Feroz:2009dv}
F.~Feroz, M.~P. Hobson, L.~Roszkowski, R.~Ruiz~de Austri, and R.~Trotta, ``{Are
  $BR(\bar{B} \to X_s \gamma)$ and $(g-2)_\mu$ consistent within the
  Constrained MSSM?},''
\href{http://arxiv.org/abs/0903.2487}{{\ttfamily arXiv:0903.2487 [hep-ph]}}.

\bibitem{Cabrera:2010xx}
M.~E. Cabrera, J.~Casas, R.~Ruiz~de Austri, and R.~Trotta, ``{Quantifying the
  tension between the Higgs mass and $(g-2)_\mu$ in the CMSSM},''
  \href{http://dx.doi.org/10.1103/PhysRevD.84.015006}{{\em Phys.Rev.}
  {\bfseries D84} (2011) 015006},
\href{http://arxiv.org/abs/1011.5935}{{\ttfamily arXiv:1011.5935 [hep-ph]}}.

\bibitem{Pierini:2011yf}
M.~Pierini, H.~Prosper, S.~Sekmen, and M.~Spiropulu, ``{Model Inference with
  Reference Priors},''
\href{http://arxiv.org/abs/1107.2877}{{\ttfamily arXiv:1107.2877 [hep-ph]}}.

\bibitem{mackay2003information}
D.~MacKay, {\em Information theory, inference, and learning algorithms}.
\newblock Cambridge Univ Pr, 2003.

\bibitem{Solomonoff19641}
R.~Solomonoff, ``A formal theory of inductive inference. part i,''
  \href{http://dx.doi.org/10.1016/S0019-9958(64)90223-2}{{\em Information and
  Control} {\bfseries 7} no.~1, (1964) 1 -- 22}.
  \url{http://www.sciencedirect.com/science/article/pii/S0019995864902232}.

\bibitem{Fichet:2012sn}
S.~Fichet, ``{Quantified naturalness from Bayesian statistics},''
\href{http://arxiv.org/abs/1204.4940}{{\ttfamily arXiv:1204.4940 [hep-ph]}}.

\bibitem{LEPHWG:1998}
{\bfseries ALEPH, DELPHI, L3, OPAL, LEP Electroweak Working Group}
  Collaboration, P.~Bock, J.~Carr, S.~De~Jong, F.~Di~Lodovico, E.~Gross,
  P.~Igo-Kemenes, P.~Janot, W.~Murray, M.~Pieri, A.~L. Read,
  V.~Ruhlmann-Kleider, and A.~Sopczak, ``Lower bound for the standard model
  higgs boson mass from combining the results of the four lep experiments,''
  tech. rep., CERN, Geneva, Apr, 1998.
\newblock \url{http://cdsweb.cern.ch/record/353201}.

\bibitem{EWWG:2010vi}
{\bfseries ALEPH, CDF, D0, DELPHI, L3, OPAL, SLD, LEP Electroweak Working
  Group, Tevatron Electroweak Working Group, SLD Electroweak and Heavy Flavour
  Groups} Collaboration, ``{Precision Electroweak Measurements and Constraints
  on the Standard Model},''
\href{http://arxiv.org/abs/1012.2367}{{\ttfamily arXiv:1012.2367 [hep-ex]}}.

\bibitem{ATLASaa:2012sk}
{\bfseries ATLAS} Collaboration, G.~Aad {\em et~al.}, ``{Search for the
  Standard Model Higgs boson in the diphoton decay channel with 4.9
  $\text{fb}^{-1}$ of pp collisions at $\sqrt{s}=7$ TeV with ATLAS},''
\href{http://arxiv.org/abs/1202.1414}{{\ttfamily arXiv:1202.1414 [hep-ex]}}.

\bibitem{ATLAS2l2v:2011aa}
{\bfseries ATLAS} Collaboration, G.~Aad {\em et~al.}, ``{Search for the Higgs
  boson in the $H \to WW^{*} \to l\nu l\nu$ decay channel in pp collisions at
  $\sqrt{s} = 7$ TeV with the ATLAS detector},''
\href{http://arxiv.org/abs/1112.2577}{{\ttfamily arXiv:1112.2577 [hep-ex]}}.

\bibitem{ATLAS4l:2012sm}
{\bfseries ATLAS} Collaboration, G.~Aad {\em et~al.}, ``{Search for the
  Standard Model Higgs boson in the decay channel $H \to ZZ^{*} \to 4l$ with
  4.8 $\text{fb}^{-1}$ of pp collisions at $\sqrt{s}=7$ TeV with ATLAS},''
\href{http://arxiv.org/abs/1202.1415}{{\ttfamily arXiv:1202.1415 [hep-ex]}}.

\bibitem{ATLAS:2012si}
{\bfseries ATLAS} Collaboration, G.~Aad {\em et~al.}, ``{Combined search for
  the Standard Model Higgs boson using up to 4.9 $\text{fb}^{-1}$ of pp
  collision data at $\sqrt{s} = 7$ TeV with the ATLAS detector at the LHC},''
  {\em Phys.Lett.} {\bfseries B710} (2012) 49--66,
\href{http://arxiv.org/abs/1202.1408}{{\ttfamily arXiv:1202.1408 [hep-ex]}}.

\bibitem{Feroz:2008xx}
F.~Feroz, M.~P. Hobson, and M.~Bridges, ``{MultiNest: an efficient and robust
  Bayesian inference tool for cosmology and particle physics},'' {\em Mon. Not.
  Roy. Astron. Soc.} {\bfseries 398} (2009) 1601--1614,
\href{http://arxiv.org/abs/0809.3437}{{\ttfamily arXiv:0809.3437 [astro-ph]}}.

\bibitem{Feroz:2007kg}
F.~Feroz and M.~P. Hobson, ``{Multimodal nested sampling: an efficient and
  robust alternative to MCMC methods for astronomical data analysis},''
\href{http://arxiv.org/abs/0704.3704}{{\ttfamily arXiv:0704.3704 [astro-ph]}}.

\bibitem{skilling:395}
J.~Skilling, ``Nested sampling,''
  \href{http://dx.doi.org/10.1063/1.1835238}{{\em AIP Conference Proceedings}
  {\bfseries 735} no.~1, (2004) 395--405}.
  \url{http://link.aip.org/link/?APC/735/395/1}.

\bibitem{Akrami:2009hp}
Y.~Akrami, P.~Scott, J.~Edsjo, J.~Conrad, and L.~Bergstrom, ``{A Profile
  Likelihood Analysis of the Constrained MSSM with Genetic Algorithms},''
  \href{http://dx.doi.org/10.1007/JHEP04(2010)057}{{\em JHEP} {\bfseries 1004}
  (2010) 057},
\href{http://arxiv.org/abs/0910.3950}{{\ttfamily arXiv:0910.3950 [hep-ph]}}.

\bibitem{Bridges:2010de}
M.~Bridges, K.~Cranmer, F.~Feroz, M.~Hobson, R.~R. de~Austri, {\em et~al.},
  ``{A Coverage Study of the CMSSM Based on ATLAS Sensitivity Using Fast Neural
  Networks Techniques},'' \href{http://dx.doi.org/10.1007/JHEP03(2011)012}{{\em
  JHEP} {\bfseries 1103} (2011) 012},
\href{http://arxiv.org/abs/1011.4306}{{\ttfamily arXiv:1011.4306 [hep-ph]}}.

\bibitem{Paige:2003mg}
F.~E. Paige, S.~D. Protopopescu, H.~Baer, and X.~Tata, ``{ISAJET 7.69: A Monte
  Carlo event generator for p p, anti-p p, and e+ e- reactions},''
\href{http://arxiv.org/abs/hep-ph/0312045}{{\ttfamily arXiv:hep-ph/0312045}}.

\bibitem{Belanger:2010pz}
G.~Belanger, F.~Boudjema, A.~Pukhov, and A.~Semenov, ``{micrOMEGAs : a tool for
  dark matter studies},''
\href{http://arxiv.org/abs/1005.4133}{{\ttfamily arXiv:1005.4133 [hep-ph]}}.

\bibitem{Belanger:2008sj}
G.~Belanger, F.~Boudjema, A.~Pukhov, and A.~Semenov, ``{Dark matter direct
  detection rate in a generic model with micrOMEGAs2.1},''
  \href{http://dx.doi.org/10.1016/j.cpc.2008.11.019}{{\em Comput. Phys.
  Commun.} {\bfseries 180} (2009) 747--767},
\href{http://arxiv.org/abs/0803.2360}{{\ttfamily arXiv:0803.2360 [hep-ph]}}.

\bibitem{Belanger:2006is}
G.~Belanger, F.~Boudjema, A.~Pukhov, and A.~Semenov, ``{micrOMEGAs2.0: A
  program to calculate the relic density of dark matter in a generic model},''
  \href{http://dx.doi.org/10.1016/j.cpc.2006.11.008}{{\em Comput. Phys.
  Commun.} {\bfseries 176} (2007) 367--382},
\href{http://arxiv.org/abs/hep-ph/0607059}{{\ttfamily arXiv:hep-ph/0607059}}.

\bibitem{Mahmoudi:2007vz}
F.~Mahmoudi, ``{SuperIso: A program for calculating the isospin asymmetry of B
  -> K* gamma in the MSSM},''
  \href{http://dx.doi.org/10.1016/j.cpc.2007.12.006}{{\em Comput. Phys.
  Commun.} {\bfseries 178} (2008) 745--754},
\href{http://arxiv.org/abs/0710.2067}{{\ttfamily arXiv:0710.2067 [hep-ph]}}.

\bibitem{Mahmoudi:2008tp}
F.~Mahmoudi, ``{SuperIso v2.3: A Program for calculating flavor physics
  observables in Supersymmetry},''
  \href{http://dx.doi.org/10.1016/j.cpc.2009.02.017}{{\em Comput. Phys.
  Commun.} {\bfseries 180} (2009) 1579--1613},
\href{http://arxiv.org/abs/0808.3144}{{\ttfamily arXiv:0808.3144 [hep-ph]}}.

\bibitem{Djouadi:1997yw}
A.~Djouadi, J.~Kalinowski, and M.~Spira, ``{HDECAY: A Program for Higgs boson
  decays in the standard model and its supersymmetric extension},''
  \href{http://dx.doi.org/10.1016/S0010-4655(97)00123-9}{{\em
  Comput.Phys.Commun.} {\bfseries 108} (1998) 56--74},
\href{http://arxiv.org/abs/hep-ph/9704448}{{\ttfamily arXiv:hep-ph/9704448
  [hep-ph]}}.

\bibitem{Feroz:2011bj}
F.~Feroz, K.~Cranmer, M.~Hobson, R.~Ruiz~de Austri, and R.~Trotta,
  ``{Challenges of Profile Likelihood Evaluation in Multi- Dimensional SUSY
  Scans},'' \href{http://dx.doi.org/10.1007/JHEP06(2011)042}{{\em JHEP}
  {\bfseries 06} (2011) 042},
\href{http://arxiv.org/abs/1101.3296}{{\ttfamily arXiv:1101.3296 [hep-ph]}}.

\bibitem{PDG:2010}
K.~N. et~al. (Particle Data~Group), ``Review of particle physics,'' {\em
  Journal of Physics G: Nuclear and Particle Physics} {\bfseries 37} no.~7A,
  (2010, and 2011 partial update for the 2012 edition.) 075021.
  \url{http://stacks.iop.org/0954-3899/37/i=7A/a=075021}.

\bibitem{jeffreys1961theory}
H.~Jeffreys, ``Theory of probability,'' 1961.

\bibitem{barbieri1988upper}
R.~Barbieri and G.~F. Giudice, ``Upper bounds on supersymmetric particle
  masses,'' {\em Nuclear Physics B} {\bfseries 306} no.~1, (1988) 63--76.

\bibitem{Roszkowski:2009ye}
L.~Roszkowski, R.~Ruiz~de Austri, and R.~Trotta, ``{Efficient reconstruction of
  CMSSM parameters from LHC data - A case study},''
  \href{http://dx.doi.org/10.1103/PhysRevD.82.055003}{{\em Phys. Rev.}
  {\bfseries D82} (2010) 055003},
\href{http://arxiv.org/abs/0907.0594}{{\ttfamily arXiv:0907.0594 [hep-ph]}}.

\bibitem{Ghilencea:2012gz}
D.~M. Ghilencea, H.~M. Lee, and M.~Park, ``{Tuning supersymmetric models at the
  LHC: A comparative analysis at two-loop level},''
  \href{http://arxiv.org/abs/1203.0569}{{\ttfamily arXiv:1203.0569 [hep-ph]}}.
23 pages, 46 figures.

\bibitem{Ghilencea:2012qk}
D.~Ghilencea and G.~Ross, ``{The fine-tuning cost of the likelihood in SUSY
  models},''
\href{http://arxiv.org/abs/1208.0837}{{\ttfamily arXiv:1208.0837 [hep-ph]}}.

\bibitem{Komatsu:2010fb}
{\bfseries WMAP} Collaboration, E.~Komatsu {\em et~al.}, ``{Seven-Year
  Wilkinson Microwave Anisotropy Probe (WMAP) Observations: Cosmological
  Interpretation},'' \href{http://dx.doi.org/10.1088/0067-0049/192/2/18}{{\em
  Astrophys.J.Suppl.} {\bfseries 192} (2011) 18},
  \href{http://arxiv.org/abs/1001.4538}{{\ttfamily arXiv:1001.4538
  [astro-ph.CO]}}.

\bibitem{Benayoun:2011mm}
M.~Benayoun, P.~David, L.~DelBuono, and F.~Jegerlehner, ``{Upgraded Breaking Of
  The HLS Model: A Full Solution to the $\tau^-e^+e^-$ and $\phi$ Decay Issues
  And Its Consequences On g-2 VMD Estimates},''
  \href{http://dx.doi.org/10.1140/epjc/s10052-011-1848-2}{{\em Eur.Phys.J.}
  {\bfseries C72} (2012) 1848},
\href{http://arxiv.org/abs/1106.1315}{{\ttfamily arXiv:1106.1315 [hep-ph]}}.

\bibitem{HFAG:2010}
{\bfseries Heavy Flavor Averaging Group} Collaboration, D.~Asner {\em et~al.},
  ``{Averages of b-hadron, c-hadron, and tau-lepton Properties},''
\href{http://arxiv.org/abs/1010.1589}{{\ttfamily arXiv:1010.1589 [hep-ex]}}.

\bibitem{BABAR:2008cy}
{\bfseries BABAR} Collaboration, B.~Aubert {\em et~al.}, ``{Measurement of
  Branching Fractions and CP and Isospin Asymmetries in $B \to K^* \gamma$},''
\href{http://arxiv.org/abs/0808.1915}{{\ttfamily arXiv:0808.1915 [hep-ex]}}.

\bibitem{Aubert:2007dsa}
{\bfseries BABAR} Collaboration, B.~Aubert {\em et~al.}, ``{Observation of the
  semileptonic decays $B \to D^* \tau^- \bar{\nu}_\tau$ and evidence for $B \to
  D \tau^- \bar{\nu}_\tau$},''
  \href{http://dx.doi.org/10.1103/PhysRevLett.100.021801}{{\em Phys. Rev.
  Lett.} {\bfseries 100} (2008) 021801},
\href{http://arxiv.org/abs/0709.1698}{{\ttfamily arXiv:0709.1698 [hep-ex]}}.

\bibitem{Antonelli:2008jg}
{\bfseries FlaviaNet Working Group on Kaon Decays} Collaboration, M.~Antonelli
  {\em et~al.}, ``{Precision tests of the Standard Model with leptonic and
  semileptonic kaon decays},''
\href{http://arxiv.org/abs/0801.1817}{{\ttfamily arXiv:0801.1817 [hep-ph]}}.

\bibitem{PDG:2006}
W.-M. e. a. P. D.~G. {Yao}, ``{Review of Particle Physics},'' {\em {Journal of
  Physics G}} {\bfseries 33} (2006) . \url{http://pdg.lbl.gov}.

\bibitem{Barger:2006dh}
V.~Barger, P.~Langacker, H.-S. Lee, and G.~Shaughnessy, ``{Higgs Sector in
  Extensions of the MSSM},''
  \href{http://dx.doi.org/10.1103/PhysRevD.73.115010}{{\em Phys. Rev.}
  {\bfseries D73} (2006) 115010},
\href{http://arxiv.org/abs/hep-ph/0603247}{{\ttfamily arXiv:hep-ph/0603247}}.

\bibitem{Aprile:2011hi}
{\bfseries XENON100} Collaboration, E.~Aprile {\em et~al.}, ``{Dark Matter
  Results from 100 Live Days of XENON100 Data},'' {\em Phys.Rev.Lett.}
  {\bfseries 107} (2011) 131302,
  \href{http://arxiv.org/abs/1104.2549}{{\ttfamily arXiv:1104.2549
  [astro-ph.CO]}}.

\bibitem{Bettler:2011rp}
M.-O. Bettler, ``{Search for $B_{s,d}\to\mu\mu$ at LHCb with 300
  $\text{pb}^{-1}$},''
\href{http://arxiv.org/abs/1110.2411}{{\ttfamily arXiv:1110.2411 [hep-ex]}}.

\bibitem{Aad:2011ib}
{\bfseries ATLAS} Collaboration, G.~Aad {\em et~al.}, ``{Search for squarks and
  gluinos using final states with jets and missing transverse momentum with the
  ATLAS detector in $\sqrt{s} = 7$ TeV proton-proton collisions},''
\href{http://arxiv.org/abs/1109.6572}{{\ttfamily arXiv:1109.6572 [hep-ex]}}.

\bibitem{Buchmueller:2011aa}
O.~Buchmueller, R.~Cavanaugh, D.~Colling, A.~de~Roeck, M.~Dolan, {\em et~al.},
  ``{Implications of Initial LHC Searches for Supersymmetry},''
  \href{http://dx.doi.org/10.1140/epjc/s10052-011-1634-1}{{\em Eur.Phys.J.}
  {\bfseries C71} (2011) 1634},
\href{http://arxiv.org/abs/1102.4585}{{\ttfamily arXiv:1102.4585 [hep-ph]}}.

\bibitem{Aprile:2011hx}
{\bfseries XENON100} Collaboration, E.~Aprile {\em et~al.}, ``{Likelihood
  Approach to the First Dark Matter Results from XENON100},''
  \href{http://dx.doi.org/10.1103/PhysRevD.84.052003}{{\em Phys. Rev.}
  {\bfseries D84} (2011) 052003},
\href{http://arxiv.org/abs/1103.0303}{{\ttfamily arXiv:1103.0303 [hep-ex]}}.

\bibitem{cowan2011asymptotic}
G.~Cowan, K.~Cranmer, E.~Gross, and O.~Vitells, ``Asymptotic formulae for
  likelihood-based tests of new physics,'' {\em The European Physical Journal
  C-Particles and Fields} {\bfseries 71} no.~2, (2011) 1--19,
  \href{http://arxiv.org/abs/1007.1727}{{\ttfamily arXiv:1007.1727 [data-an]}}.

\bibitem{Read:CLs}
A.~L. Read, ``{Presentation of search results: the $CL_s$ technique},'' {\em
  Journal of Physics G: Nuclear and Particle Physics} {\bfseries 28} no.~10,
  (2002) 2693. \url{http://stacks.iop.org/0954-3899/28/i=10/a=313}.

\bibitem{Alarcon:2011zs}
J.~M. Alarcon, J.~Martin~Camalich, and J.~A. Oller, ``{The chiral
  representation of the $\pi N$ scattering amplitude and the pion-nucleon sigma
  term},''
\href{http://arxiv.org/abs/1110.3797}{{\ttfamily arXiv:1110.3797 [hep-ph]}}.

\bibitem{Pavan:2001wz}
M.~M. Pavan, I.~I. Strakovsky, R.~L. Workman, and R.~A. Arndt, ``{The pion
  nucleon Sigma term is definitely large: Results from a GWU analysis of pi N
  scattering data},'' {\em PiN Newslett.} {\bfseries 16} (2002) 110--115,
\href{http://arxiv.org/abs/hep-ph/0111066}{{\ttfamily arXiv:hep-ph/0111066}}.

\bibitem{Gasser1991252}
J.~Gasser, H.~Leutwyler, and M.~Sainio, ``Sigma-term update,''
  \href{http://dx.doi.org/10.1016/0370-2693(91)91393-A}{{\em Physics Letters B}
  {\bfseries 253} no.~1-2, (1991) 252 -- 259}.
  \url{http://www.sciencedirect.com/science/article/pii/037026939191393A}.

\bibitem{Koch:1982pu}
R.~Koch, ``{A New Determination of the pi N Sigma Term Using Hyperbolic
  Dispersion Relations in the (nu**2, t) Plane},''
\href{http://dx.doi.org/10.1007/BF01571999}{{\em Z. Phys.} {\bfseries C15}
  (1982) 161--168}.

\bibitem{Giedt:2009mr}
J.~Giedt, A.~W. Thomas, and R.~D. Young, ``{Dark matter, the CMSSM and lattice
  QCD},'' \href{http://dx.doi.org/10.1103/PhysRevLett.103.201802}{{\em Phys.
  Rev. Lett.} {\bfseries 103} (2009) 201802},
\href{http://arxiv.org/abs/0907.4177}{{\ttfamily arXiv:0907.4177 [hep-ph]}}.

\bibitem{Young:2009zb}
R.~D. Young and A.~W. Thomas, ``{Octet baryon masses and sigma terms from an
  SU(3) chiral extrapolation},''
  \href{http://dx.doi.org/10.1103/PhysRevD.81.014503}{{\em Phys. Rev.}
  {\bfseries D81} (2010) 014503},
\href{http://arxiv.org/abs/0901.3310}{{\ttfamily arXiv:0901.3310 [hep-lat]}}.

\bibitem{Gasser:1982ap}
J.~Gasser and H.~Leutwyler, ``{Quark Masses},''
\href{http://dx.doi.org/10.1016/0370-1573(82)90035-7}{{\em Phys. Rept.}
  {\bfseries 87} (1982) 77--169}.

\bibitem{Borasoy:1996bx}
B.~Borasoy and U.-G. Meissner, ``{Chiral expansion of baryon masses and
  sigma-terms},'' \href{http://dx.doi.org/10.1006/aphy.1996.5630}{{\em Annals
  Phys.} {\bfseries 254} (1997) 192--232},
\href{http://arxiv.org/abs/hep-ph/9607432}{{\ttfamily arXiv:hep-ph/9607432}}.

\bibitem{Sainio:2001bq}
M.~E. Sainio, ``{Pion nucleon sigma-term: A review},'' {\em PiN Newslett.}
  {\bfseries 16} (2002) 138--143,
\href{http://arxiv.org/abs/hep-ph/0110413}{{\ttfamily arXiv:hep-ph/0110413}}.

\bibitem{Knecht:1999dp}
M.~Knecht, ``{Working group summary: pi N sigma term},'' {\em PiN Newslett.}
  {\bfseries 15} (1999) 108--113,
\href{http://arxiv.org/abs/hep-ph/9912443}{{\ttfamily arXiv:hep-ph/9912443}}.

\bibitem{Ellis:2008hf}
J.~R. Ellis, K.~A. Olive, and C.~Savage, ``{Hadronic Uncertainties in the
  Elastic Scattering of Supersymmetric Dark Matter},''
  \href{http://dx.doi.org/10.1103/PhysRevD.77.065026}{{\em Phys. Rev.}
  {\bfseries D77} (2008) 065026},
\href{http://arxiv.org/abs/0801.3656}{{\ttfamily arXiv:0801.3656 [hep-ph]}}.

\bibitem{Aad:2008zzm}
{\bfseries ATLAS} Collaboration, G.~Aad {\em et~al.}, ``{The ATLAS Experiment
  at the CERN Large Hadron Collider},''
\href{http://dx.doi.org/10.1088/1748-0221/3/08/S08003}{{\em JINST} {\bfseries
  3} (2008) S08003}.

\bibitem{cms:2008zzk}
{\bfseries CMS} Collaboration, R.~Adolphi {\em et~al.}, ``{The CMS experiment
  at the CERN LHC},''
\href{http://dx.doi.org/10.1088/1748-0221/3/08/S08004}{{\em JINST} {\bfseries
  3} (2008) S08004}.

\bibitem{Aad:2011zj}
{\bfseries ATLAS} Collaboration, G.~Aad {\em et~al.}, ``{Search for Diphoton
  Events with Large Missing Transverse Momentum in 1 fb$^{-1}$ of 7 TeV
  Proton-Proton Collision Data with the ATLAS Detector},''
\href{http://arxiv.org/abs/1111.4116}{{\ttfamily arXiv:1111.4116 [hep-ex]}}.

\bibitem{Aad:2011cw}
{\bfseries ATLAS} Collaboration, G.~Aad {\em et~al.}, ``{Searches for
  supersymmetry with the ATLAS detector using final states with two leptons and
  missing transverse momentum in $\sqrt{s} = 7$ TeV proton-proton
  collisions},''
\href{http://arxiv.org/abs/1110.6189}{{\ttfamily arXiv:1110.6189 [hep-ex]}}.

\bibitem{Aad:2011qa}
{\bfseries ATLAS} Collaboration, G.~Aad {\em et~al.}, ``{Search for new
  phenomena in final states with large jet multiplicities and missing
  transverse momentum using $\sqrt{s}=7$ TeV pp collisions with the ATLAS
  detector},'' \href{http://dx.doi.org/10.1007/JHEP11(2011)099}{{\em JHEP}
  {\bfseries 1111} (2011) 099},
\href{http://arxiv.org/abs/1110.2299}{{\ttfamily arXiv:1110.2299 [hep-ex]}}.

\bibitem{ATLAS:2011ad}
{\bfseries ATLAS} Collaboration, G.~Aad {\em et~al.}, ``{Search for
  supersymmetry in final states with jets, missing transverse momentum and one
  isolated lepton in $\sqrt{s} = 7$ TeV pp collisions using 1 $\text{fb}^{-1}$
  of ATLAS data},''
\href{http://arxiv.org/abs/1109.6606}{{\ttfamily arXiv:1109.6606 [hep-ex]}}.

\bibitem{Chatrchyan:1381201}
{\bfseries CMS} Collaboration, S.~Chatrchyan {\em et~al.}, ``{Search for
  Supersymmetry at the LHC in Events with Jets and Missing Transverse
  Energy},''. \url{http://cdsweb.cern.ch/record/1381201}.

\bibitem{CMS-PAS-SUS-11-005}
{\bfseries CMS} Collaboration, S.~Chatrchyan {\em et~al.}, ``{Search for
  supersymmetry in all-hadronic events with MT2},''.
  \url{http://cdsweb.cern.ch/record/1377032}.

\bibitem{CMS-PAS-SUS-11-004}
{\bfseries CMS} Collaboration, S.~Chatrchyan {\em et~al.}, ``{Search for
  supersymmetry in all-hadronic events with missing energy},''.
  \url{http://cdsweb.cern.ch/record/1378478}.

\bibitem{Desai:2011th}
N.~Desai and B.~Mukhopadhyaya, ``{Constraints on supersymmetry with light third
  family from LHC data},''
\href{http://arxiv.org/abs/1111.2830}{{\ttfamily arXiv:1111.2830 [hep-ph]}}.

\bibitem{Beskidt:2011qf}
C.~Beskidt, W.~de~Boer, D.~Kazakov, F.~Ratnikov, E.~Ziebarth, {\em et~al.},
  ``{Constraints from the decay $B_s^0 \to \mu^+ \mu^-$ and LHC limits on
  Supersymmetry},''
  \href{http://dx.doi.org/10.1016/j.physletb.2011.10.053}{{\em Phys.Lett.}
  {\bfseries B705} (2011) 493--497},
\href{http://arxiv.org/abs/1109.6775}{{\ttfamily arXiv:1109.6775 [hep-ex]}}.

\bibitem{Allanach:2011qr}
B.~Allanach, T.~Khoo, and K.~Sakurai, ``{Interpreting a 1 $\text{fb}^{-1}$
  ATLAS Search in the Minimal Anomaly Mediated Supersymmetry Breaking Model},''
\href{http://arxiv.org/abs/1110.1119}{{\ttfamily arXiv:1110.1119 [hep-ph]}}.

\bibitem{Gieseke:2011na}
S.~Gieseke, D.~Grellscheid, K.~Hamilton, A.~Papaefstathiou, S.~Platzer, {\em
  et~al.}, ``{Herwig++ 2.5 Release Note},''
\href{http://arxiv.org/abs/1102.1672}{{\ttfamily arXiv:1102.1672 [hep-ph]}}.

\bibitem{Ovyn:2009tx}
S.~Ovyn, X.~Rouby, and V.~Lemaitre, ``{DELPHES, a framework for fast simulation
  of a generic collider experiment},''
\href{http://arxiv.org/abs/0903.2225}{{\ttfamily arXiv:0903.2225 [hep-ph]}}.

\bibitem{Beenakker:1996ed}
W.~Beenakker, R.~Hopker, and M.~Spira, ``{PROSPINO: A Program for the
  production of supersymmetric particles in next-to-leading order QCD},''
\href{http://arxiv.org/abs/hep-ph/9611232}{{\ttfamily arXiv:hep-ph/9611232
  [hep-ph]}}.

\bibitem{Agostinelli:2002hh}
{\bfseries GEANT4} Collaboration, S.~Agostinelli {\em et~al.}, ``{GEANT4: A
  simulation toolkit},''
\href{http://dx.doi.org/10.1016/S0168-9002(03)01368-8}{{\em Nucl. Instrum.
  Meth.} {\bfseries A506} (2003) 250--303}.

\bibitem{Buckley:2011kc}
A.~Buckley, A.~Shilton, and M.~J. White, ``{Fast supersymmetry phenomenology at
  the Large Hadron Collider using machine learning techniques},''
\href{http://arxiv.org/abs/1106.4613}{{\ttfamily arXiv:1106.4613 [hep-ph]}}.

\bibitem{Hocker:2007ht}
A.~Hoecker, P.~Speckmayer, J.~Stelzer, J.~Therhaag, E.~von Toerne, and H.~Voss,
  ``{TMVA: Toolkit for Multivariate Data Analysis},'' {\em PoS} {\bfseries
  ACAT} (2007) 040,
\href{http://arxiv.org/abs/physics/0703039}{{\ttfamily arXiv:physics/0703039}}.

\bibitem{Zhong:2011xm}
J.-H. Zhong, R.-S. Huang, S.-C. Lee, R.-S. Huang, and S.-C. Lee, ``{A program
  for the Bayesian Neural Network in the ROOT framework},''
  \href{http://dx.doi.org/10.1016/j.cpc.2011.07.019}{{\em Comput.Phys.Commun.}
  {\bfseries 182} (2011) 2655--2660},
\href{http://arxiv.org/abs/1103.2854}{{\ttfamily arXiv:1103.2854
  [physics.data-an]}}.

\bibitem{ATLAS-CONF-2012-033}
{\bfseries ATLAS} Collaboration, G.~Aad {\em et~al.}, ``Search for squarks and
  gluinos using final states with jets and missing transverse momentum with the
  atlas detector in √s = 7 tev proton-proton collisions,'' tech. rep., CERN,
  Geneva, Mar, 2012.
\newblock \url{http://cdsweb.cern.ch/record/1432199}.

\bibitem{CMS-PAS-SUS-12-005}
{\bfseries CMS} Collaboration, S.~Chatrchyan {\em et~al.}, ``Search for
  supersymmetry with the razor variables at cms,''.
  \url{http://cdsweb.cern.ch/record/1430715}.

\bibitem{ATLAS-CONF-2011-163}
{\bfseries ATLAS} Collaboration, G.~Aad {\em et~al.}, ``{Combination of Higgs
  Boson Searches with up to 4.9 $\text{fb}^{-1}$ of pp Collisions Data Taken at
  a center-of-mass energy of 7 TeV with the ATLAS Experiment at the LHC},''
  Tech. Rep. ATLAS-CONF-2011-163, CERN, Geneva, Dec, 2011.
\newblock \url{http://cdsweb.cern.ch/record/1406358}.

\bibitem{CMS-PAS-HIG-11-032}
{\bfseries CMS} Collaboration, S.~Chatrchyan {\em et~al.}, ``Combination of sm
  higgs searches,''. \url{http://cdsweb.cern.ch/record/1406347}.

\bibitem{Chatrchyan:2012tx}
{\bfseries CMS} Collaboration, S.~Chatrchyan {\em et~al.}, ``{Combined results
  of searches for the standard model Higgs boson in pp collisions at $\sqrt{s}
  = 7$ TeV},''
\href{http://arxiv.org/abs/1202.1488}{{\ttfamily arXiv:1202.1488 [hep-ex]}}.

\bibitem{Akula:2011aa}
S.~Akula, B.~Altunkaynak, D.~Feldman, P.~Nath, and G.~Peim, ``{Higgs Boson Mass
  Predictions in SUGRA Unification, Recent LHC-7 Results, and Dark Matter},''
  {\em Phys.Rev.} {\bfseries D85} (2012) 075001,
\href{http://arxiv.org/abs/1112.3645}{{\ttfamily arXiv:1112.3645 [hep-ph]}}.

\bibitem{Kadastik:2011aa}
M.~Kadastik, K.~Kannike, A.~Racioppi, and M.~Raidal, ``{Implications of the 125
  GeV Higgs boson for scalar dark matter and for the CMSSM phenomenology},''
\href{http://arxiv.org/abs/1112.3647}{{\ttfamily arXiv:1112.3647 [hep-ph]}}.

\bibitem{Baer:2012uy}
H.~Baer, V.~Barger, and A.~Mustafayev, ``{Neutralino dark matter in
  mSUGRA/CMSSM with a 125 GeV light Higgs scalar},''
\href{http://arxiv.org/abs/1202.4038}{{\ttfamily arXiv:1202.4038 [hep-ph]}}.

\bibitem{Azatov:2012bz}
A.~Azatov, R.~Contino, and J.~Galloway, ``{Model-Independent Bounds on a Light
  Higgs},''
\href{http://arxiv.org/abs/1202.3415}{{\ttfamily arXiv:1202.3415 [hep-ph]}}.

\bibitem{Hoecker:2010qn}
A.~Hoecker, ``{The Hadronic Contribution to the Muon Anomalous Magnetic Moment
  and to the Running Electromagnetic Fine Structure Constant at MZ - Overview
  and Latest Results},''
  \href{http://dx.doi.org/10.1016/j.nuclphysbps.2011.06.031}{{\em
  Nucl.Phys.Proc.Suppl.} {\bfseries 218} (2011) 189--200},
\href{http://arxiv.org/abs/1012.0055}{{\ttfamily arXiv:1012.0055 [hep-ph]}}.

\bibitem{Goecke:2010if}
T.~Goecke, C.~S. Fischer, and R.~Williams, ``{Hadronic light-by-light
  scattering in the muon g-2: a Dyson-Schwinger equation approach},''
  \href{http://dx.doi.org/10.1103/PhysRevD.83.094006}{{\em Phys.Rev.}
  {\bfseries D83} (2011) 094006},
\href{http://arxiv.org/abs/1012.3886}{{\ttfamily arXiv:1012.3886 [hep-ph]}}.

\bibitem{Hagiwara:2011af}
K.~Hagiwara, R.~Liao, A.~D. Martin, D.~Nomura, and T.~Teubner, ``{$(g-2)_\mu$
  and alpha($M_Z^2$) re-evaluated using new precise data},''
  \href{http://dx.doi.org/10.1088/0954-3899/38/8/085003}{{\em J.Phys.G}
  {\bfseries G38} (2011) 085003},
\href{http://arxiv.org/abs/1105.3149}{{\ttfamily arXiv:1105.3149 [hep-ph]}}.

\bibitem{Bodenstein:2011qy}
S.~Bodenstein, C.~Dominguez, and K.~Schilcher, ``{Hadronic contribution to the
  muon g-2: A Theoretical determination},''
  \href{http://dx.doi.org/10.1103/PhysRevD.85.014029}{{\em Phys.Rev.}
  {\bfseries D85} (2012) 014029},
\href{http://arxiv.org/abs/1106.0427}{{\ttfamily arXiv:1106.0427 [hep-ph]}}.

\bibitem{Goecke:2011bm}
T.~Goecke, C.~S. Fischer, and R.~Williams, ``{Hadronic contribution to the muon
  g-2: a Dyson-Schwinger perspective},''
\href{http://arxiv.org/abs/1111.0990}{{\ttfamily arXiv:1111.0990 [hep-ph]}}.

\bibitem{:2012gk}
{\bfseries ATLAS Collaboration} Collaboration, G.~Aad {\em et~al.},
  ``{Observation of a new particle in the search for the Standard Model Higgs
  boson with the ATLAS detector at the LHC},''
  \href{http://dx.doi.org/10.1016/j.physletb.2012.08.020}{{\em Phys.Lett.B}
  (2012) },
\href{http://arxiv.org/abs/1207.7214}{{\ttfamily arXiv:1207.7214 [hep-ex]}}.

\bibitem{:2012gu}
{\bfseries CMS Collaboration} Collaboration, S.~Chatrchyan {\em et~al.},
  ``{Observation of a new boson at a mass of 125 GeV with the CMS experiment at
  the LHC},'' \href{http://dx.doi.org/10.1016/j.physletb.2012.08.021}{{\em
  Phys.Lett.B} (2012) },
\href{http://arxiv.org/abs/1207.7235}{{\ttfamily arXiv:1207.7235 [hep-ex]}}.

\end{thebibliography}\endgroup


\appendix


\section{Fast approximation to combined CLs limits for correlated likelihoods}
\label{app:approxCLs}
In this appendix we offer a brief justification of the simplified method used to combine the ATLAS $CL_s$ limits on sparticle production in our analysis. In this method an approximate combined confidence limit is obtained for a specified model point by simply taking the most powerful observed (lowest $CL_s$ value) limit from one of several signal regions, or search channels. Our aim is to demonstrate a set of minimal conditions under which this procedure is conservative. This will be done by demonstrating conditions under which the following inequality holds:
\begin{equation}
\label{eq:goal}
\min\left(CL_{s_1},CL_{s_2}\right) \ge CL_{s_{1,2}}
\end{equation}
where $CL_{s_1}$ is the value of the $CL_s$ statistic for some signal model, derived from a dataset which we may call `channel 1'; $CL_{s_2}$ is the value of $CL_s$ under the same signal model but derived from a correlated dataset `channel 2'; and $CL_{s_{1,2}}$ is the value of $CL_s$ for this signal model derived from the full combination of the two datasets, accounting rigorously for correlations between datasets. This inequality does not hold in general, but if the experimental situation is such that it does hold, it means that the combined dataset results in a more powerful limit than either of the individual datasets alone, or conversely that considering only the most constraining of the two individual dataset limits is conservative. In the course of this exercise we will make use of the asymptotic results obtained in ref. \cite{cowan2011asymptotic}.

We remind the reader that the $CL_s$ statistic is defined as
\begin{equation}
\label{eq:A0}
CL_{s} = \frac{p_{s+b}}{1-p_b}
\end{equation}
where $p_{s+b}$ and $p_b$ are p-values derived using the null hypotheses `$s+b$' and `$b$' respectively. $s+b$ is the hypothesis that the data is generated from the nominal signal plus background model, while $b$ supposes that the data contains background events only. In the $CL_s$ method these p-values are computed using the likelihood ratio statistic 
\begin{equation}
\label{eq:A1}
q = -2\ln\frac{L_{s+b}}{L_b}=-2\ln\frac{L(\mu=1,\hat{\theta}(1))}{L(\mu=0,\hat{\theta}(0))},
\end{equation}
where $L_{s+b}$ and $L_{b}$ are the likelihoods of the `$s+b$' and `$b$' models respectively. The second equality defines the background model as one which can be obtained by scaling the signal model by an appropriate `signal strength' parameter $\mu$, which is set to zero. $\hat{\theta}(1)$ and $\hat{\theta}(0)$ are the profiled values of any nuisance parameters. In the asymptotic limit (which requires sufficiently many candidate events) this statistic is given by the Wald approximation, with $\mu$ as the parameter of interest, as
\begin{equation}
\label{eq:A2}
q = \frac{(\hat{\mu}-1)^2}{\sigma^2} - \frac{\hat{\mu}^2}{\sigma^2} = \frac{1-2\hat{\mu}}{\sigma^2},
\end{equation}
where $\hat\mu$ is the best fit value of $\mu$ given some dataset, and $\sigma^2$ is the variance of $\hat\mu$ (which is normally distributed) under either the `$s+b$' or the `$b$' models, that is $\sigma$ takes the values $\sigma_{s+b}$ and $\sigma_{b}$ when the $\mu=1$ and $\mu=0$ models are assumed to be generating the data respectively. Using so-called `Asimov' data sets, which when observed cause $\hat\mu$ to adopt its true value (either 1 or 0; see ref. \cite{cowan2011asymptotic}) we can obtain $\sigma^2$ as
\begin{align}
\label{eq:A3}
\sigma^2 = \frac{1-2\mu'}{q_A},\hspace{0.4cm}\text{so that}&~ \
  & \sigma^2_{s+b} &= \frac{1}{\left|q_{A_{s+b}}\right|} \\
  \text{and}& ~ & \sigma^2_{b} &= \frac{1}{\left|q_{A_b}\right|}
\end{align}
where $\mu'$ is the assumed true value of $\mu$ and $q_A$ is the value of $q$ obtained using the relevant Asimov dataset.

The asymptotic distribution $f$ of the statistic $q$ is normal with mean $(1-2\mu')/\sigma^2$ and variance $4/\sigma^2$ so the p-values in eq. (\ref{eq:A0}) can be computed by
\begin{align}
\label{eq:A5}
p_{s+b} = \int_{q_\text{obs}}^{\infty} f(q|s+b) dq &= \
    1 - \Phi\left(\frac{q_\text{obs} + 1/\sigma^2_{s+b}}{2/\sigma_{s+b}}\right) \nonumber \\
  &= 1 - \Phi\left(\frac{q_\text{obs} - q_{A_{s+b}}}{2\sqrt{|q_{A_{s+b}}|}}\right)
\end{align}
and
\begin{align}
p_{b} = \int_{-\infty}^{q_\text{obs}} f(q|b) dq &= \
    \Phi\left(\frac{q_\text{obs} -1/\sigma^2_{b}}{2/\sigma_{b}}\right) \nonumber \\
  &= \Phi\left(\frac{q_\text{obs} - q_{A_b}}{2\sqrt{|q_{A_b}|}}\right)
\end{align}
Let us now go to the case where the observed events in all channels are in accordance with
the background hypothesis, such that $\hat{\mu}\sim0$. Then $q_\text{obs}\sim q_{A_b}$. Furthermore, in this case the $95\%$ $CL_s$ limit lies near model points which predict low signals, so we may further take $\sigma\sim \sigma_{s+b}\sim\sigma_b$ (since the distribution $f$ under both $s+b$ and $b$ hypotheses will be very similar). Also note that in this limit $q_{A_{s+b}}=-q_{A_b}$. Our p-values can thus be simplified to 
\begin{equation}
\label{eq:pvals}
p_{s+b} =  1 - \Phi\left(\sqrt{|q_A|}\right) \hspace{0.3cm}\text{and}\hspace{0.3cm}
p_{b}   =  \Phi\left(0\right) = \frac{1}{2}
\end{equation}
(where we have also used the knowledge that $\text{sign}(q_{A_{s+b}}) = -1$). We can thus write the inequality of eq. (\ref{eq:goal}) as:
\begin{align}
\frac{p_{s+b;1}}{1-p_{b;1}} &\ge \frac{p_{s+b;1,2}}{1-p_{b;1,2}} \nonumber \\
\rightarrow\hspace{0.3cm}
1 - \Phi\left(\sqrt{|q_{1_{A}}|}\right) \
&\ge \
1 - \Phi\left(\sqrt{|q_{1,2_{A}}|}\right)
\end{align}
where we have assumed WLOG that $CL_{s_1}\le CL_{s_2}$. The function $\Phi(x)$ is monotonically increasing with $x$, so our inequality will hold if
\begin{equation}
\label{eq:simpineq}
\left|q_{1_{A}}\right| \le \left|q_{1,2_{A}}\right|
\end{equation}
To determine when this is the case, we need to express $q_{1,2_{A}}$ in terms of the parameters describing $q_{1_{A}}$ and $q_{2_{A}}$. We can do this by obtaining the two parameter Wald expansion for the combined test statistic $q_{1,2}$ (i.e. taking a Taylor expansion of q about the best fit values of $\mu_1$ and $\mu_2$, up to second order):
\multlinegap=20pt
\begin{multline}
\label{eq:qcomb}
q_{1,2}= \frac{1}{1-\rho}\bigg(\frac{1-2\hat{\mu}_1}{\sigma_1^2} + \frac{1-2\hat{\mu}_2}{\sigma_2^2} \\
 -2\rho\frac{1-\hat{\mu}_1-\hat{\mu}_2}{\sigma_1\sigma_2} \bigg),
\end{multline}
\multlinegap=0pt
where $\rho$ characterises linear correlations between the two channels, taking values in the domain $(-1,1)$, and $\hat{\mu}_1,\hat{\mu}_2$ and $\sigma_1^2,\sigma_2^2$ are the best fit $\mu$ values and their variances, as obtained above for each individual channel. Again we use the Asimov dataset for the background hypothesis, which sets $\hat{\mu}_1=\hat{\mu}_2=0$, to find $q_{1,2_{A,b}}$:
\begin{equation}
\label{eq:qcombA}
q_{1,2_{A,b}} = \frac{1}{1-\rho}\left(\frac{1}{\sigma_1^2} + \frac{1}{\sigma_2^2} -2\rho\frac{1}{\sigma_1\sigma_2} \right),
\end{equation}
which, like $q_{1_{A,b}}$, is strictly positive. Using this expression together with eq. (\ref{eq:A3}) we can rewrite the inequality of eq. (\ref{eq:simpineq}) as
\begin{equation}
\label{eq:condition}
\frac{1}{\sigma^2_1} \le \frac{1}{1-\rho}\left(\frac{1}{\sigma_1^2} + \frac{1}{\sigma_2^2} -2\rho\frac{1}{\sigma_1\sigma_2} \right)
\end{equation}
One can readily see that eq. (\ref{eq:condition}) holds in the case $\rho=0$, i.e. when no correlations exist between channels. Knowing this, we may vary $\rho$ from this point and see where the equality is achieved in order to check if the inequality may be violated. Setting the equality we solve for $\sigma_1$, finding the two general solutions
\begin{equation}
\label{eq:solutions}
\sigma_1=\sigma_2\left(\rho\pm\sqrt{(\rho-1)\rho}\right),
\end{equation}
from which it is apparent that no real solutions exist for $0<\rho<1$, while such solutions do exist for $-1<\rho<0$. We could convert this to a bound on the allowed values of $\sigma_1/\sigma_2$, since only the positive root solution can give a positive $\sigma_1$, but negative correlations are not relevant for our signal regions, which are correlated due to shared events, so we are done.

We can thus conclude that if channel correlations are linear and positive, the observed event counts are not far from the expected background, the nominal signal hypothesis at the limit is small, and enough events are observed for asymptotic formulae to hold, then we can safely take the most powerful limit from among several channels as an estimate of the full combination, without overestimating the combined limit. Violations of these conditions may result in the target inequality of eq. (\ref{eq:goal}) being violated, with a particular concern being that this can occur as the observed events differ from the background expectation; however, it is difficult to determine the general conditions under which this happens. Certainly if one channel sees an excess above the background while another does not then in general the combined limit will be weaker than one obtained using only the more constraining (background-like) channel.

Nevertheless, in our special case we may be confident that our method remains approximately valid thanks to the procedure used by ATLAS to produce their official limits  (in ref. \cite{Aad:2011ib}), to which our approximate limits are fitted. ATLAS also do not attempt to rigorously account for the correlations between channels; they follow a similar procedure to us and, for each point in the CMSSM parameter space, take the limit from the channel with the best \emph{expected} limit. We, on the other hand, take the channel with the best \emph{observed} limit, which, following the discussion of this appendix, can be expected to less reliably approximate the rigorous combination.

We follow our more approximate procedure because ATLAS do not provide the expected limits on the mean signal for each channel; however, it is possible to estimate these using the asympotic formulae discussed in this appendix, and so we use these estimates to gauge the seriousness of the difference between our method and the one used by ATLAS.

To do this a model of the likelihood in each signal region is needed. Taking the random variable to be the best fit signal strength $\hat{\mu}$, the simplest option is the normal limit of a Poisson likelihood, with standard deviation $\sigma$  modified by convolution with normal signal and background systematics $\sigma_s$ and $\sigma_b$. The mean and variance are then simply
\begin{align}
\label{eq:pvals}
\mu &= \frac{n-b}{s} \\
\sigma^2 &= (\mu'(s + \sigma_s^2) + \sigma_b^2)/s^2 
\end{align}
where $n=\mu's + b$ is the expected total number of events (and $\mu'=1$ or $0$ as before). ATLAS provide estimates of $\sigma_b$ so we use these, however $\sigma_s$ is not provided since it varies point to point. This variation would require a large effort to model so we simply fit a single value for $\sigma_s$ for each channel, ensuring that the observed 95\% $CL_s$ limits obtained from our simplified likelihood agree with ATLAS (we have also checked that varying this value has little effect on our results).

\begin{figure}
\begin{center}
\includegraphics[width=\columnwidth]{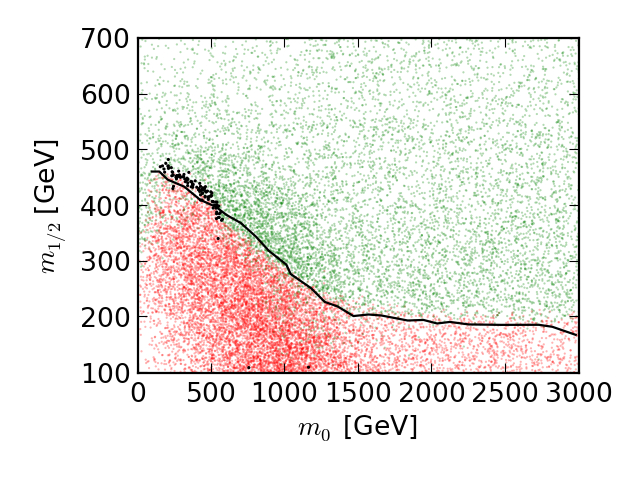}
\end{center}
\caption{Classification of training data for the ATLAS 1 fb$^{-1}$ jets+MET search used in the main analysis. Two methods for combining the ATLAS limits for each search channel are used: the method used in this analysis uses the most constraining observed $CL_s$ value from the set of channels at each training data point to determine its classification, while ATLAS use the observed $CL_s$ value from the signal region with the most powerful expected exclusion. We have estimated the limit that would be obtained from the ATLAS method using asymptotic approximations for the signal likelihood. Training data model points which are excluded at $95\%$ $CL_s$ by both limits are coloured red, while model points not excluded by either are coloured green. Points where conflict exists are coloured black. The official ATLAS limit is overlaid for comparison. Points are sampled from the full CMSSM parameter space as described in the text, but are projected onto the ($m_0$,$m_{1/2}$) plane for visualisation.}
\label{fig:disagreement}
\end{figure}

We then use this model likelihood to estimate the expected limits on the signal yield in each channel for each point in our training data set, and obtain an estimate of the ATLAS combined observed limit by taking the observed $CL_s$ value of each training data point to be the one obtained from the channel with the lowest expected $CL_s$ value for that point (i.e. following ATLAS's method). We find the difference between this estimate of the ATLAS limit and the one used in our analysis to be very small: of the 26491 training points there are 100 which are classified (into excluded/not excluded) differently by the two limits. We show these points in figure \ref{fig:disagreement}; they predominantly occur in a group clustered at low $m_0$, and for most of them the observed strongest limit comes from R1, while we estimate that the expected strongest limit comes from R2.  

\section{Plots of CMSSM profile likelihoods and marginalised posteriors}
\label{app:plots}

This appendix contains the figures referred to in section~\ref{sec:results}. We refer the reader to that section for further information.

\begin{figure*}[f]
\begin{center}
\includegraphics[width=0.40\paperwidth]{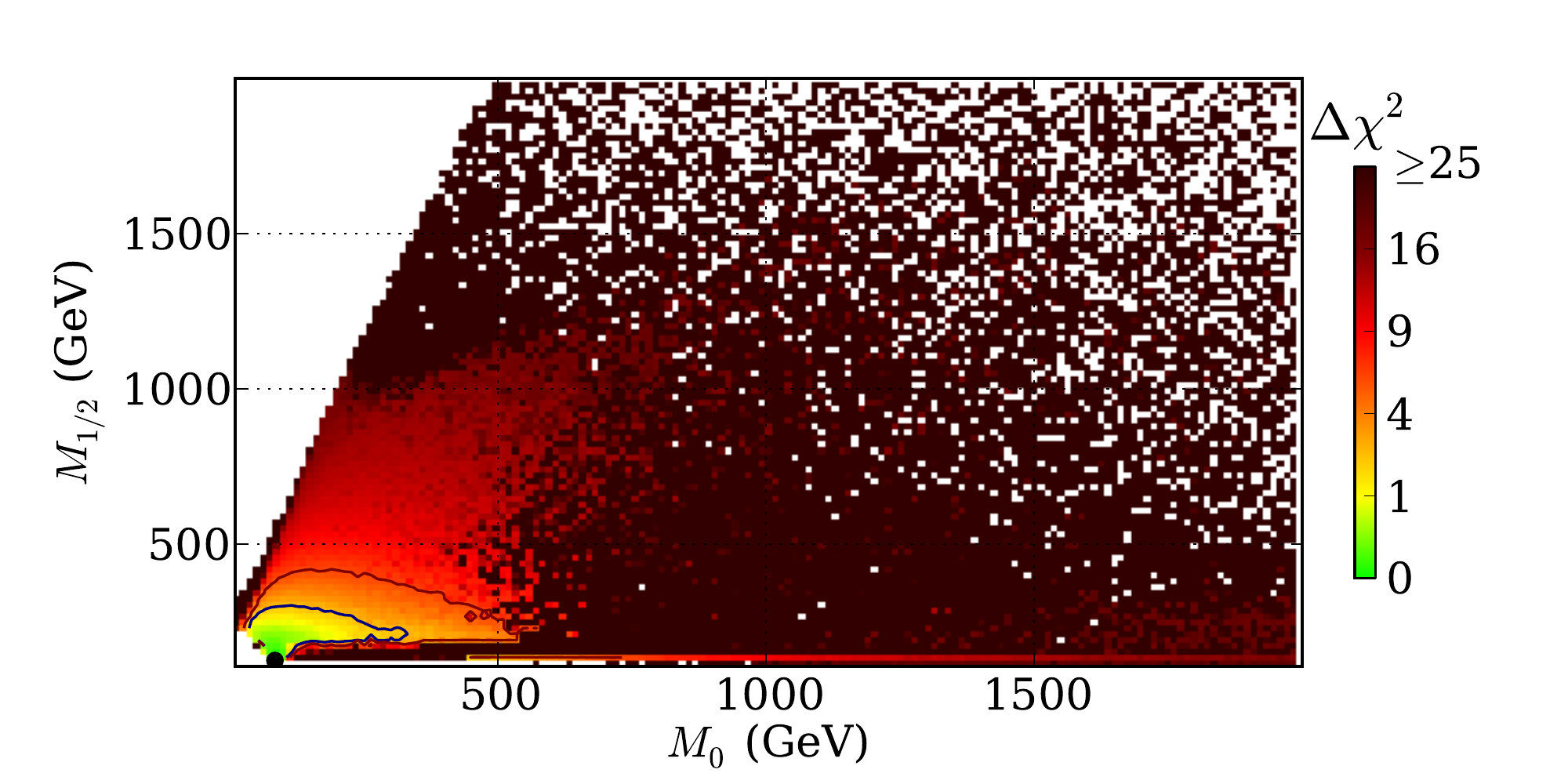}
\hspace{-16mm}
\includegraphics[width=0.40\paperwidth]{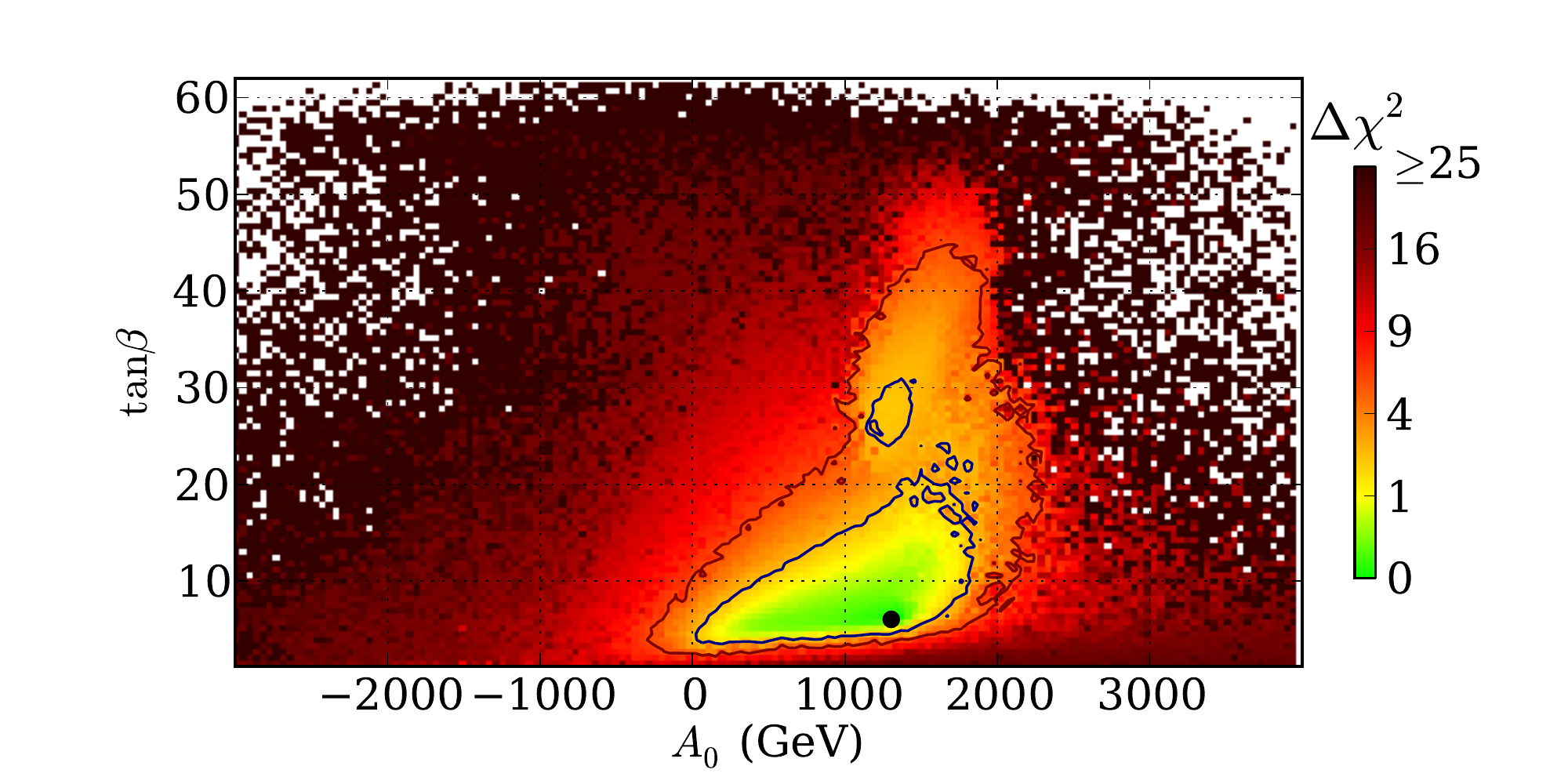}
\includegraphics[width=0.40\paperwidth]{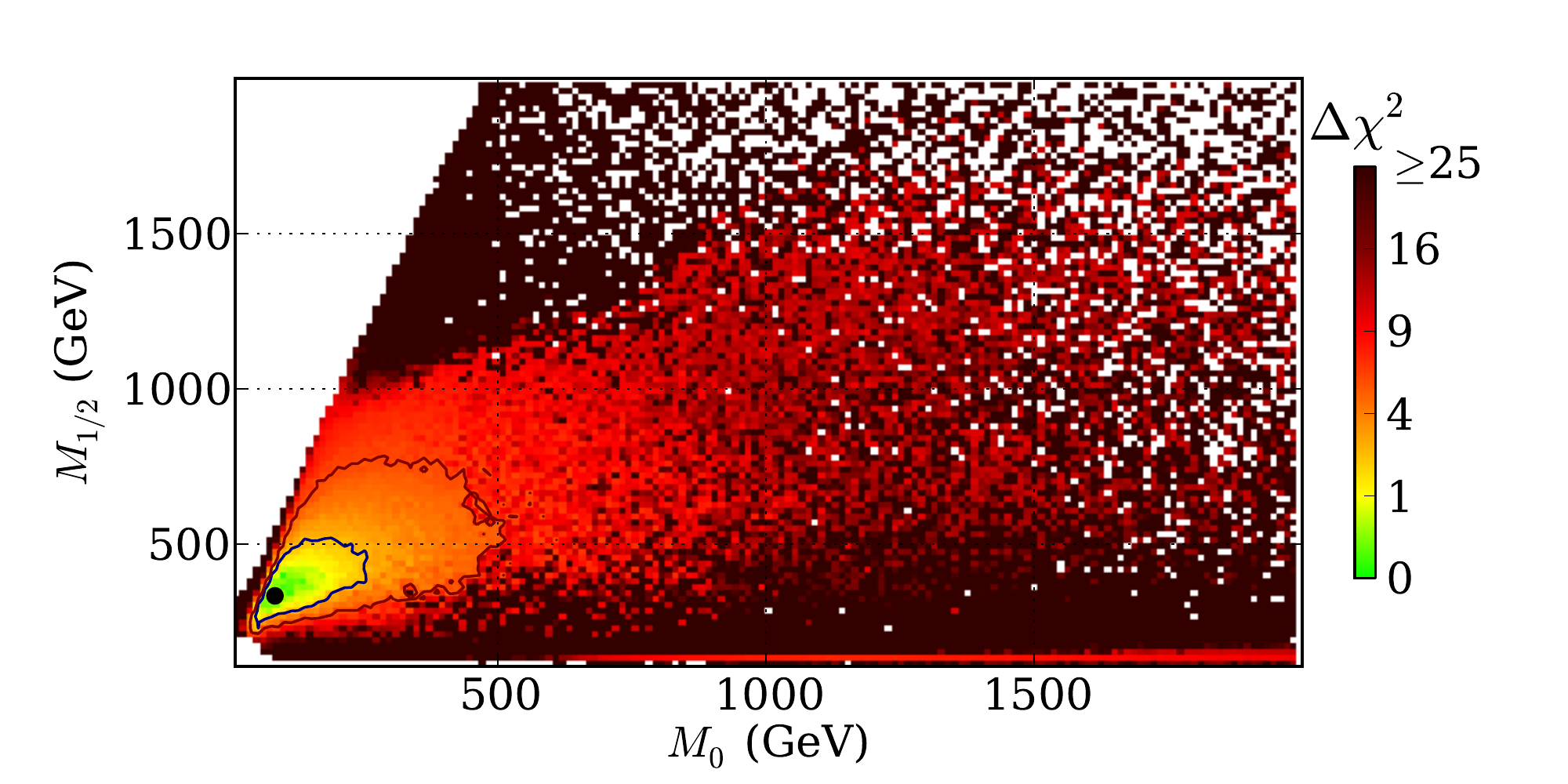}
\hspace{-16mm}
\includegraphics[width=0.40\paperwidth]{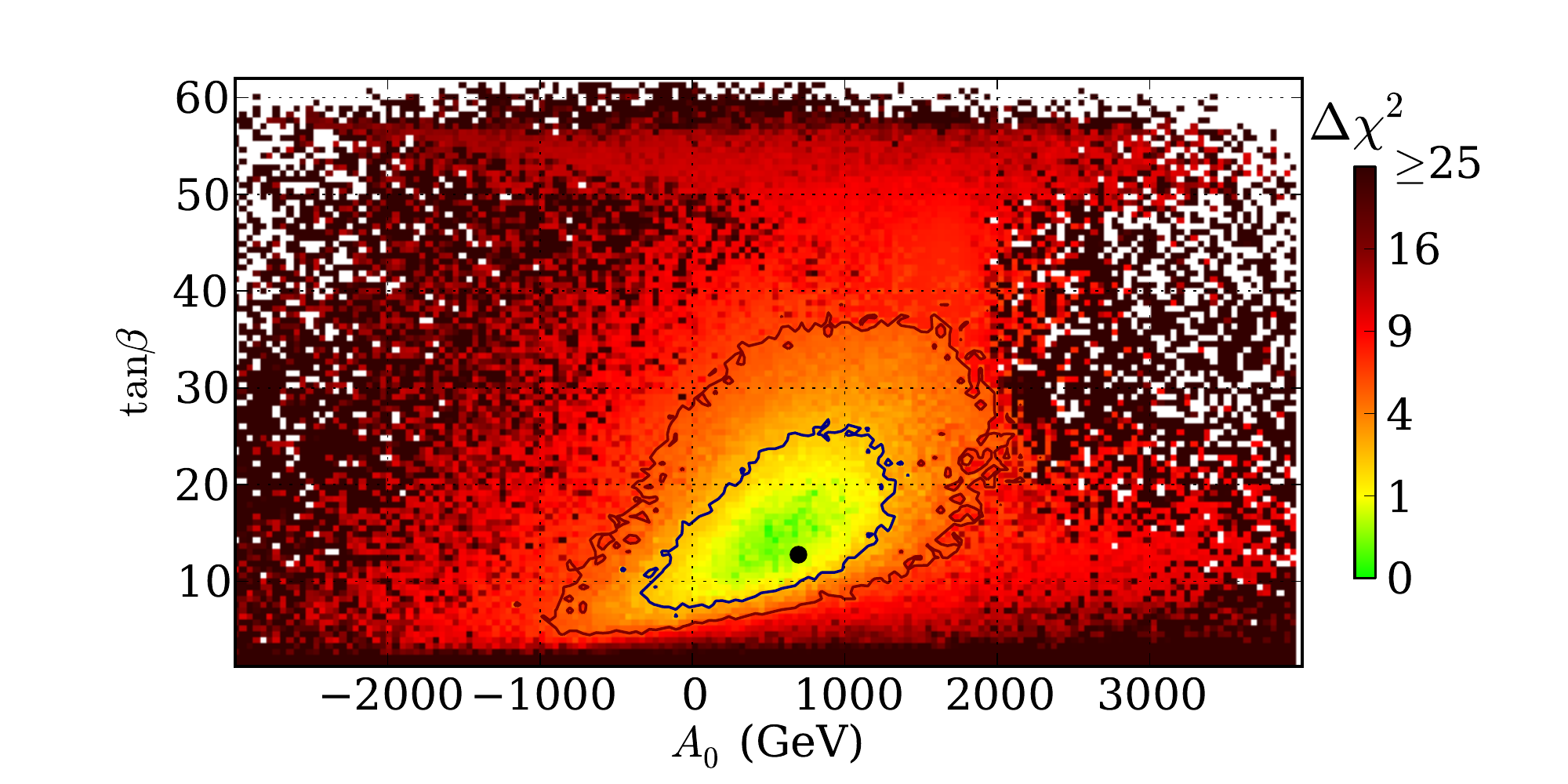}
\includegraphics[width=0.40\paperwidth]{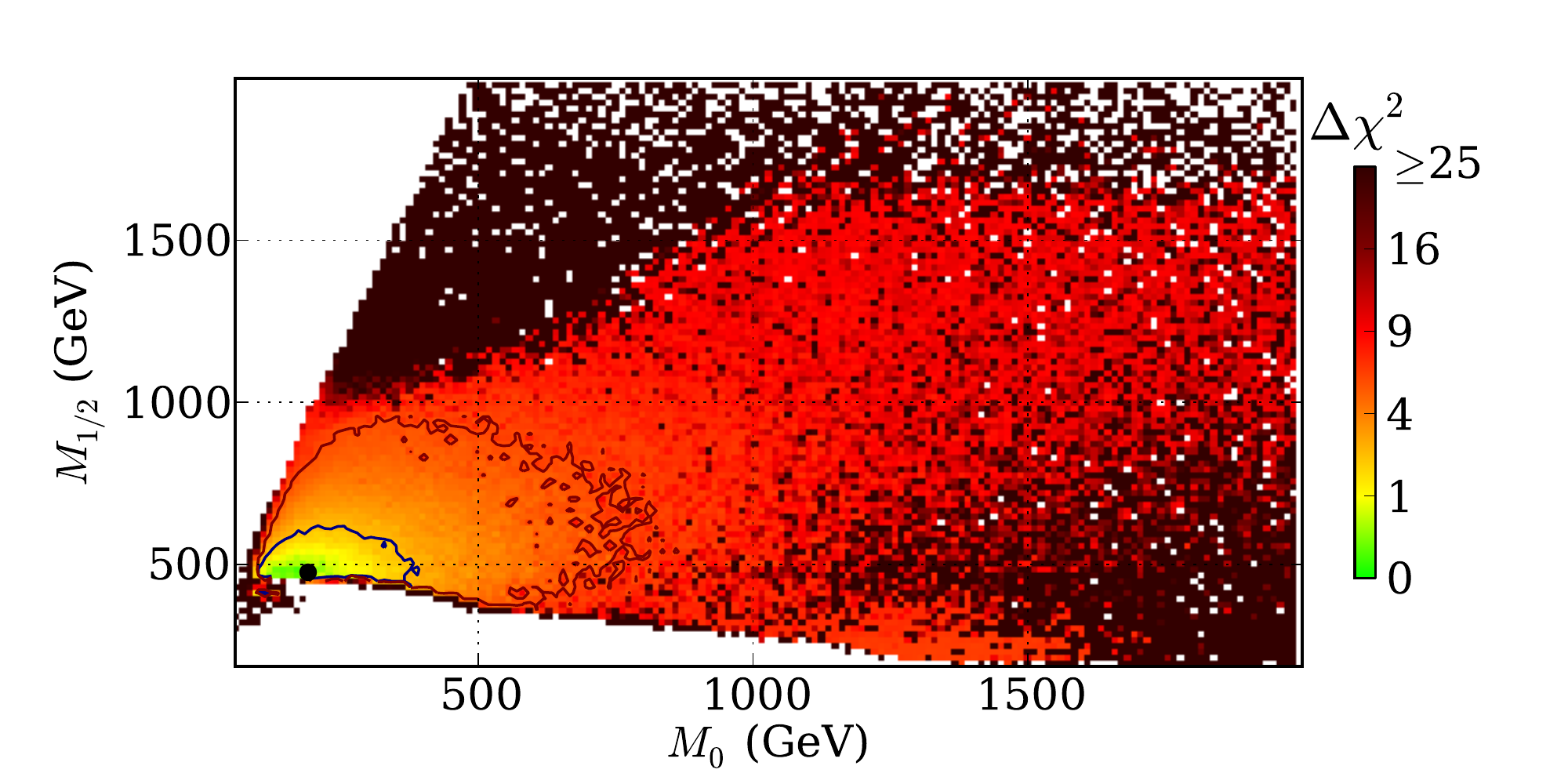}
\hspace{-16mm}
\includegraphics[width=0.40\paperwidth]{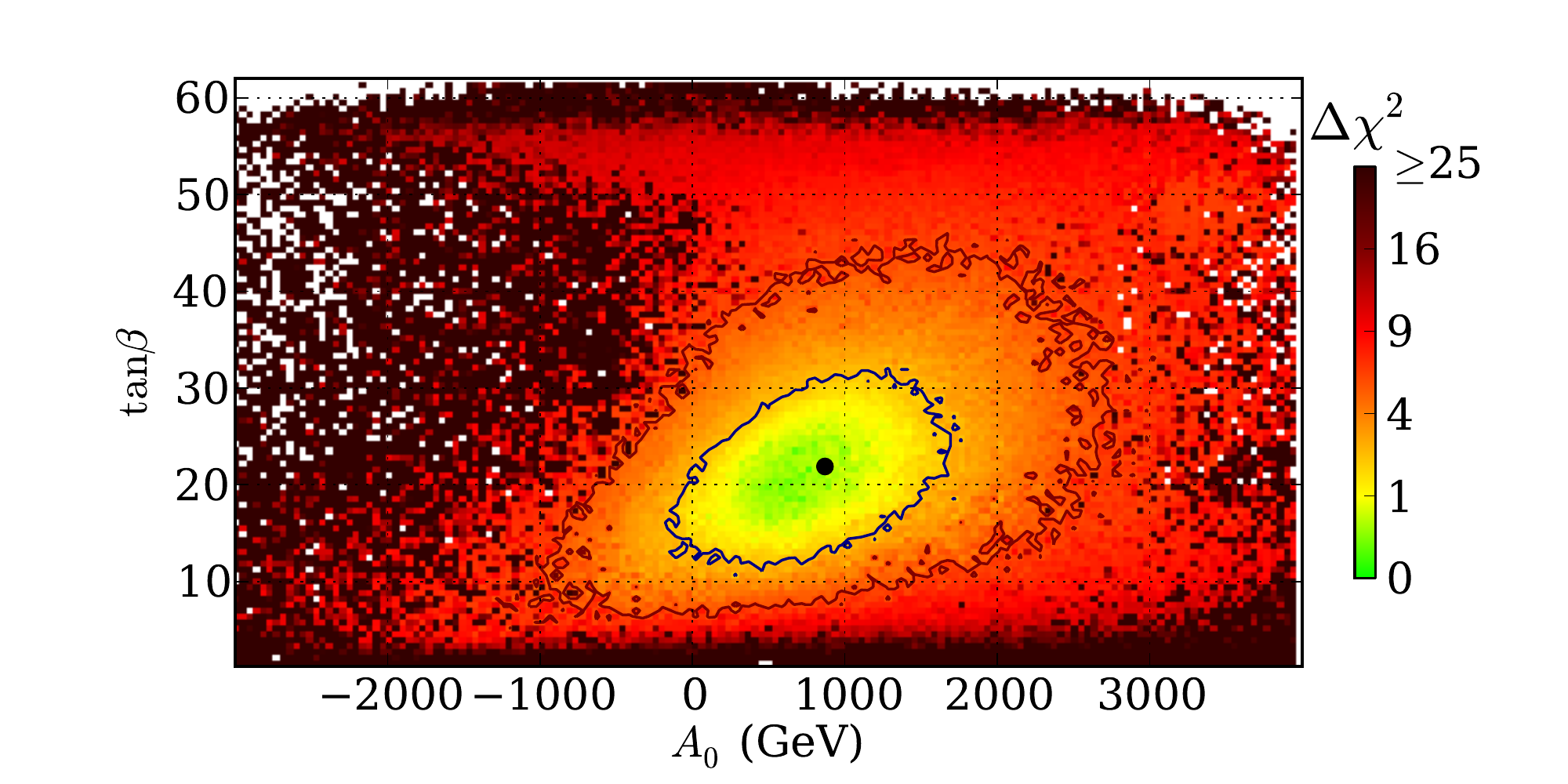}
\includegraphics[width=0.40\paperwidth]{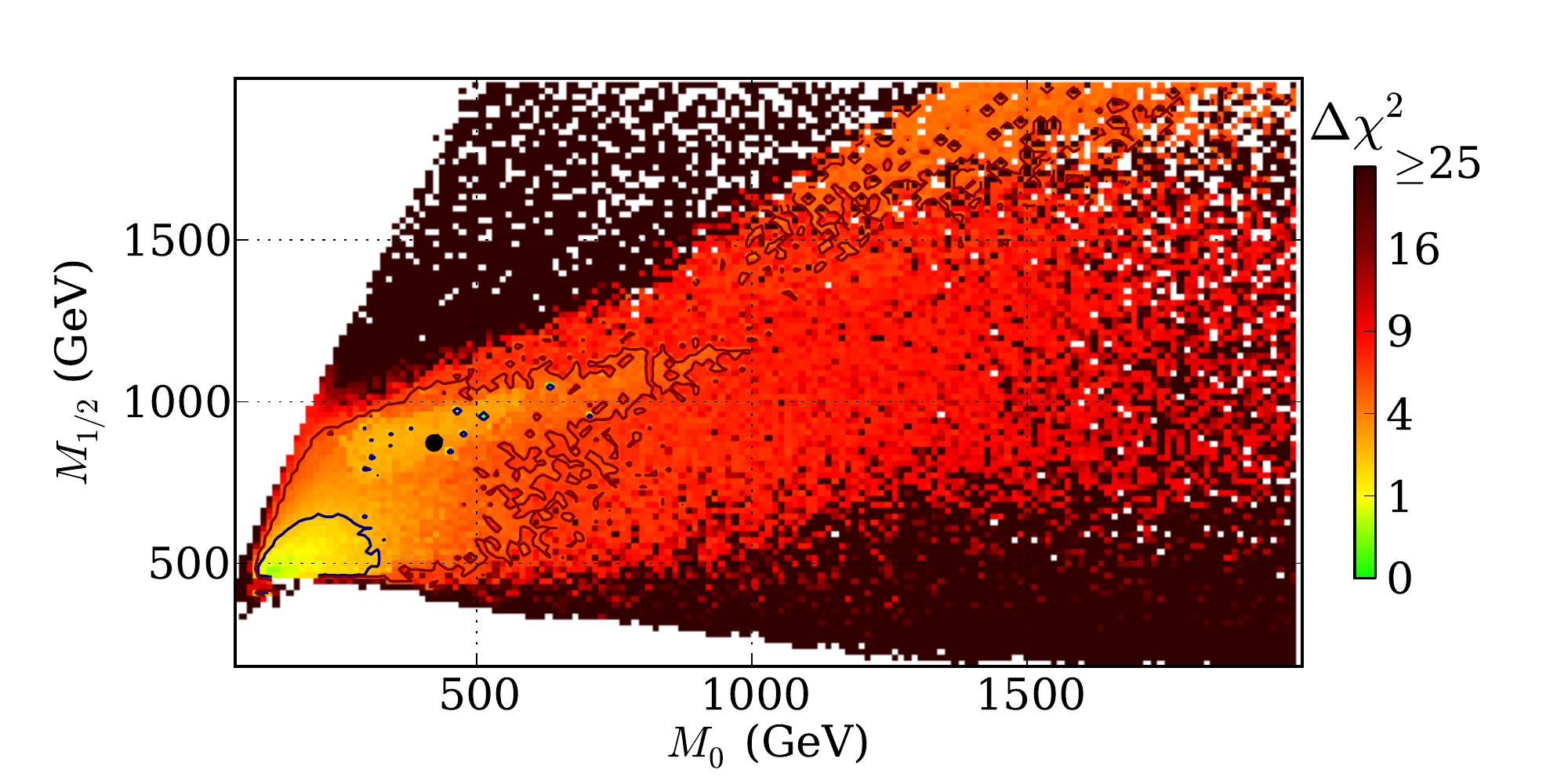}
\hspace{-16mm}
\includegraphics[width=0.40\paperwidth]{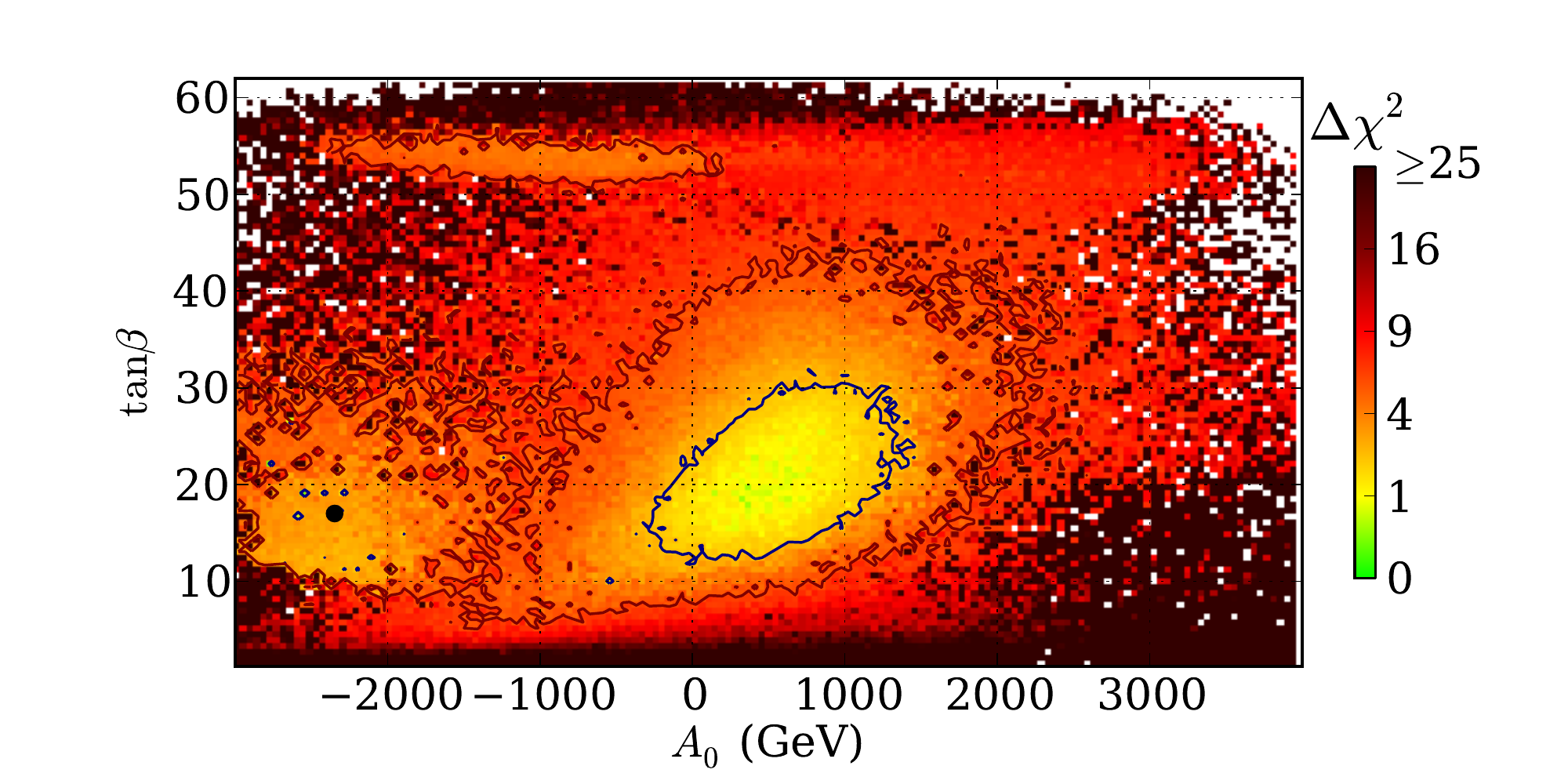}
\end{center}
\caption{The evolution of the profile of the (log-)likelihood function from the ``pre-LEP'' situation (first row), to including the LEP Higgs search and XENON100 data (second row), to adding the 1 $\text{fb}^{-1}$ LHC sparticle searches (third row), to folding in the 2012 February Higgs search results. Contours containing 68\% and 95\% confidence regions are shown. The above results were obtained using the log prior. Results obtained using the CCR prior (not shown) show variations consistent with the different sampling density but are qualitatively similar.
}
\label{fig:likelog}
\end{figure*}

\begin{figure*}[f]
\begin{center}
\includegraphics[width=0.40\paperwidth]{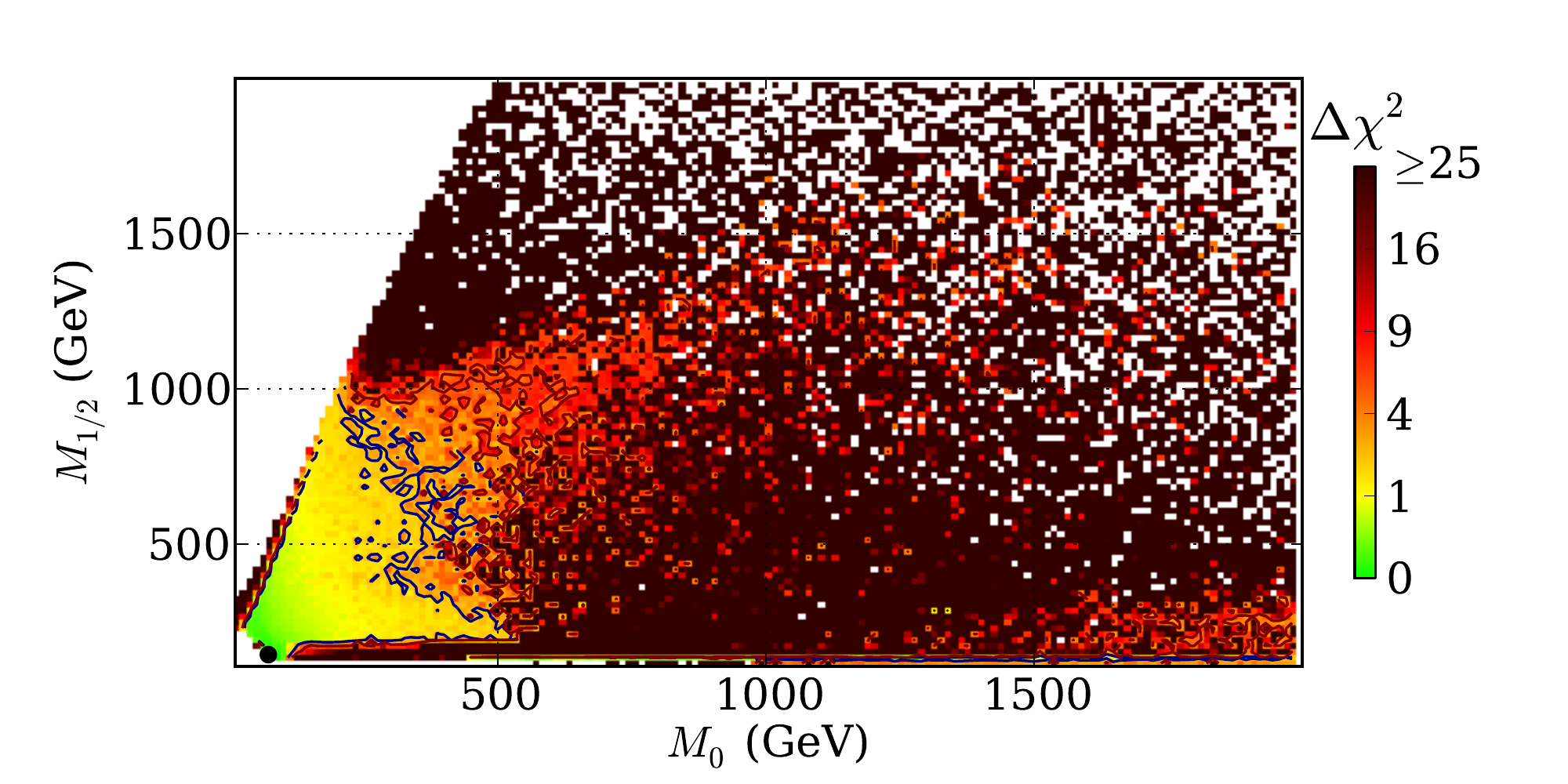}
\hspace{-16mm}
\includegraphics[width=0.40\paperwidth]{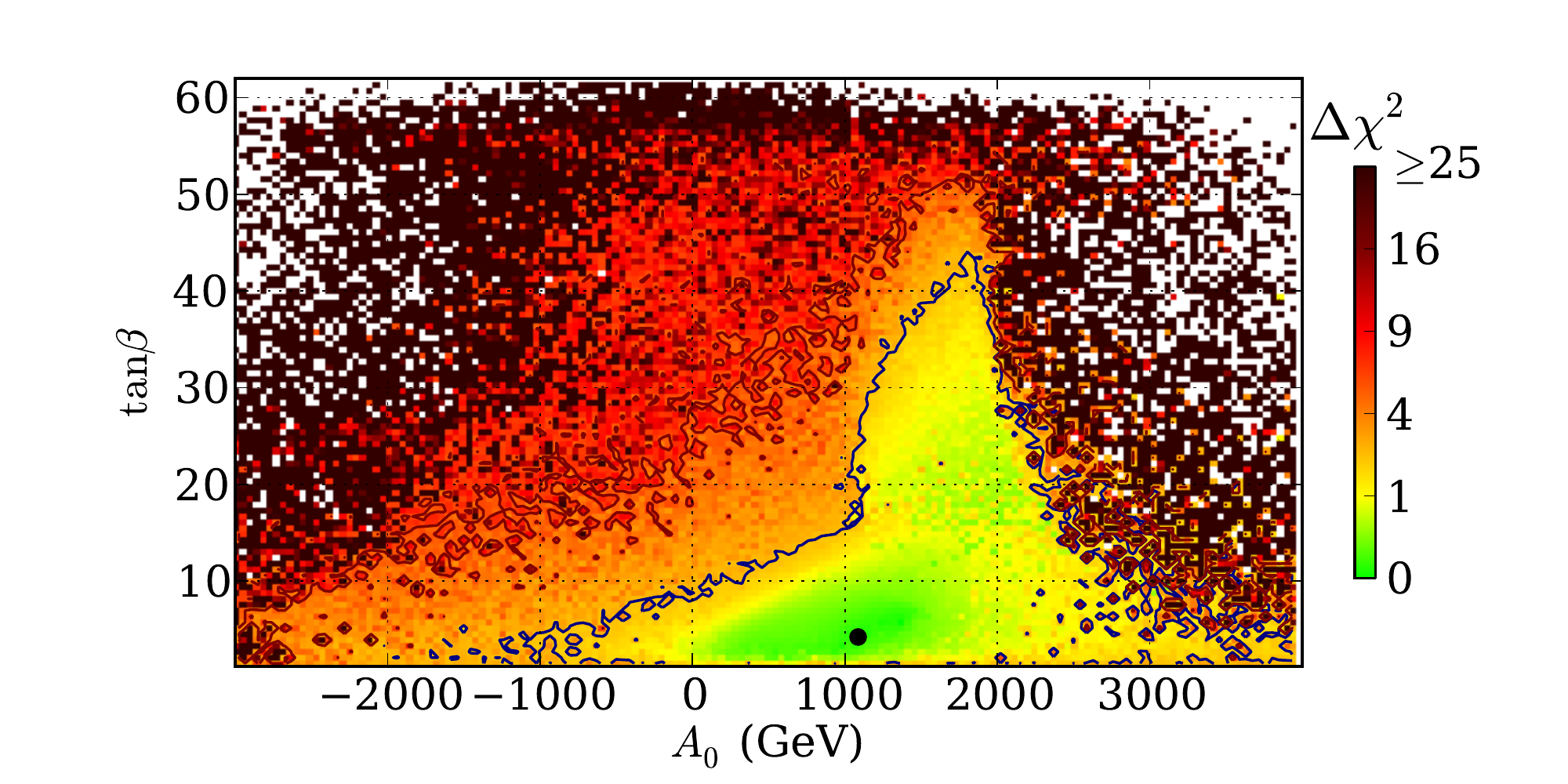}
\includegraphics[width=0.40\paperwidth]{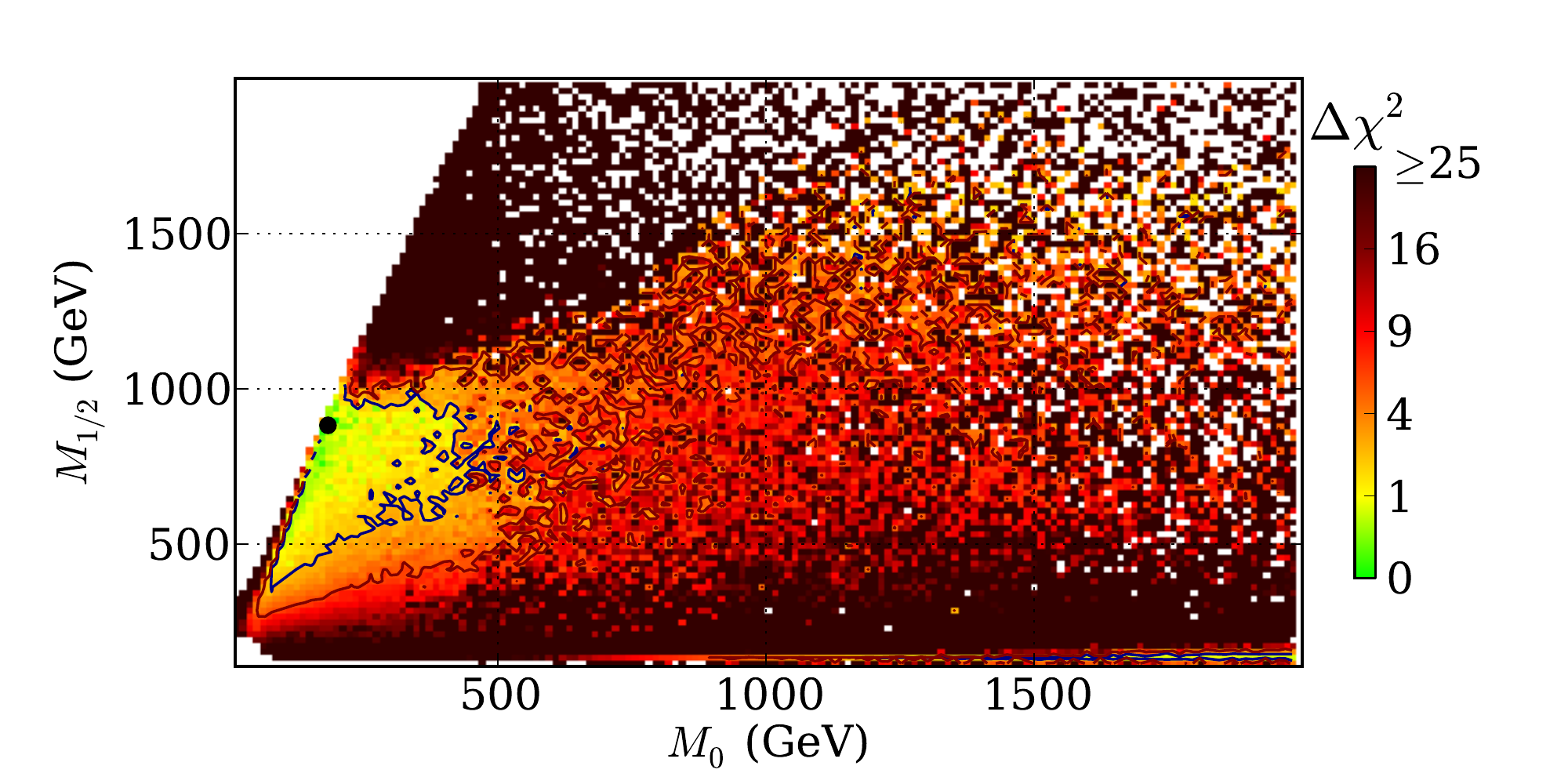}
\hspace{-16mm}
\includegraphics[width=0.40\paperwidth]{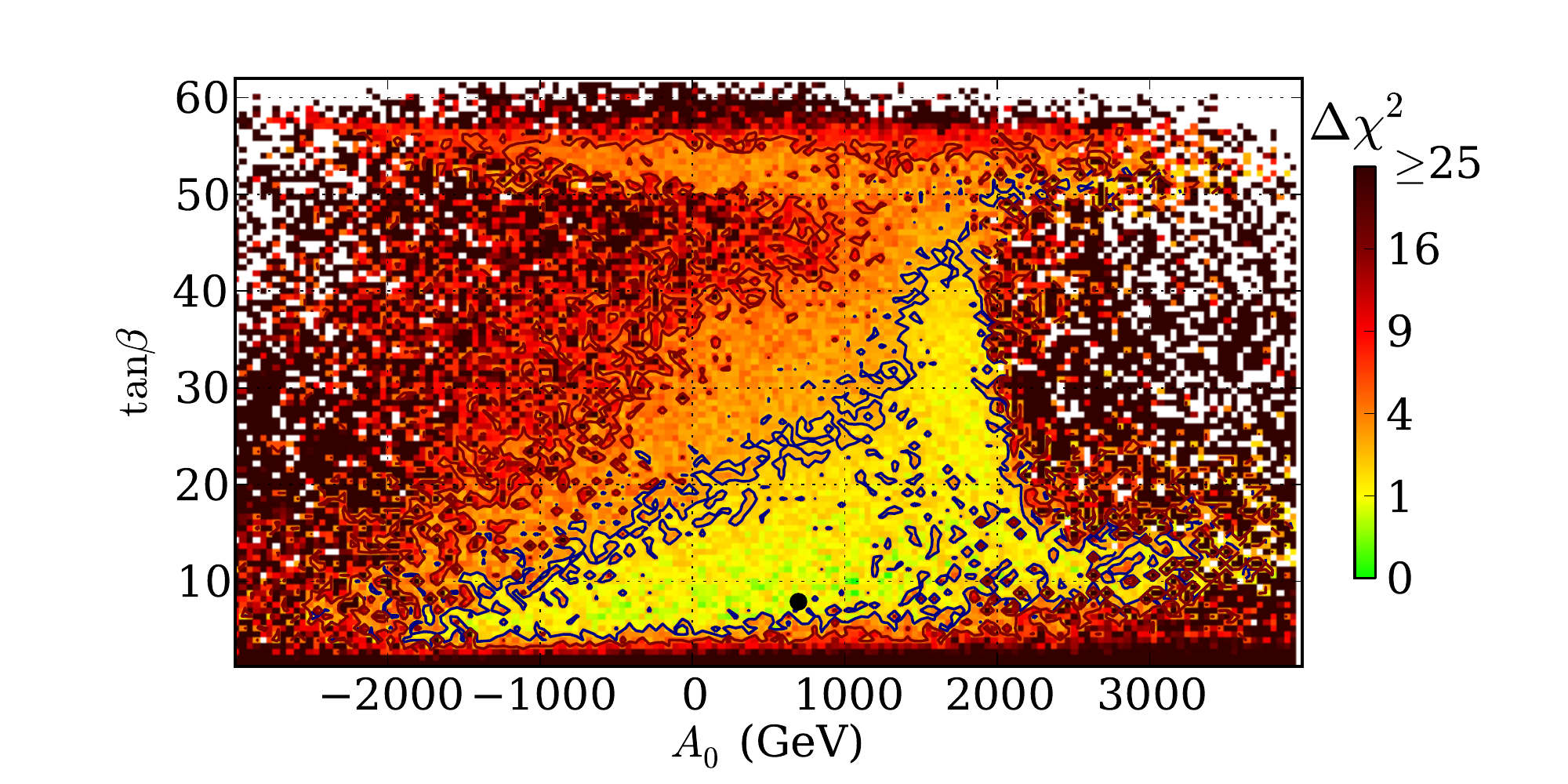}
\includegraphics[width=0.40\paperwidth]{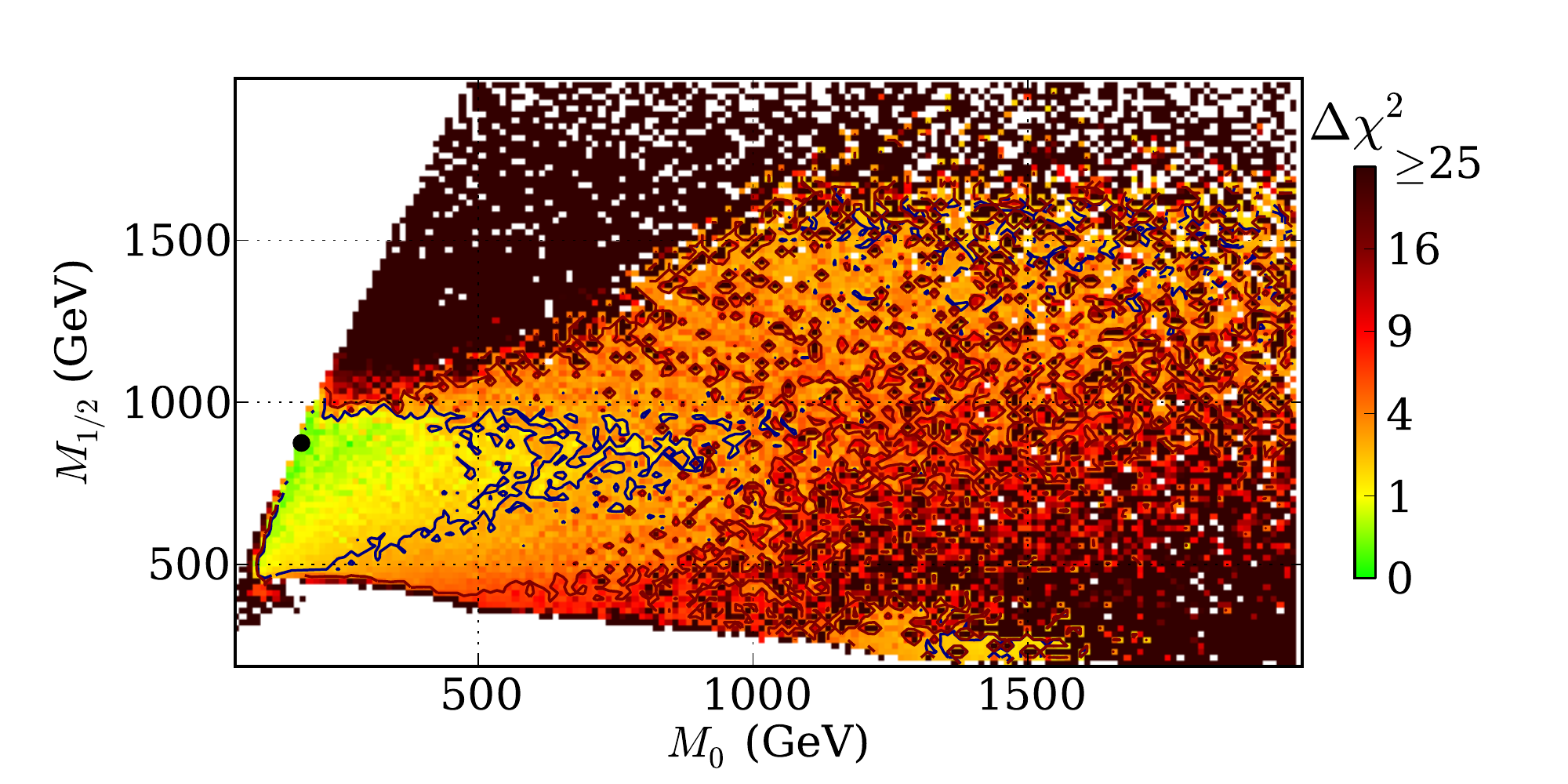}
\hspace{-16mm}
\includegraphics[width=0.40\paperwidth]{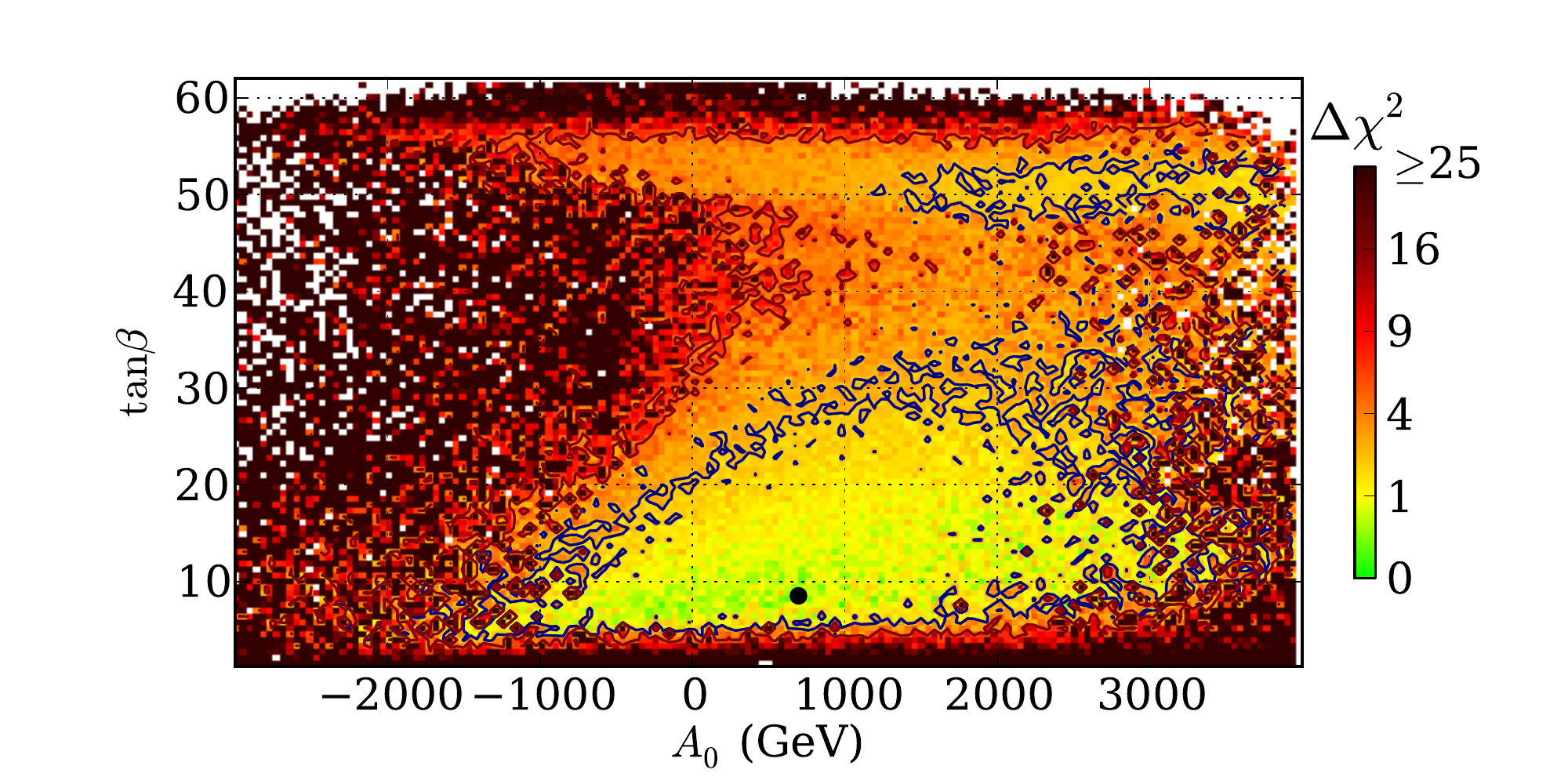}
\includegraphics[width=0.40\paperwidth]{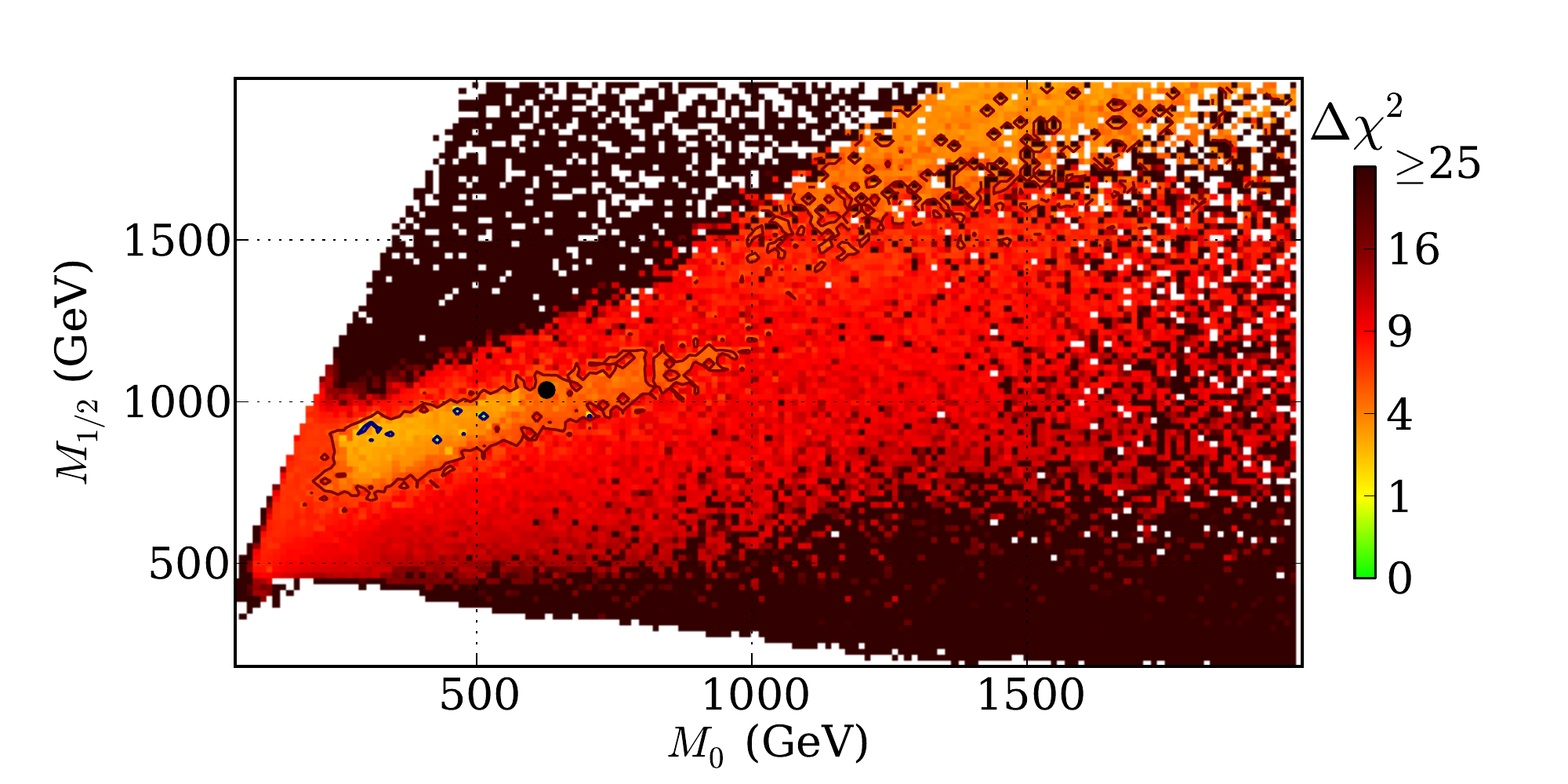}
\hspace{-16mm}
\includegraphics[width=0.40\paperwidth]{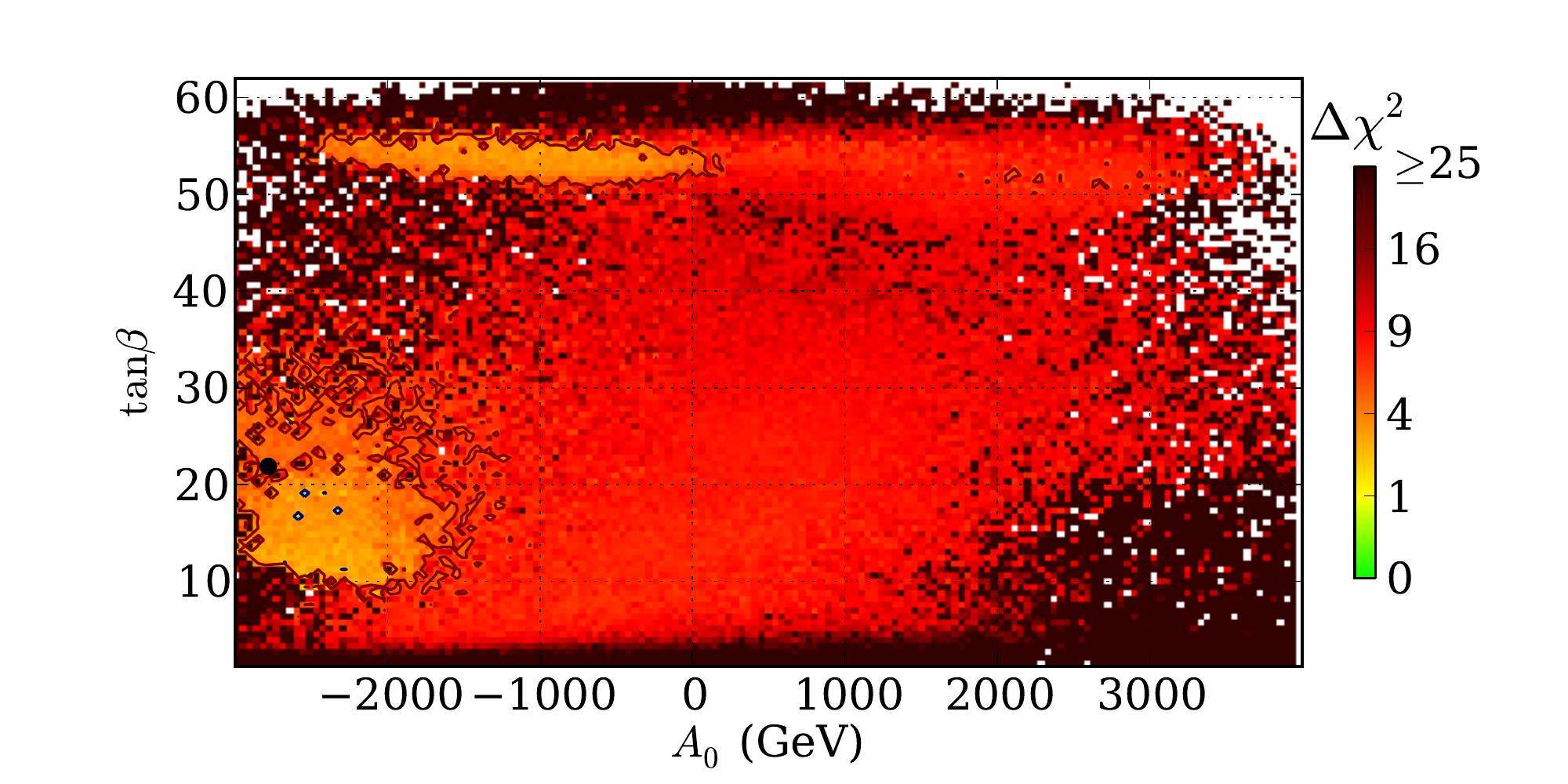}
\end{center}
\caption{The evolution of the profile of the (log-)likelihood function from the ``pre-LEP'' situation (first row), to including the LEP Higgs search and XENON100 data (second row), to adding the LHC sparticle searches (third row), to folding in the 2012 February Higgs search results. Contours containing 68\% and 95\% confidence regions are shown. The above results were obtained using the log prior and have been reweighted to estimate the effect of removing the $\delta a_\mu$ constraint. Significant deterioration of the sampling is seen due to the shift in the preferred regions away from the originally sampled regions, however the general impact of removing the $\delta a_\mu$ constraint can be seen in the motion of the preferred regions upwards in the mass parameters. Of particular note is the very strong shift to high $A_0$ when the ATLAS Higgs search results are imposed, which is much less pronounced in figure~\ref{fig:likelog}, indicating very strong tension between the ATLAS Higgs search results and the $\delta a_\mu$ constraint. Results obtained using the CCR prior (not shown) show variations consistent with the different sampling density but are qualitatively similar.
}
\label{fig:likelogno(g-2)}
\end{figure*}

\begin{figure*}[f]
\begin{center}
\includegraphics[width=0.40\paperwidth]{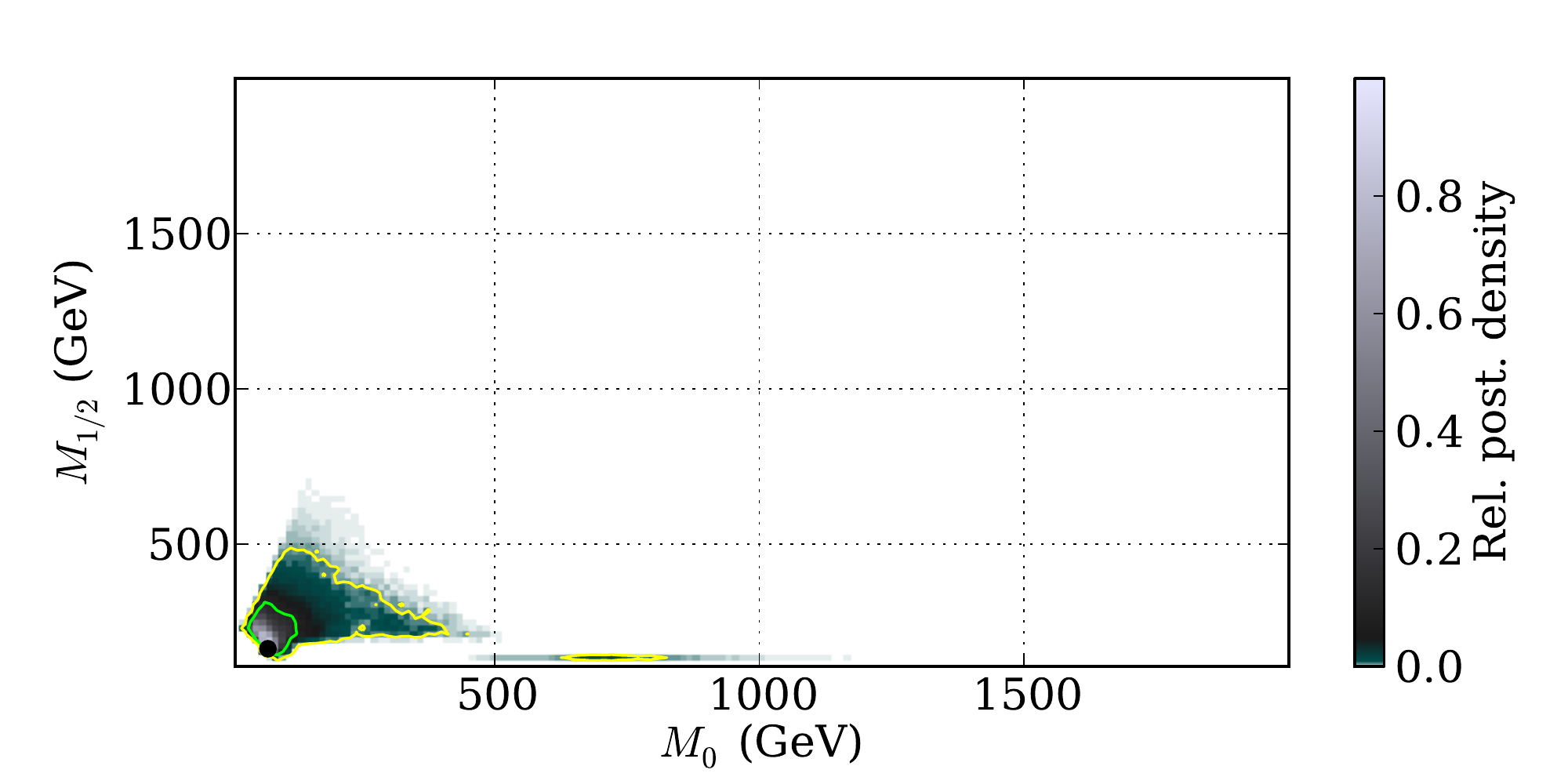}
\hspace{-16mm}
\includegraphics[width=0.40\paperwidth]{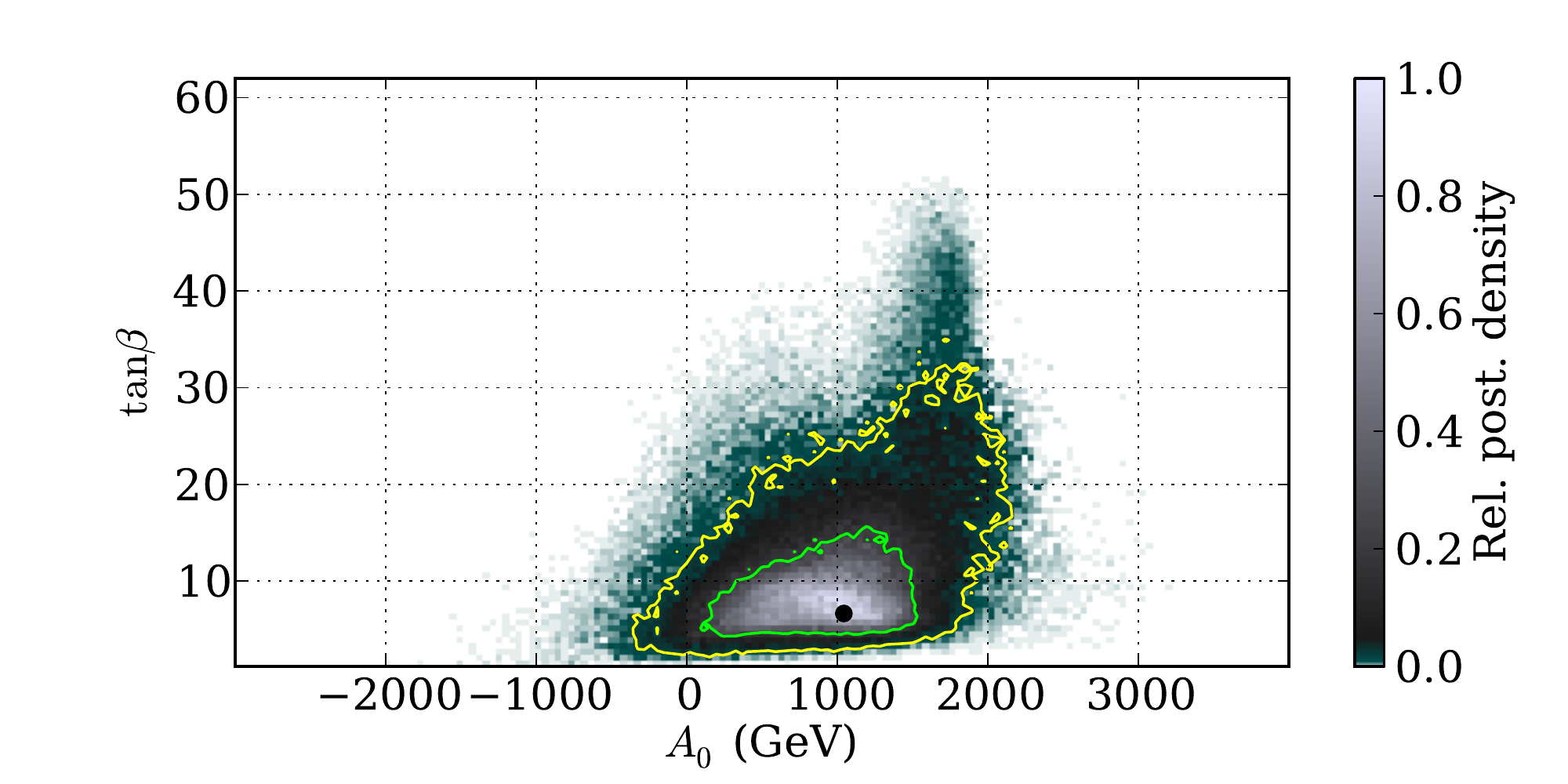}
\includegraphics[width=0.40\paperwidth]{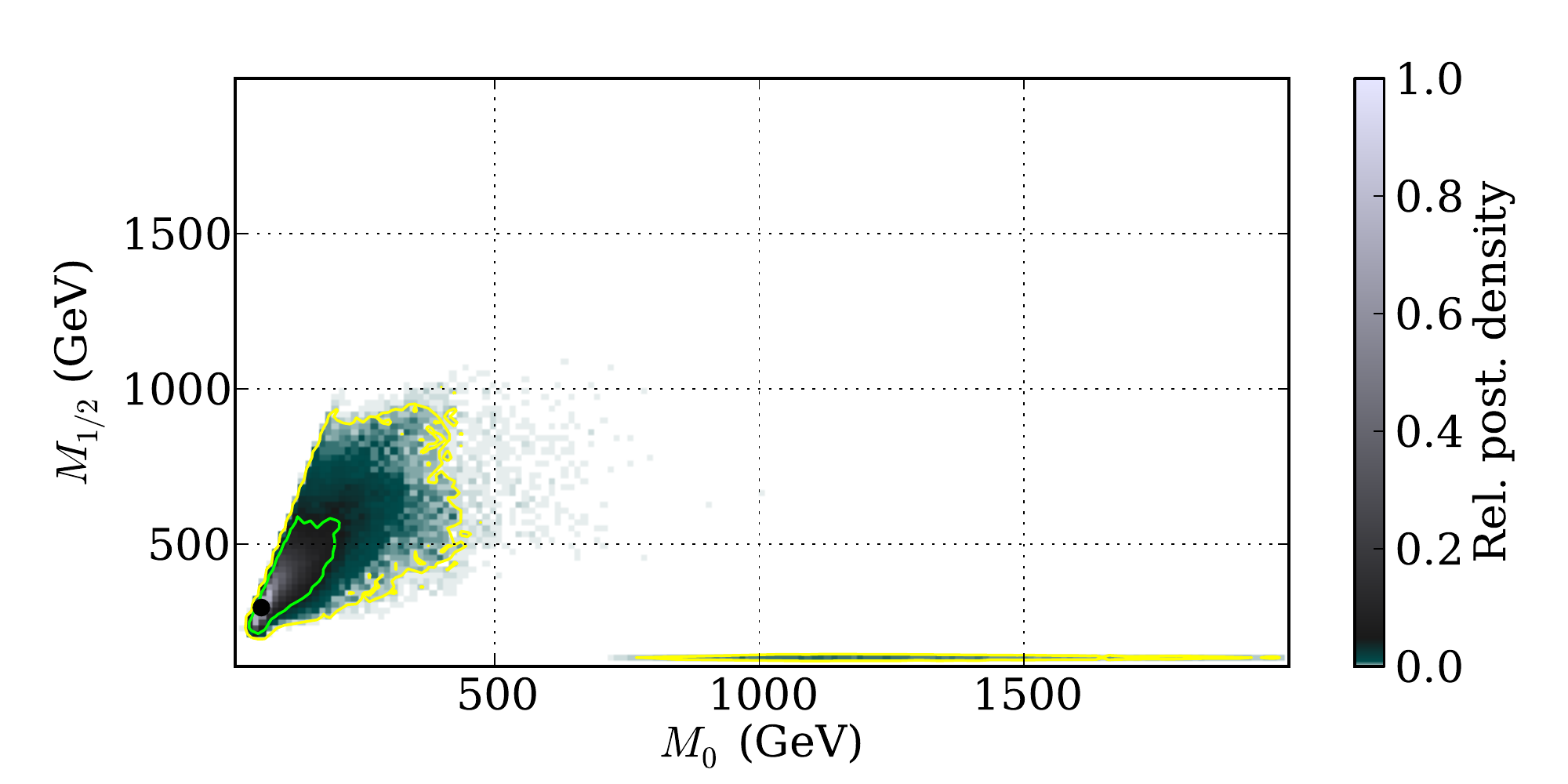}
\hspace{-16mm}
\includegraphics[width=0.40\paperwidth]{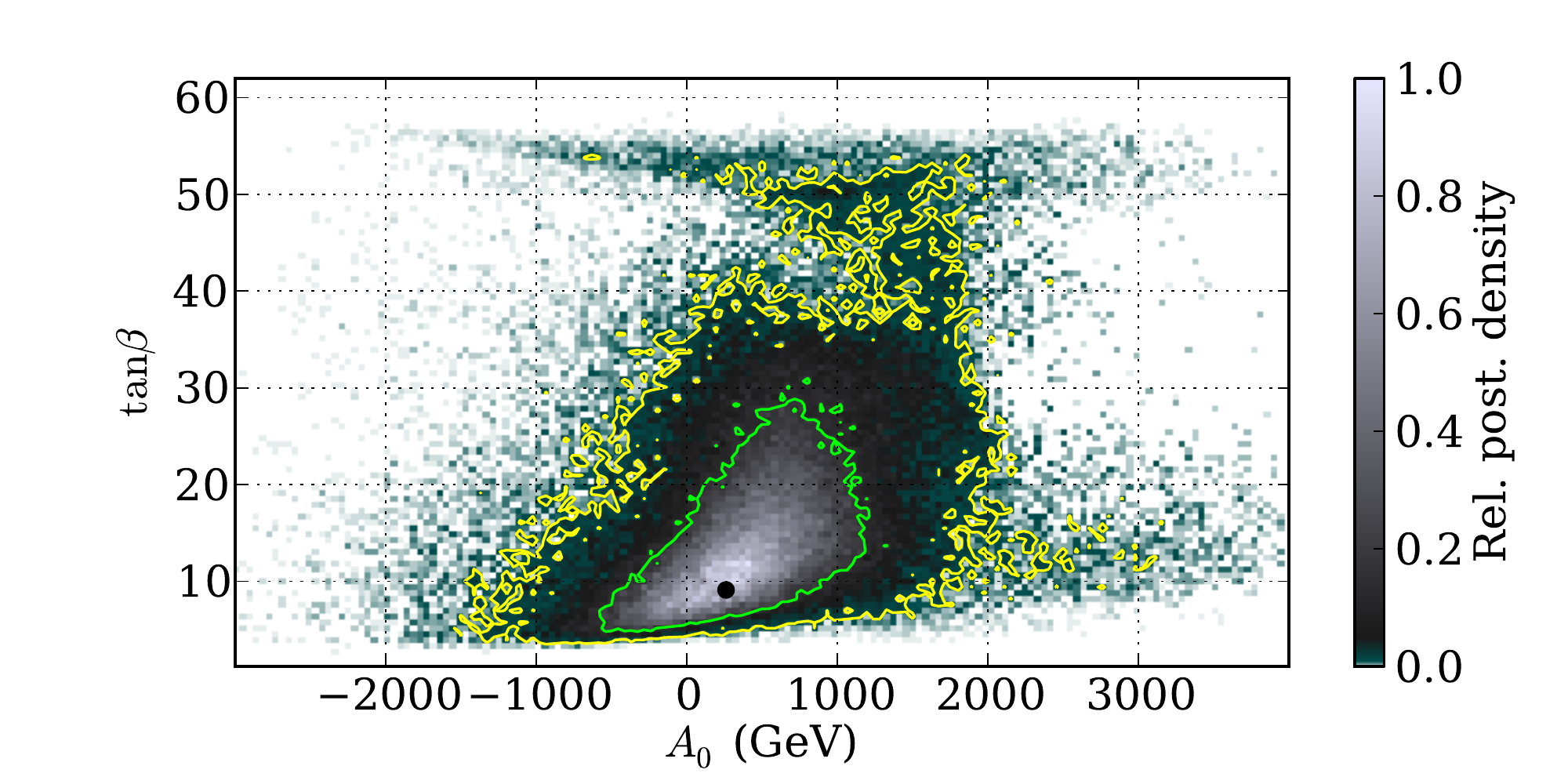}
\includegraphics[width=0.40\paperwidth]{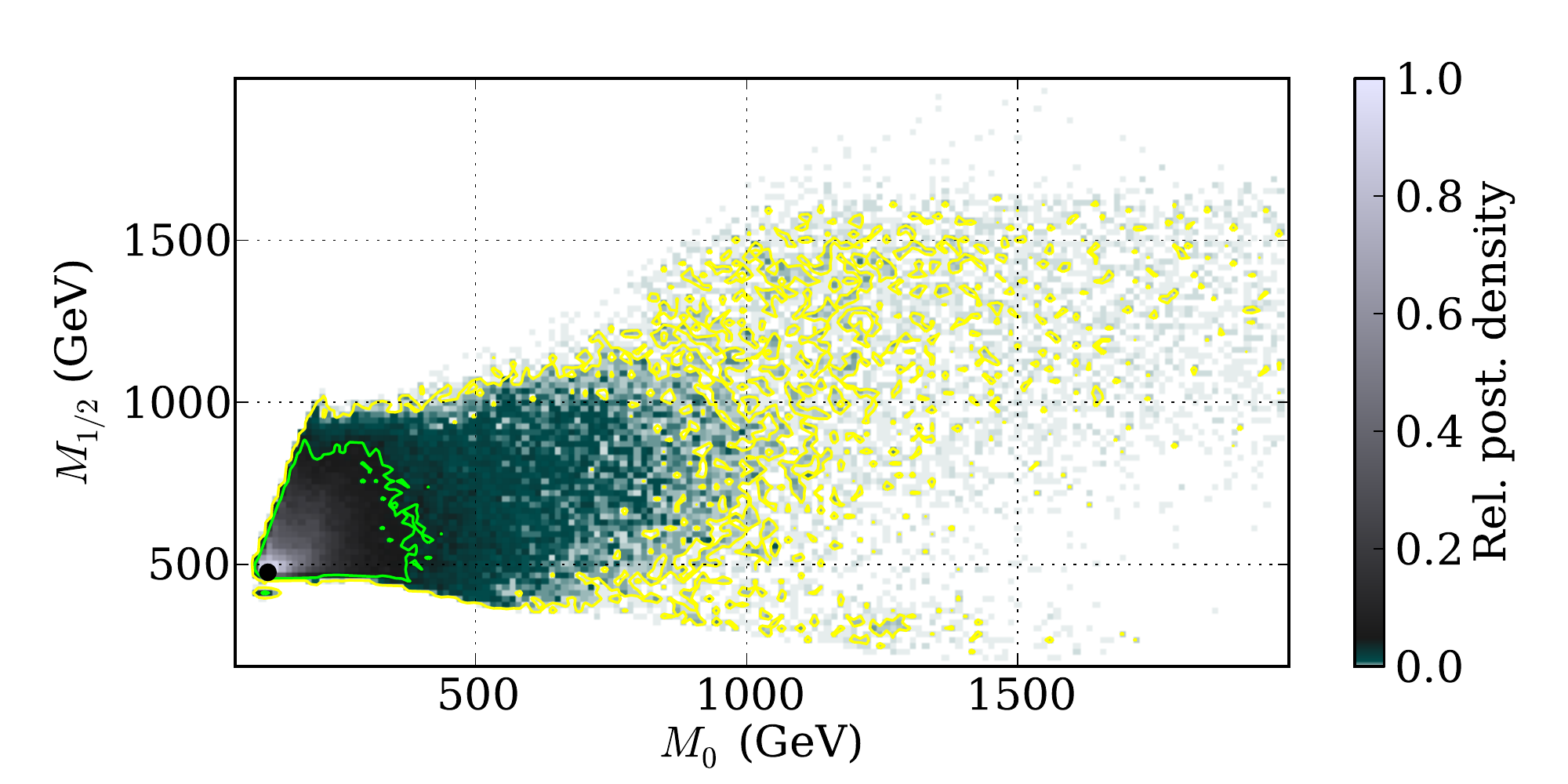}
\hspace{-16mm}
\includegraphics[width=0.40\paperwidth]{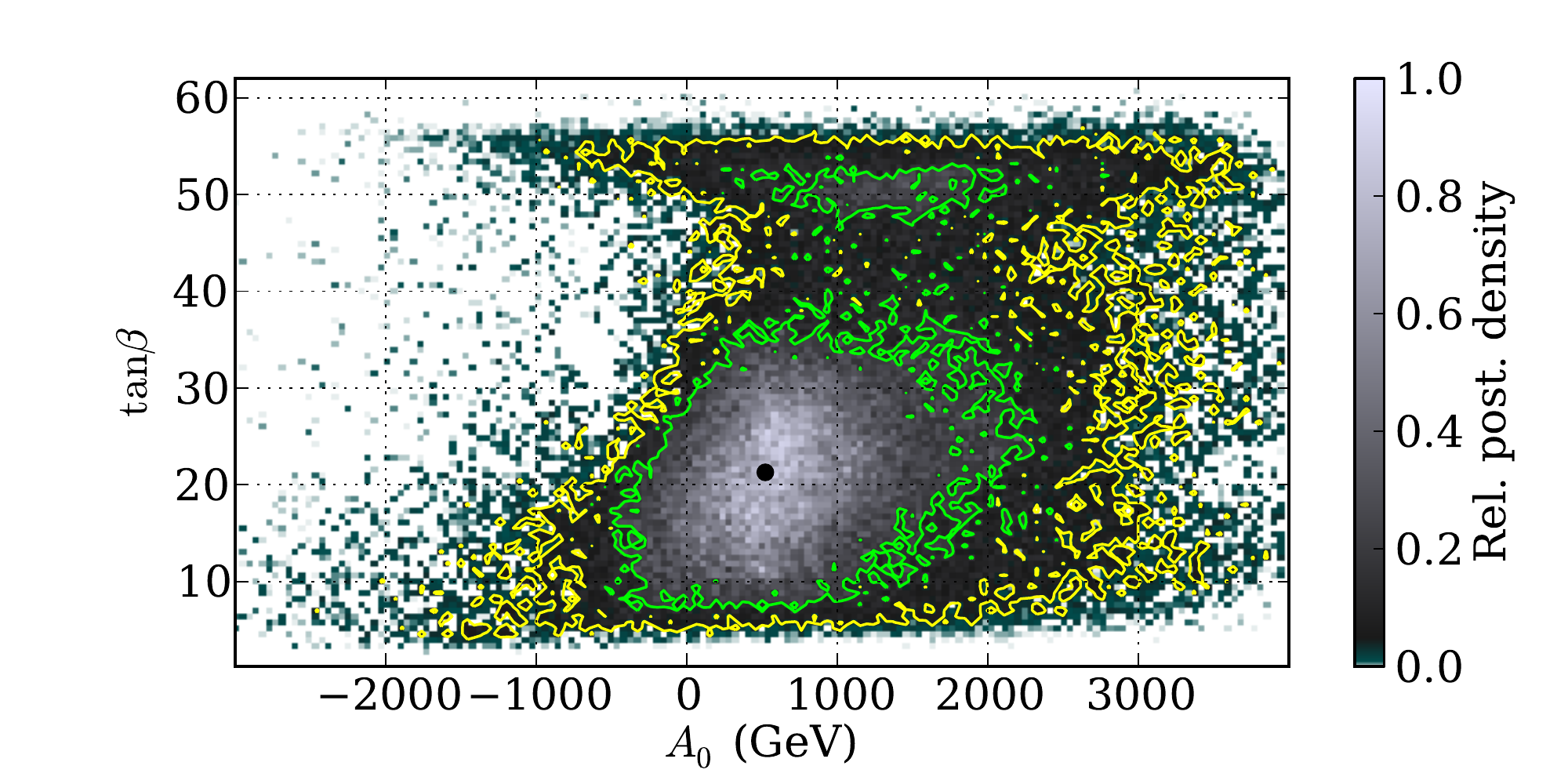}
\includegraphics[width=0.40\paperwidth]{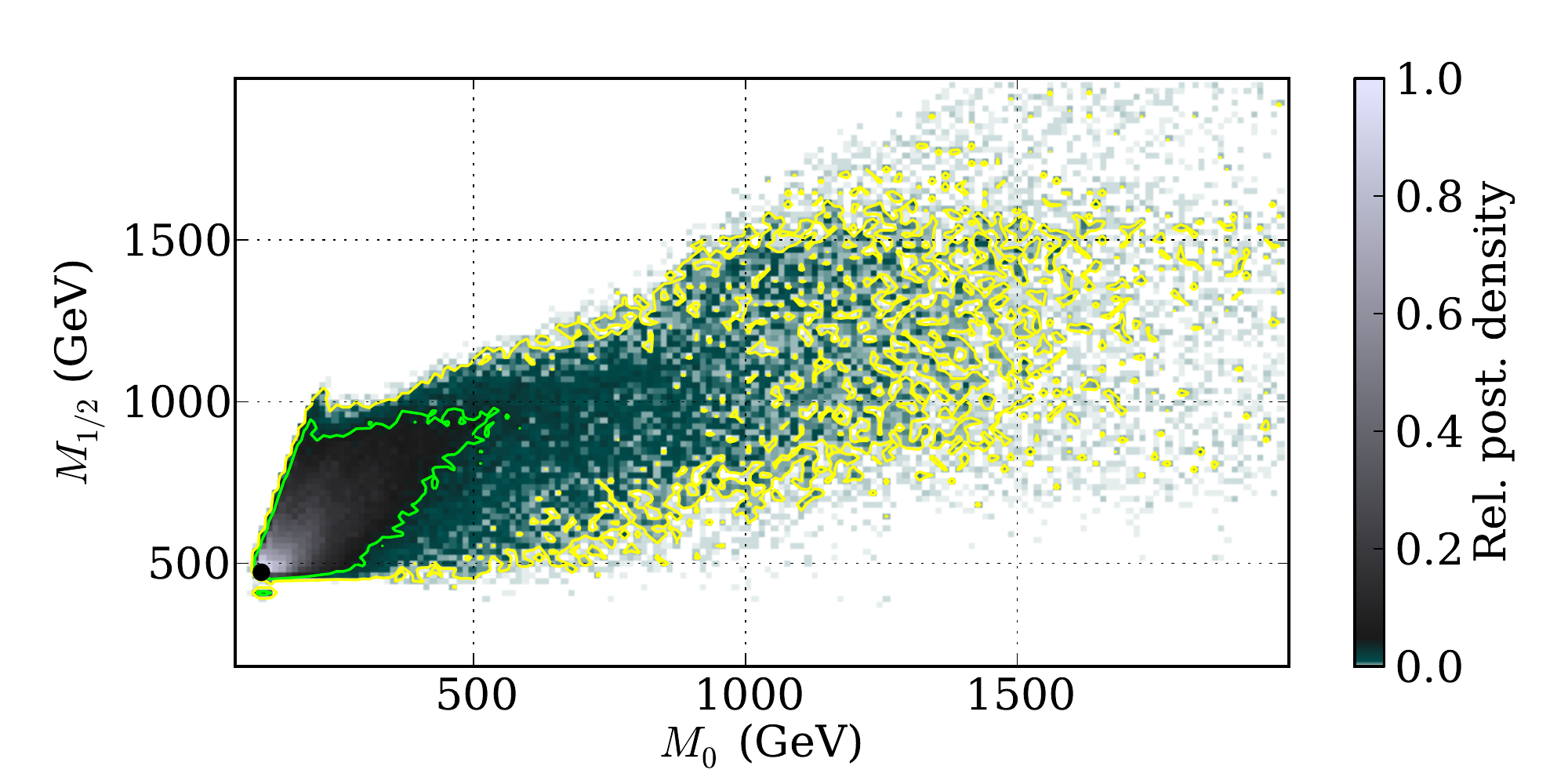}
\hspace{-16mm}
\includegraphics[width=0.40\paperwidth]{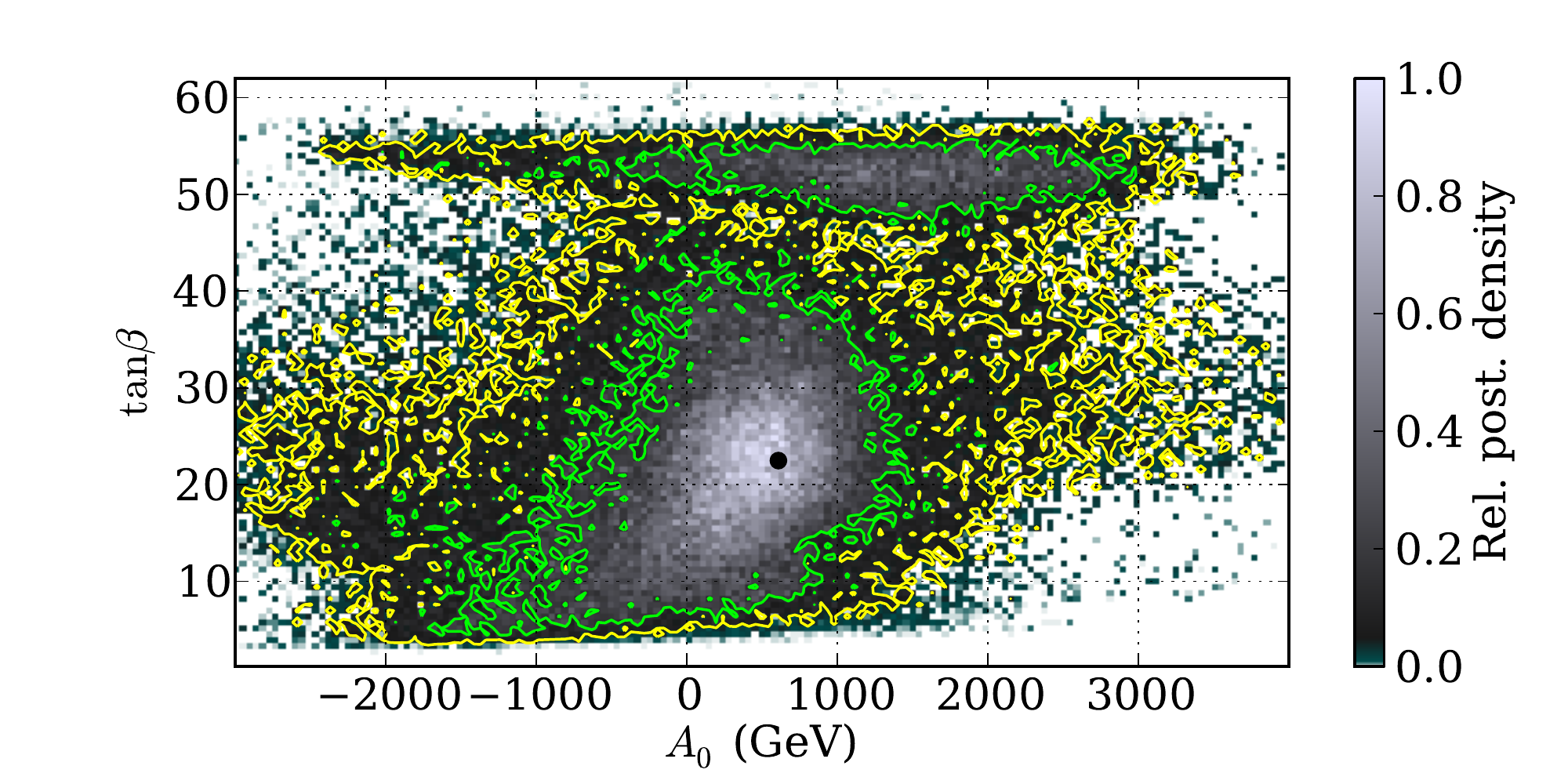}
\end{center}
\caption{The evolution of the CMSSM marginalised posterior probability distributions from the ``pre-LEP'' situation (first row), to including the LEP Higgs search and XENON100 data (second row), to adding the LHC sparticle searches (third row), to folding in the 2012 February Higgs search results.  Log priors are used and 68\% and 95\% credible regions are shown.
}
\label{fig:pos1log}
\end{figure*}

\begin{figure*}[f]
\begin{center}
\includegraphics[width=0.40\paperwidth]{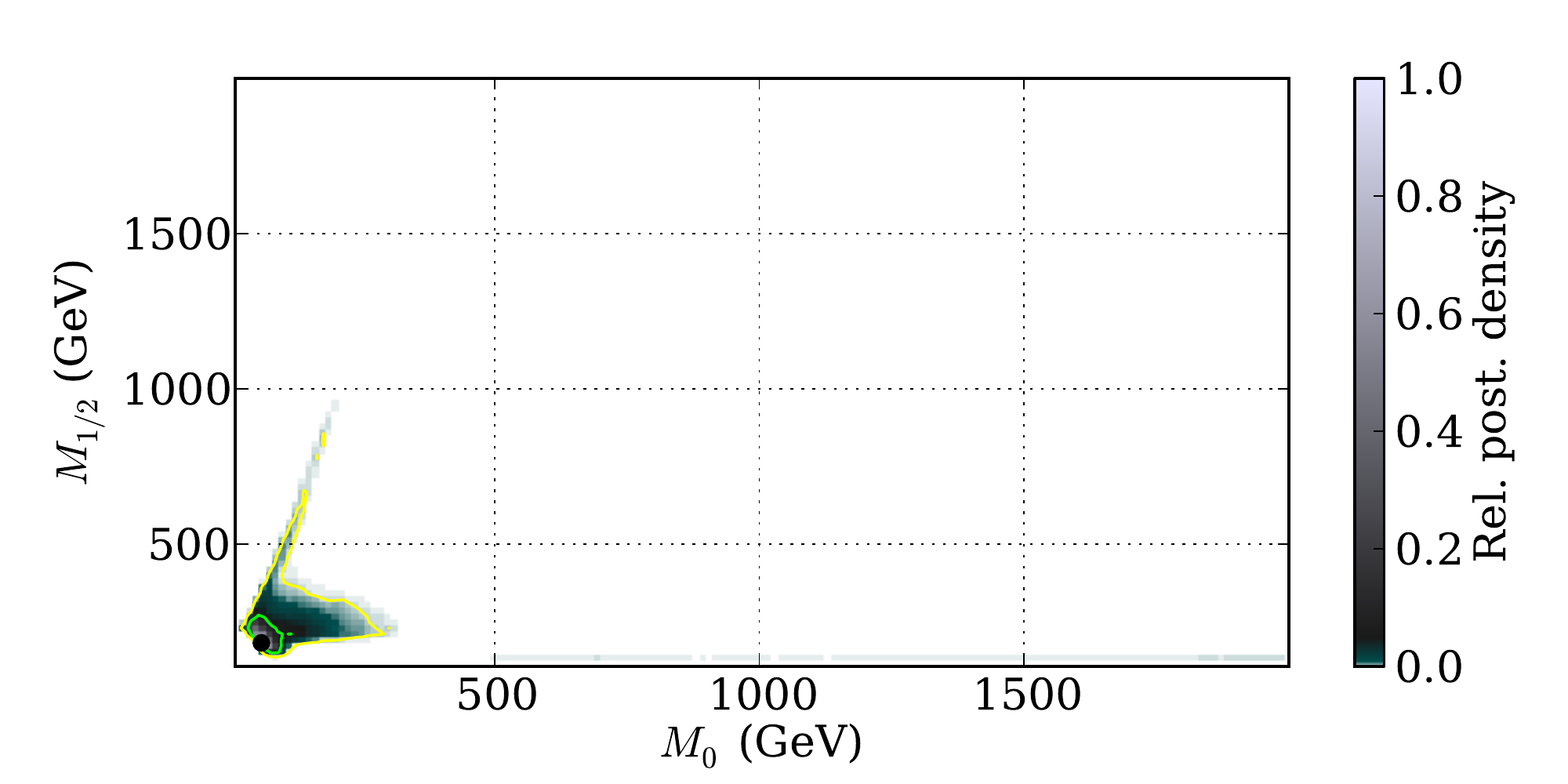}
\hspace{-16mm}
\includegraphics[width=0.40\paperwidth]{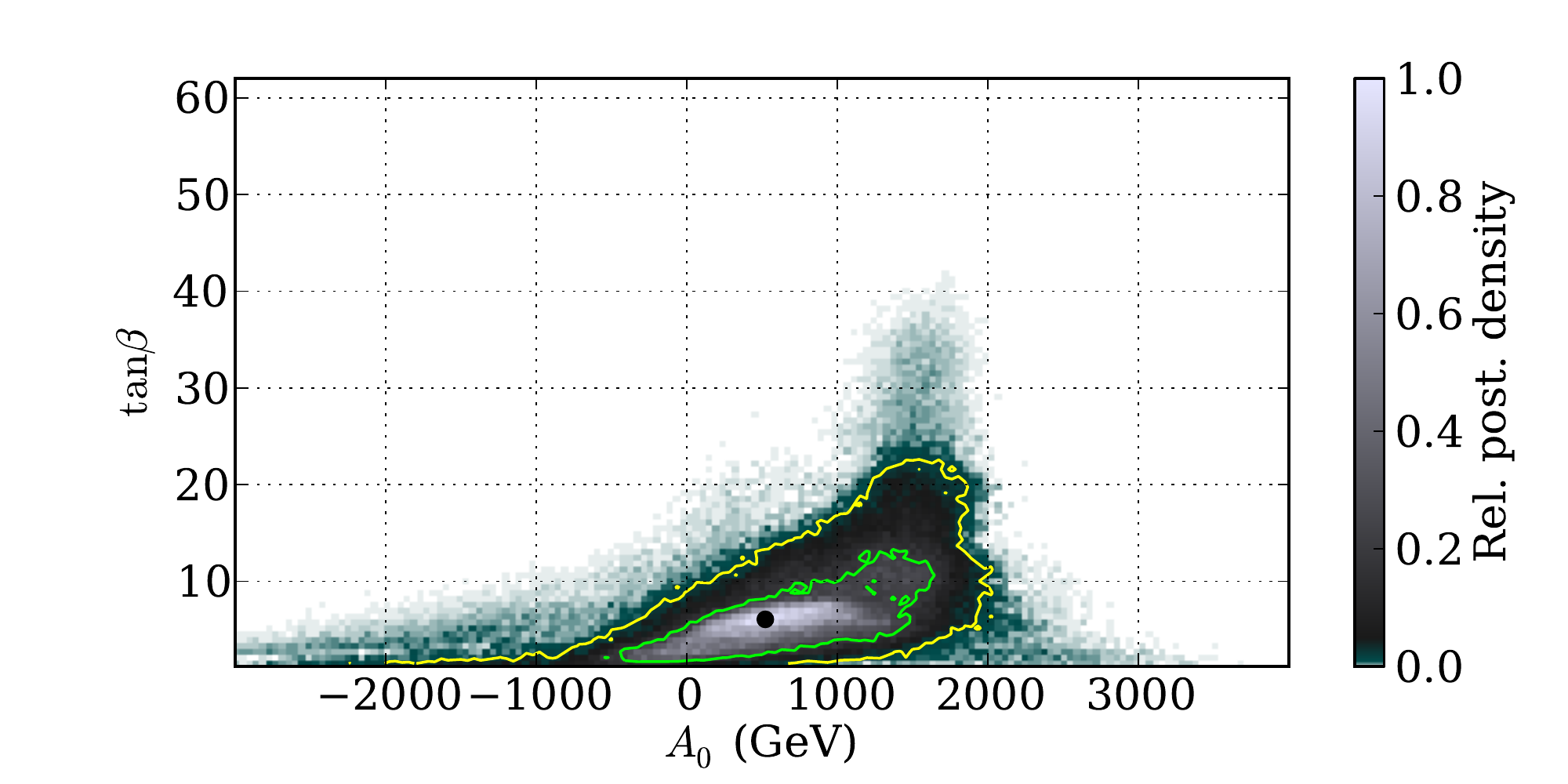}
\includegraphics[width=0.40\paperwidth]{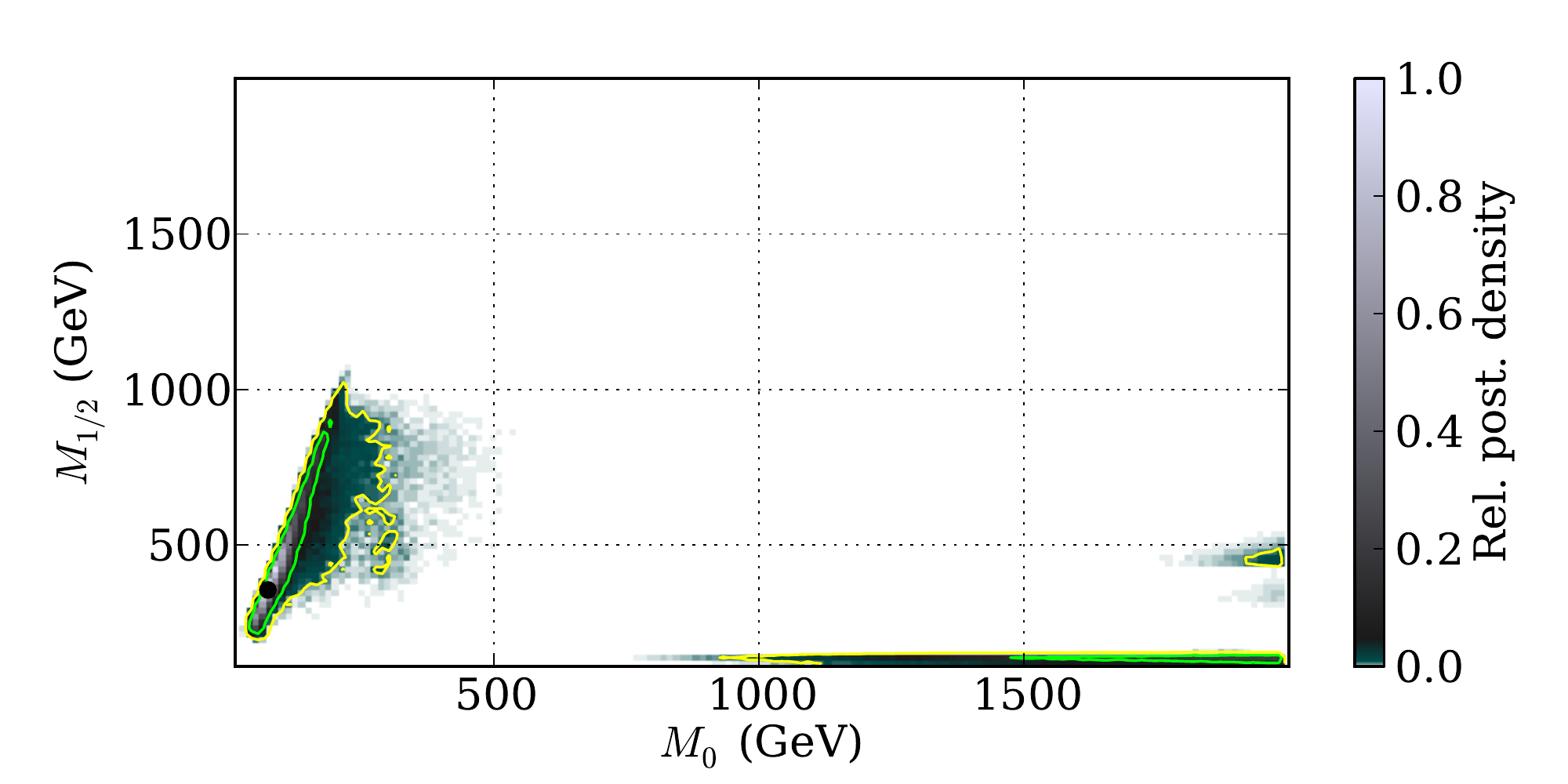}
\hspace{-16mm}
\includegraphics[width=0.40\paperwidth]{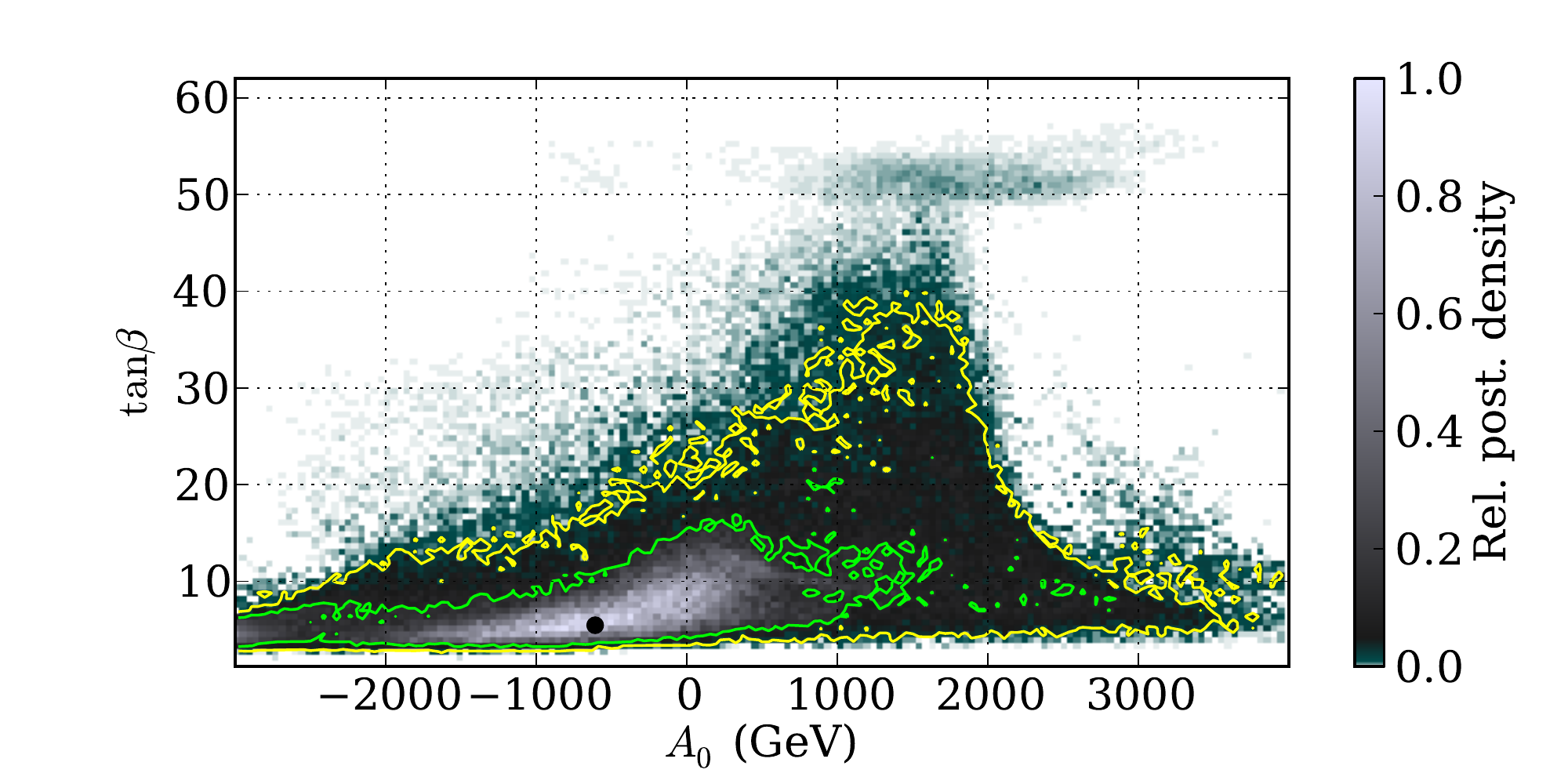}
\includegraphics[width=0.40\paperwidth]{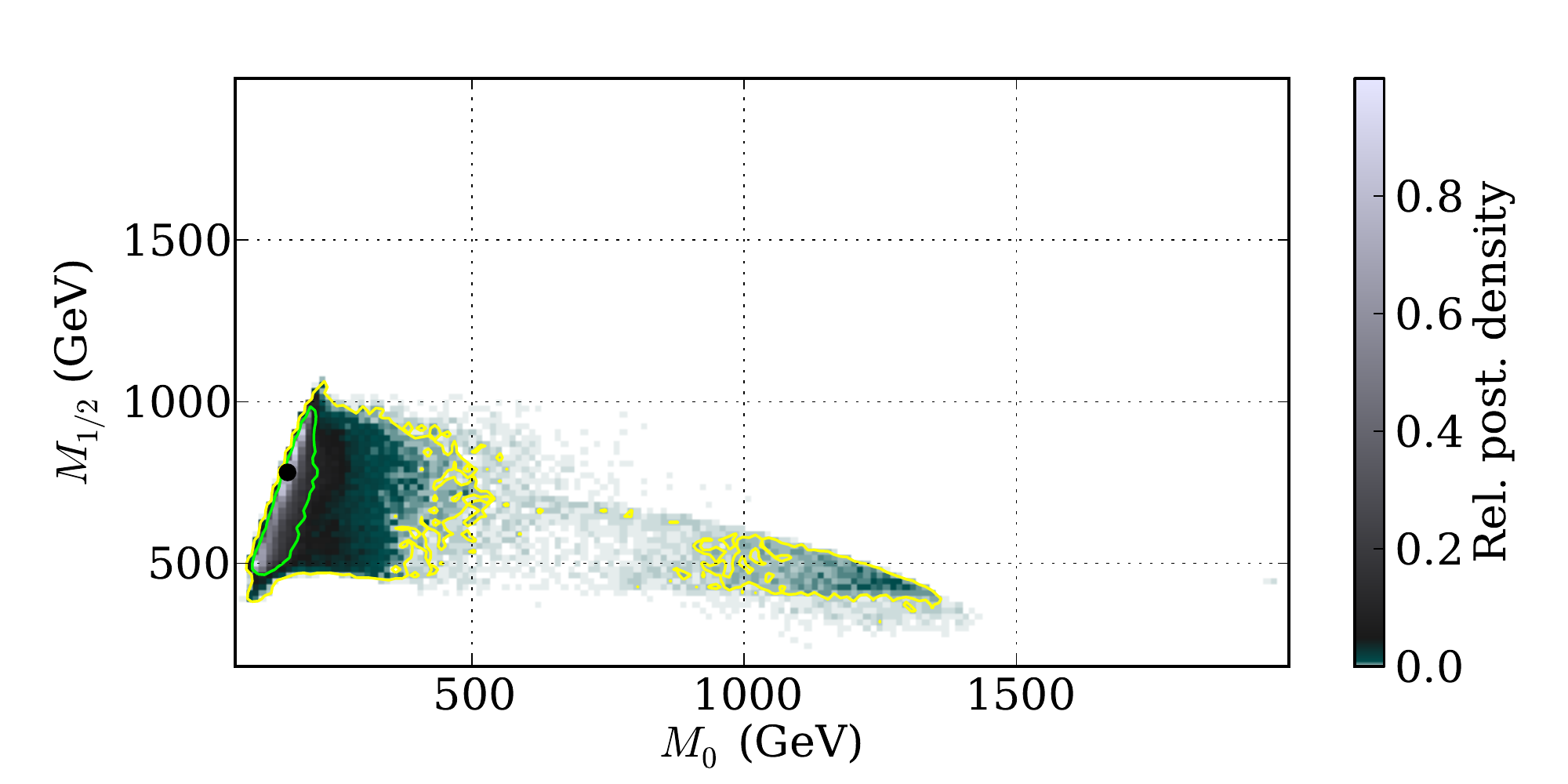}
\hspace{-16mm}
\includegraphics[width=0.40\paperwidth]{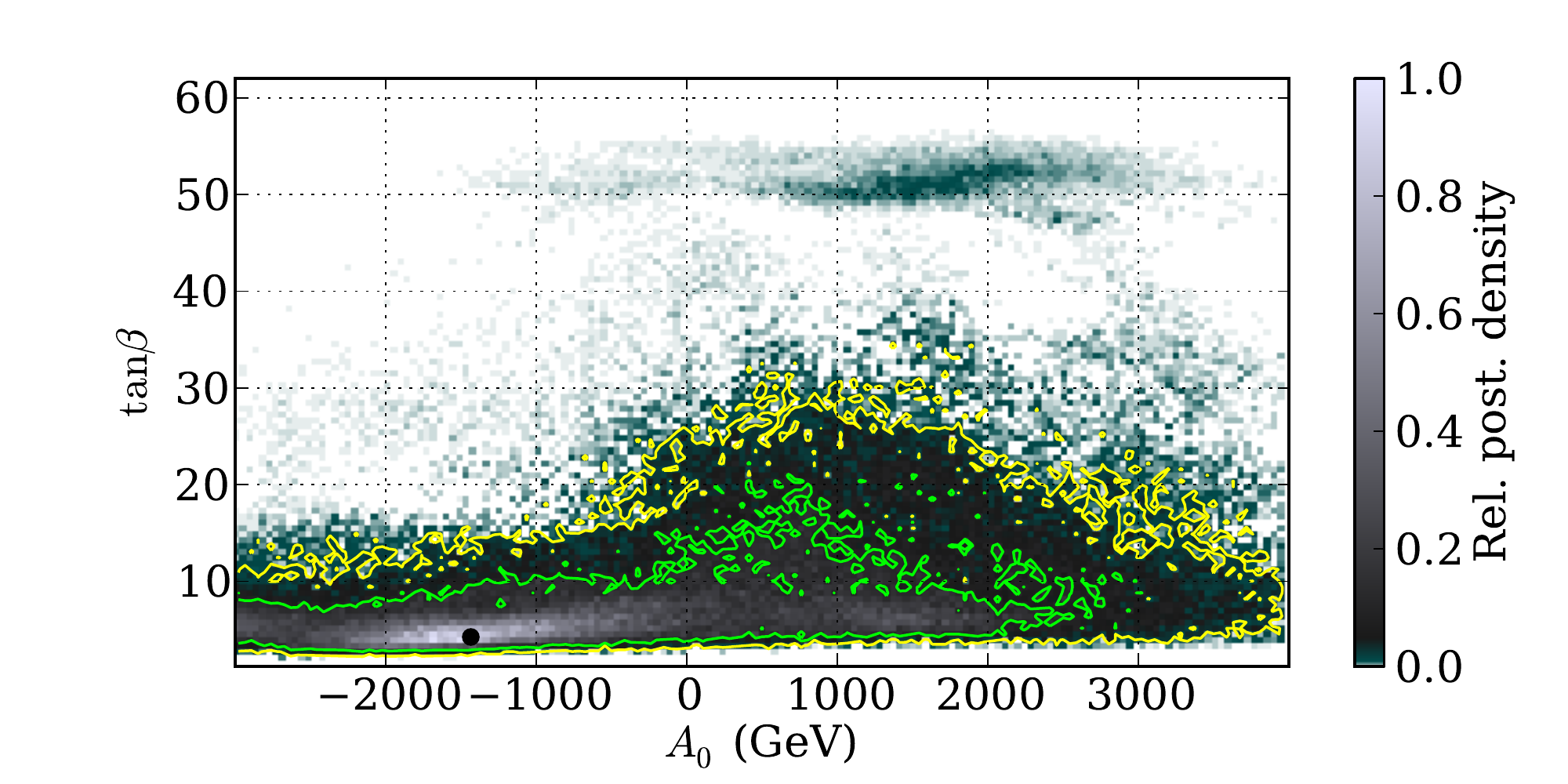}
\includegraphics[width=0.40\paperwidth]{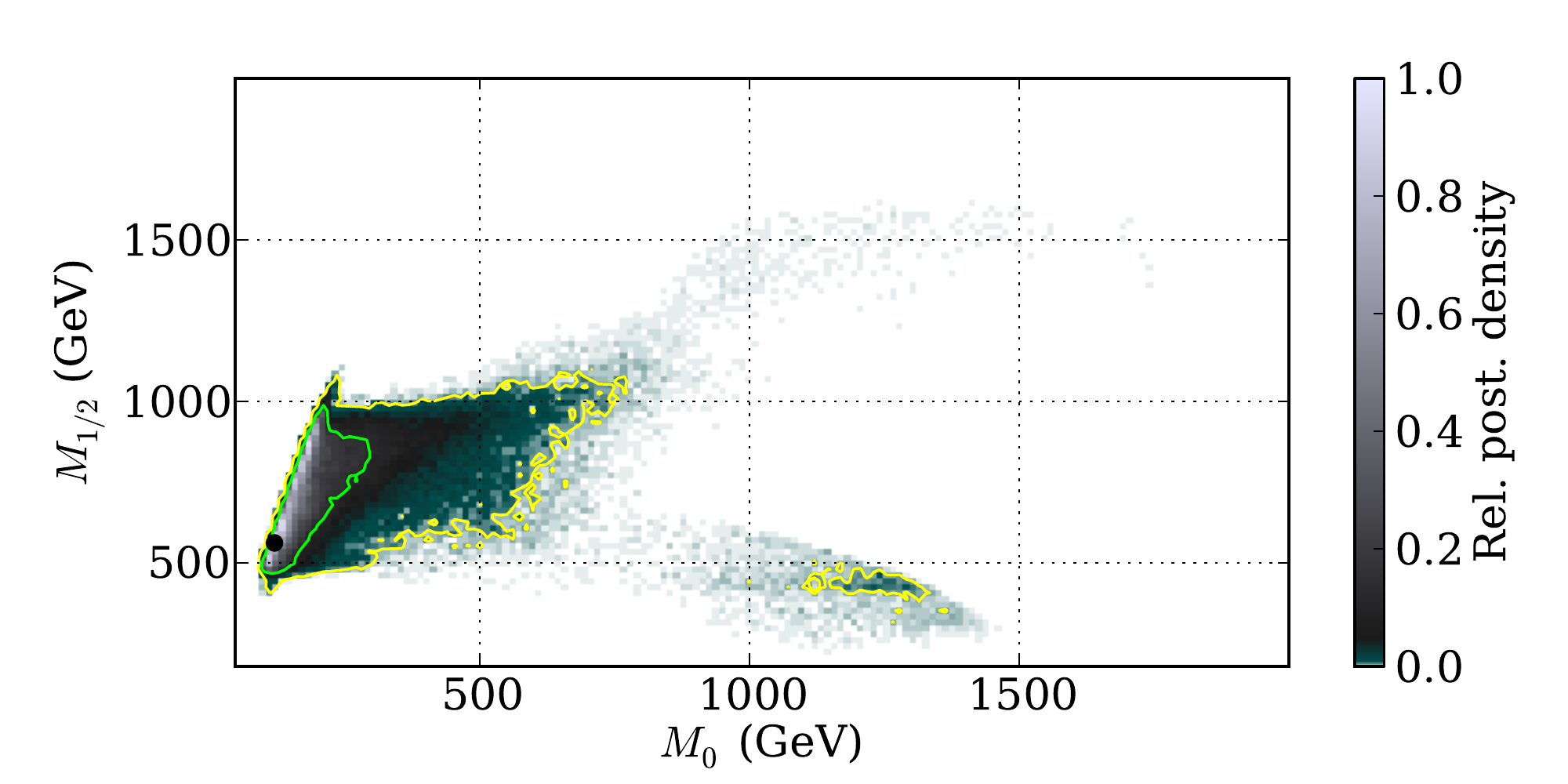}
\hspace{-16mm}
\includegraphics[width=0.40\paperwidth]{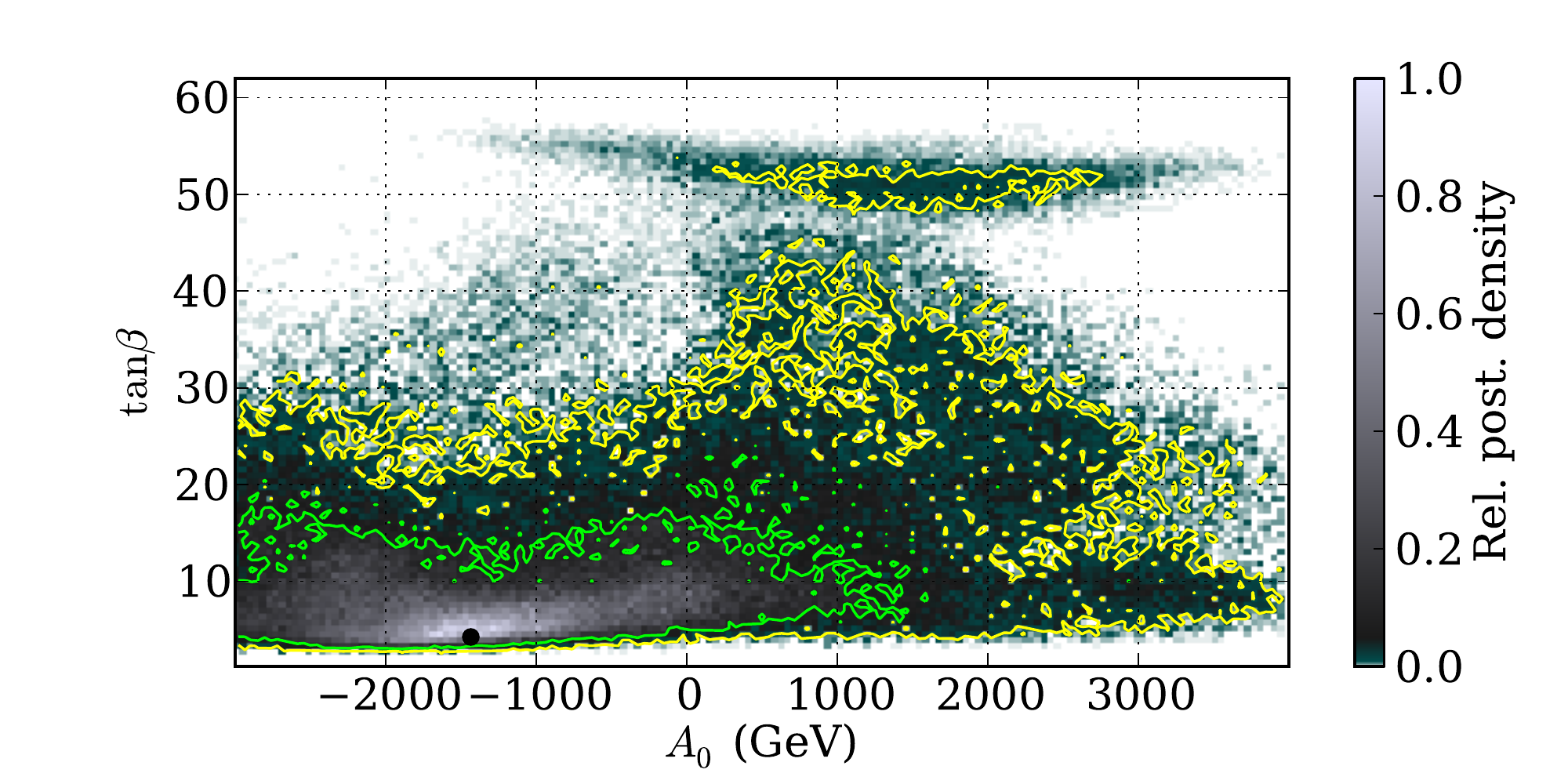}
\end{center}
\caption{The evolution of the CMSSM marginalised posterior probability distributions from the ``pre-LEP'' situation (first row), to including the LEP Higgs search and XENON100 data (second row), to adding the LHC sparticle searches (third row), to folding in the 2012 February Higgs search results.  Natural (``CCR'') priors are used and 68\% and 95\% credible regions are shown. The natural prior can be seen to favour lower $M_0$ and $\tan\beta$ than the log prior.
}
\label{fig:pos1CCR}
\end{figure*}

\end{document}